\documentclass[a4paper,10pt,fleqn]{article}
\usepackage{jheppub}
\usepackage[T1]{fontenc}
\usepackage{setspace}
\usepackage{enumitem}
\usepackage{amsmath}

\usepackage{amsthm}
\usepackage{xcolor} % adds color options
\usepackage{bm}
\usepackage[normalem]{ulem}
\usepackage{mathtools}
\usepackage{caption}
\usepackage{slashed}
\usepackage{subcaption}
\usepackage{tikz}
\usepgflibrary{arrows}
\usetikzlibrary{calc}
\usepackage{braket}

\definecolor{jlab_red}{RGB}{192,39,45}
\definecolor{jlab_orange}{RGB}{249,102,0}
\definecolor{jlab_blue}{RGB}{47,122,121}
\definecolor{jlab_green}{RGB}{65,125,10}

\hypersetup{%
pdftitle = {},
pdfsubject = {QCD},
pdfauthor = {},
colorlinks = {true},
filecolor = {black},
linkcolor = {jlab_blue},
menucolor = {black},
citecolor = {jlab_green},
urlcolor = {jlab_green},
}{}

\newcommand{\LowLyingSpectrum}[0]{Dudek:2012xn,Dudek:2014qha,Fahy:2014jxa,Wilson:2014cna,Green:2014dea,Wilson:2015dqa,Briceno:2015dca,Bolton:2015psa,Bulava:2015qjz,Dudek:2016bxq,Dudek:2016wcf,Junnarkar:2015jyf,Dudek:2016esq,Dudek:2016cru,Morningstar:2016arm,Briceno:2016kkp,Bolton:2016ptw,Bulava:2016mks,Briceno:2016mjc,Moir:2016srx,Doring:2016bdr,Wilson:2016rid,Briceno:2017max,Briceno:2017qmb,Brett:2017yhm,Andersen:2017una,Woss:2018irj,Brett:2018jqw,Molina:2018otc,Francis:2018qch,Brett:2018fdb,Andersen:2018mau,Hanlon:2018yfv,Woss:2019hse,Wilson:2019wfr,Bulava:2019hpz,Blanton:2019vdk,Erben:2019nmx,Andersen:2019ktw,Fischer:2020yvw,Cheung:2020mql,Woss:2020ayi,Hansen:2020otl,Johnson:2020ilc,Green:2021qol,Gayer:2021xzv,Blanton:2021llb,Mai:2021nul,Morningstar:2021ewk,PadmanathMadanagopalan:2021exb,Green:2021sxb,Padmanath:2022cvl,Radhakrishnan:2022ubg,Lang:2022elg,Bulava:2022vpq,Green:2022rjj,Bulava:2023wrz,Lyu:2023xro,Draper:2023boj,Rodas:2023gma,Rodas:2023nec,BaryonScatteringBaSc:2023ori,BaryonScatteringBaSc:2023zvt,Wilson:2023anv,Wilson:2023hzu,Bulava:2023uma,Skinner:2023wwb,Bulava:2024bsi,Collins:2024sfi,Yeo:2024chk,Boyle:2024hvv,Whyte:2024ihh,Dudek:2024roh,Boyle:2024grr,Yan:2024gwp,Dawid:2024dgy,Vujmilovic:2024snz,Francis:2024fwf,Erben:2025zph,Dawid:2025doq,Lang:2025pjq,Dawid:2025zxc}

\newcommand{\AllThreeBody}[0]{Polejaeva:2012ut,Hansen:2014eka,Hansen:2015zga,Briceno:2017tce,Guo:2017ism,Hammer:2017uqm,Hammer:2017kms,Mai:2017bge,Doring:2018xxx,Briceno:2018mlh,Klos:2018sen,Guo:2018ibd,Briceno:2018aml,Blanton:2019igq,Pang:2019dfe,Romero-Lopez:2019qrt,Hansen:2020zhy,Blanton:2020gha,Blanton:2020jnm,Guo:2020spn,Romero-Lopez:2020rdq,Pang:2020pkl,Blanton:2020gmf,Muller:2020vtt,Muller:2020wjo,Hansen:2021ofl,Blanton:2021mih,Muller:2021uur,Blanton:2021eyf,Muller:2022oyw,Severt:2022jtg,Jackura:2022xml,Baeza-Ballesteros:2023ljl,Draper:2023xvu,Bubna:2023oxo}

\newcommand{\MultiHadronAmps}[0]{Pelaez:2021dak,Accardi:2023chb}

\newcommand{\OldFV}[0]{Huang:1957im}

\newcommand{\AllTwoBodyFormal}[0]{Rummukainen:1995vs,Feng:2004ua,He:2005ey,Christ:2005gi,Kim:2005gf,Lage:2009zv,Bernard:2010fp,Fu:2011xz,Leskovec:2012gb,Briceno:2012yi,Hansen:2012tf,Guo:2012hv,Briceno:2013lba,Briceno:2014oea}

\newcommand{\nn}[0]{\nonumber}

\newcommand{\Mc}{\mathcal{M}}
\newcommand{\Yc}{\mathcal{Y}}
\newcommand{\Kc}{\mathcal{K}}
\newcommand{\Rc}{\mathcal{R}}
\newcommand{\Lc}{\mathcal{L}}
\newcommand{\Cc}{\mathcal{C}}
\newcommand{\Bc}{\mathcal{B}}

\newcommand{\Ec}{\mathcal{E}}
\newcommand{\Tc}{\mathcal{T}}
\newcommand{\Uc}{\mathcal{U}}
\newcommand{\Ac}{\mathcal{A}}

\definecolor{jlab_red}{RGB}{192,39,45}
\definecolor{jlab_orange}{RGB}{249,102,0}
\definecolor{jlab_blue}{RGB}{47,122,121}
\definecolor{jlab_green}{RGB}{65,125,10}

\usepackage{mfirstuc} % to uppercase the name
\newcommand{\addReviewer}[2]{
\expandafter\newcommand\csname #1\endcsname[1]{{\sf \color{#2} {#1}:\,##1}}
\expandafter\newcommand\csname #1cor\endcsname[2]{{\color{#2} {#1}:\,\st{##1}{\sf ##2}}}
\expandafter\newcommand\csname #1color\endcsname{#2}
}
\addReviewer{ar}{jlab_green}
\addReviewer{rb}{jlab_blue}
\addReviewer{mh}{jlab_orange}
\addReviewer{aj}{jlab_red}

\title{
Extracting scattering amplitudes for arbitrary two-particle systems with one-particle left-hand cuts via lattice QCD
}

\author[a]{Andr\'e Bai\~ao Raposo}
\author[b,c]{, Raul A. Brice\~no}
\author[d]{, Maxwell~T.~Hansen}
\author[e]{, and Andrew W. Jackura}

\affiliation[a]{Institut f\"ur Theoretische Physik II, Fakult\"at f\"ur Physik und Astronomie, Ruhr-Universität Bochum, 44780 Bochum, Germany}
\affiliation[b]{Department of Physics, University of California, Berkeley, CA 94720, USA}
\affiliation[c]{Nuclear Science Division, Lawrence Berkeley National Laboratory, Berkeley, CA 94720, USA}
\affiliation[d]{Higgs Centre for Theoretical Physics, School of Physics and Astronomy, The University of Edinburgh, Edinburgh EH9 3FD, UK}
\affiliation[e]{Department of Physics,
William \& Mary,
Williamsburg, VA 23187, USA}

\emailAdd{andre.baiaoraposo@ruhr-uni-bochum.de}
\emailAdd{rbriceno@berkeley.edu}
\emailAdd{maxwell.hansen@ed.ac.uk}
\emailAdd{awjackura@wm.edu}

\abstract{We derive a general formalism that relates the spectrum of two-particle systems in a finite volume to physical scattering amplitudes, taking into account the presence of any left-hand branch cuts due to single-particle exchanges. The method first relates the finite-volume spectrum to an infinite-volume short-range quantity, denoted $\Mc_0$, and then relates the latter to the physical scattering amplitudes via known integral equations. The derivation of both relations is performed using all-orders perturbation theory and is exact up to neglected exponentially suppressed volume dependence. The relations hold for arbitrary two-particle systems with any number of coupled channels, non-identical and non-degenerate particles, and any intrinsic spin.}

\begin{document}
\maketitle
\flushbottom
\abovedisplayskip 11pt
\belowdisplayskip 11pt

\section{Introduction}
\label{sec:intro}

In recent decades, remarkable progress has been made in applying numerical lattice quantum chromodynamics (QCD) to systematically and rigorously determine properties of the low-lying spectrum directly from the fundamental theory. To reach this understanding, the field has had to overcome two major obstacles:

First, because of the non-perturbative nature of QCD, analytic attempts to systematically and quantitatively understand its low-energy properties are limited, only applicable for certain observables, and with sources of systematic uncertainty that can be difficult to quantify. The lattice paradigm circumvents this challenge by placing the theory in a finite, discretized, Euclidean spacetime, where reliable numerical determinations of correlation functions can be achieved by applying Monte Carlo importance sampling on high-performance computers.

Second, the vast majority of states in QCD are unstable resonances that decay to two or more hadrons via the strong force, and their existence must be reconstructed from the analytic properties of multi-hadron scattering amplitudes.\footnote{See for example refs.~\cite{\MultiHadronAmps}, for recent reviews on the application of this to experimental scattering data.} Such amplitudes, in turn, are not directly accessible in the finite Euclidean spacetime inherent to numerical lattice QCD but can be accessed indirectly using a mathematical framework that relates them to the discrete energies of the theory in a finite periodic spatial volume. This approach was pioneered by L\"uscher \cite{Luscher:1986pf} following early work in other contexts, e.g.~ref.~\cite{\OldFV}. The method has since been generalized to treat any number of two-particle channels, allowing each channel to contain either identical or non-identical particles, non-degenerate masses, and arbitrary intrinsic spin~\cite{\AllTwoBodyFormal}. The relation between energies and amplitudes has also been extensively developed for three-particle states in refs.~\cite{\AllThreeBody}. These methods have been extensively applied in lattice QCD calculations of scattering amplitudes and resonance properties~\cite{\LowLyingSpectrum}.

This work extends the aforementioned two-particle results by treating one-particle exchanges (OPE) that can generate branch cuts (often called left-hand cuts) in angular-momentum-projected amplitudes. This is achieved by generalizing the formalism recently derived by two of us in ref.~\cite{Raposo:2023oru} for a single channel of identical particles. In addition to extending the applicability, we provide an alternative derivation and alternative equivalent forms of the main result. The derivation we present here is simpler and more direct than that given in ref.~\cite{Raposo:2023oru}, although relying on technical aspects that were thoroughly investigated in that work. Before elaborating further on the new framework and reviewing relevant previous work, we first consider a few possible applications.

Two notable examples that have received significant attention in recent years are the H-dibaryon~\cite{Jaffe:1976yi} and the tetraquark candidate $T_{cc}$~\cite{LHCb:2021vvq}:

To our knowledge, the gap in the literature concerning the role of OPE branch cuts in finite-volume scattering formulae was first explicitly pointed out in ref.~\cite{Green:2021qol} in the context of a systematic search for the H-dibaryon at unphysically heavy pion masses, $m_\pi\approx420$~MeV. In that work, the authors numerically extracted finite-volume energies of the $\Lambda \Lambda$ system both near and on top of the OPE left-hand cut generated by $\eta$-meson exchange and concluded that these had to be discarded since the formalism available at the time was only applicable in the range $(2 m_\Lambda)^2 - m_\eta^2< s < (2 m_\Lambda + m_\eta)^2$, where $s = (E^\star)^2$ is the squared center-of-momentum-frame (CMF) energy, and $m_\eta$ and $m_\Lambda$ are the $\eta$ meson and $\Lambda$ baryon masses, respectively.

The same issue is relevant for the $T_{cc}$, a narrow resonance recently observed by the LHCb experiment. This doubly-charmed resonance, which lies below the $DD^\star$ threshold and decays strongly to $DD\pi$ states, is of great interest as it is one of the best candidates for an unambiguous four-quark state. Although for physical pion masses the $D^\star$ itself is an unstable resonance decaying to $D \pi$, only a $\sim5\%$ increase in $m_\pi$ is sufficient to render it stable such that one can treat energies in the $DD^\star$ region as two-particle states. Motivated by a desire to confirm and understand the $T_{cc}$ in this simplified two-particle picture, various lattice QCD collaborations have performed calculations of the finite-volume spectrum to constrain the $DD^\star$ amplitude~\cite{Padmanath:2022cvl,Lyu:2023xro,Whyte:2024ihh, Collins:2024sfi}. These works, with the exception of ref.~\cite{Lyu:2023xro}, relied on the finite-volume paradigm outlined above. However, as was pointed out in ref.~\cite{Du:2023hlu}, the application of the two-particle formalism available at the time was only valid in the range $(m_D+m_{D^\star})^2-m_\pi^2 < s < (2m_D+m_\pi )^2$, where the lower bound of validity arises from the OPE left-hand cut arising from pion exchange in the $D D^\star \to D D^\star$ amplitude. Just as with the H-dibaryon, a consequence of this is that a set of the subthreshold finite-volume energies cannot be analyzed without first generalizing the formalism.

The impact of OPE branch cuts has been well understood for two-body scattering amplitudes for a longer time,\footnote{For a discussion on the potential impact of the left-hand cut on the $DD^\star$ amplitude in the $T_{cc}$ channel, see ref.~\cite{Du:2023hlu}.} and only the finite-volume context had not been well explored until recently. An early work related to this is ref.~\cite{Sato:2007ms}, which emphasizes the effect of single-pion exchanges in exponentially suppressed volume corrections to the L\"uscher quantization condition, using non-relativistic nuclear chiral EFT. Unlike our approach, it employs a Bethe-Salpeter kernel defined solely by the single-pion exchange potential. The identified volume effects arise from on-shell vs. off-shell differences in this potential, leading to real-valued corrections. Their study neglects partial-wave mixing and focuses on the two-nucleon system. More recently, ref.~\cite{Meng:2021uhz} proposed a plane-wave basis (instead of the usual partial-wave approach) to the L\"uscher formula, subsequently applied to $DD^*$ scattering data in ref.~\cite{Meng:2023bmz}. In ref.~\cite{Bubna:2024izx}, an approach based on non-relativistic EFT was also proposed to treat the issue of left-hand branch cuts and the associated problem of long-range potentials. This has conceptual similarities with the work presented here, and previously in ref.~\cite{Raposo:2023oru}, namely in the way that short- and long-range contributions are separated and treated. Very recently, ref.~\cite{Yu:2025gzg} has proposed using Hamiltonian Effective Field Theory to tackle the same issue of long-range potentials in numerical lattice calculations. Numerical validation and further theoretical exploration of equivalences between the proposed formalisms remain open for future work.

This formalism will not only impact the first-principles characterization of QCD resonances, but will also play an important role in future constraints on two- and three-nucleon interactions, which will help to constrain ab initio nuclear methods. A closely related application is the prediction of single and double beta decay rates, which can play a role in precision tests of the Standard Model. Currently, the lattice QCD community is still resolving the nucleon-nucleon spectrum at unphysical heavy quark masses, with the most reliable calculations in the range $400 \, {\rm MeV} \lesssim m_\pi \lesssim 700 \, {\rm MeV}$. As these calculations begin to approach the physical point, the OPE branch point at $s  = (2m_N)^2 - m_\pi^2$, where $m_N$ is the nucleon mass, will move closer to the kinematic region where finite-volume energies can be extracted. As shown in ref.~\cite{Raposo:2023oru}, this leads to either enhanced exponential finite-volume effects (near the OPE cut) or power-like finite-volume effects (on the cut), and these must be treated explicitly to obtain reliable results.

Reference~\cite{Raposo:2023oru} solves the OPE cut problem for elastic scattering of a single channel of identical particles, using $NN$ as an example. The derivation is based on a skeleton expansion which represents a finite-volume correlation function as a geometric series of Feynman diagrams with irreducible vertex functions. The loops in these diagrams are summed over the discrete set of finite-volume momentum modes and integrated over the time components; as a result, the vertex functions are sampled at off-shell momenta, i.e. four-momenta \(k\) that do not satisfy \(k^2 = m^2\). In the usual derivations of finite-volume scattering formulas, the vertex functions can be expanded around their on-shell points, and the off-shell dependence can be absorbed into new infinite-volume quantities up to exponentially suppressed volume effects. However, as highlighted in ref.~\cite{Raposo:2023oru}, this on-shell placement creates problems on the OPE cut. By using partially off-shell kinematics to avoid the corresponding singularities, an alternative relation is derived that remains valid down to the two-pion exchange branch cut starting at $s = (2m_N)^2 - (2m_\pi)^2$. The new relation allows to constrain an intermediate quantity \( \overline{\mathcal K}{}^{\sf os}(s) \), which can be related to the scattering amplitude via integral equations also derived in the published work.

In addition to the extension to any number and type of two-particle channels, in this work, we also provide some technical results that are important for the practical implementation of the formalism. In particular, we use a recently developed formalism for the partial-wave projection of integral equations for three-particle systems~\cite{Jackura:2023qtp, Briceno:2024ehy} to derive improved integral equations for the two-particle sector that are partial-wave projected.

The need to generalize the formalism of ref.~\cite{Raposo:2023oru} is not a mere academic exercise. There is already a lattice QCD calculation~\cite{Whyte:2024ihh} of a coupled-channel system ($DD^*-D^* D^* $) with energy levels near the nearest left-hand cut. Although this analysis did not explicitly account for OPE effects, it did find evidence for the presence of the $T_{cc}$ as a virtual bound state, as well as the first evidence for its excited state, the $T_{cc}'$.

We also note that for some systems, such as the $T_{cc}$, an alternative approach to resolving the presence of the left cut has been proposed. In particular, it is possible to describe such systems in terms of three-particle states that are analytically continued below threshold~\cite{Romero-Lopez:2019qrt,Briceno:2024txg,Hansen:2024ffk,Briceno:2024ehy,Dawid:2024dgy}. This approach has been shown to automatically treat the OPE cuts explicitly. An additional advantage is that it may allow a global analysis in a larger kinematic region beyond the nearest three-particle thresholds. For other systems, such as $NN$ and $\Lambda \Lambda$, it is less obvious whether it will be as advantageous to apply the three-particle perspective, and the work presented here offers a useful and perhaps simpler alternative.

This manuscript is organized as follows. In section~\ref{sec:QC_derivation}, we present a derivation of the quantization condition for any number of channels, without intrinsic spin. In section~\ref{sec:QC_spin}, we generalize this to include spin. In section~\ref{sec:equivalence}, we prove that the quantization condition presented here is equivalent to one in ref.~\cite{Raposo:2023oru}, when one restricts to the case of a single channel of identical particles. The quantization condition requires only one set of dynamical inputs, namely the couplings associated with the OPE, and the finite-volume spectrum. From this, one can constrain an infinite-volume intermediate quantity denoted $\Mc_0$. In section~\ref{sec:scattering_amps}, we explain the relation between $\Mc_0$ and the physical amplitude. In the limit that the OPE couplings are fixed to 0, $\Mc_0$ is the scattering amplitude. Otherwise, the relation requires solving a class of integral equations. We give exact expressions for the integral equations for any two-body scattering amplitude that has been partial-wave projected in terms of the partial-wave projected OPE. Finally, in section~\ref{sec:OPE}, we give a prescription for partial-wave projecting the OPE and give exact expressions for two classes of systems, spinless-spinless, and spinless-vector.

\section{Finite-volume quantization conditions}
\label{sec:QC}

\begin{figure}[t]
\centering
\includegraphics[width=.7\textwidth]{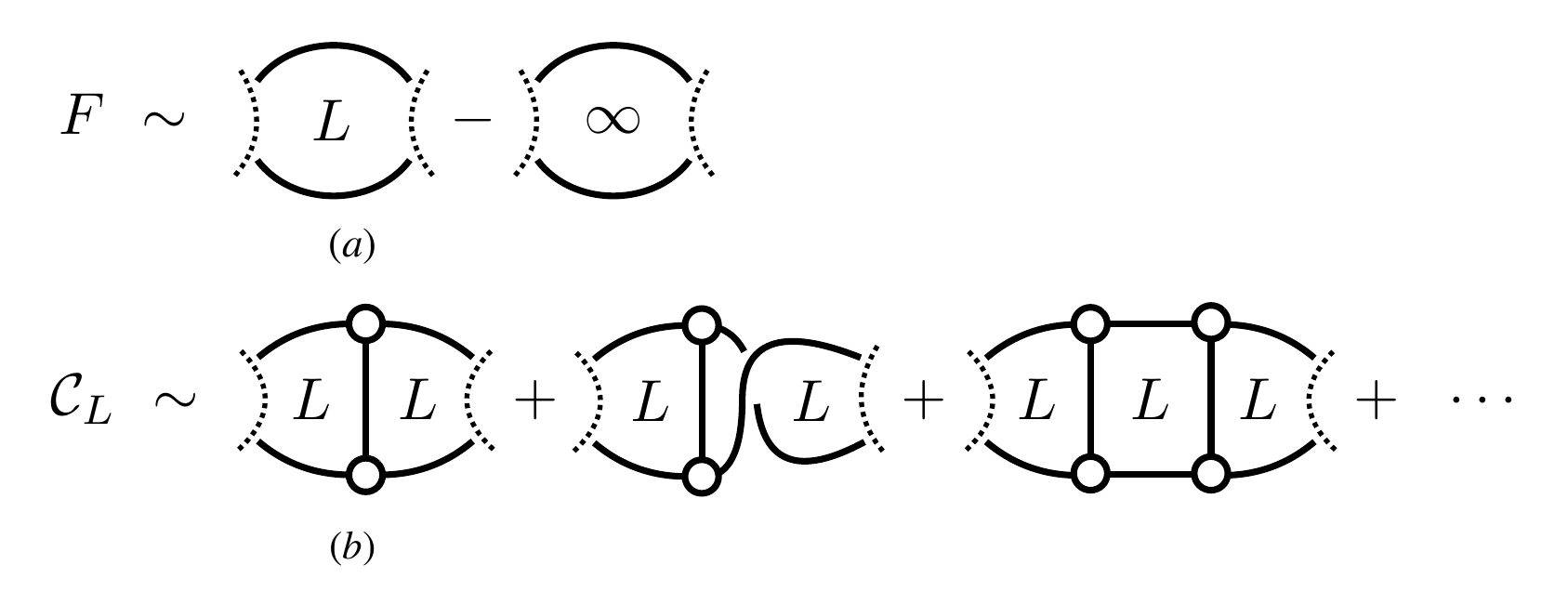}
\caption{Schematic representations of the finite-volume functions appearing in the quantization condition of eq.~\eqref{eq:QC_cc_spin_v0}. $F$ is the standard L\"uscher function encoding the difference between finite- and infinite-volume two-particle loops, labeled $L$ and $\infty$, respectively. $\Cc_L$ is a finite-volume quantity corresponding to the sum of any number of OPEs ($t$- and $u$-channel) connected by two-particle loops. Detailed definitions of $F$, $\Cc_L$, and their building blocks are given later in this section.}
\label{fig:QC_objects}
\end{figure}

We begin by stating the first of our two main results of this work, the quantization condition relating finite-volume energies to intermediate infinite-volume quantities. We find it useful to lead with this to emphasize the essential message before going into details of the derivation. Our main result for a generic system with coupled two-particle channels is that, up to neglected exponentially suppressed effects, the finite-volume energies satisfy
\begin{align}
\det_{aJ m_J \ell S} \big[\Mc_{0}^{-1} + F + \Cc_L \big] = 0
\label{eq:QC_cc_spin_v0} \,.
\end{align}
As with many previous results, our quantization condition takes the form of a determinant over a combined index space, defined by the degrees of freedom that are not fixed with the total energy.
We choose the $a J m_J \ell S$-basis, where $a$ is the flavor channel, $J m_J$ is the total angular momentum, $\ell$ is the orbital angular momentum, and $S$ is the total intrinsic spin.

The determinant depends on three essential quantities: (i) $\Mc_0$ is the intermediate infinite-volume scattering quantity to be determined from the finite-volume spectrum. While its general definition is technical, we note that it coincides with the standard physical scattering amplitude in the limit that all OPE couplings vanish. (ii) $F$ is the previously derived generalization of the L\"uscher finite-volume function for arbitrary channels including spin~\cite{Briceno:2014oea}. Therefore, all modifications due to the OPE are embedded in (iii) the new finite-volume function $\Cc_L$, which also vanishes when all OPE couplings go to zero. We give a schematic definition of $F$ and $\Cc_L$ in figure~\ref{fig:QC_objects}. The exact definitions of each one of these objects are in the main body of the subsections below, in particular in eqs.~\eqref{eq:F_def} and \eqref{eq:CL}.

\subsection{Derivation for coupled-channel systems with no intrinsic spin}
\label{sec:QC_derivation}

In this section, we derive a quantization condition in a complementary but equivalent fashion to ref.~\cite{Raposo:2023oru}, extending the results of that paper to non-identical particles and multiple scattering channels.

Our main task in this section is to obtain an expression for the finite-volume analog of the two-body scattering amplitude, $\Mc_L$. In the context of a generic effective theory of hadrons, this object is defined through the same expansion in terms of Feynman diagrams as the physical infinite-volume amplitude, $\Mc$, but evaluated instead in a periodic cubic finite spatial volume of side length $L$. In a finite volume, loop integration over continuous spatial momentum is replaced by a sum over the discrete set of spatial momenta allowed by the periodic boundary conditions, $\mathbf k = (2\pi/L) \mathbf n$ with $\mathbf n \in \mathbb Z^3$:
\begin{equation}
\int \! \frac{{\rm d}^3 {\bf k}}{(2\pi)^3} ~ \longrightarrow ~ \frac{1}{L^3} \sum_{\mathbf k} \,.
\end{equation}
The infinite-volume amplitude $\Mc$ is recovered from $\Mc_L$ by introducing an $i\epsilon$ prescription on the energy $E \to E + i\epsilon$, and taking the infinite-volume limit $L \to \infty$, so that the sums over discrete momentum become Feynman integrals.

In processes involving multiple open two-particle scattering channels, we look at the finite- and infinite-volume amplitudes as matrices in the space of channels. For example, we take $\Mc$ to have entries $\Mc_{a'a}$, where the indices $a$ and $a'$ specify the channels of the incoming and outgoing states, respectively.

By first deriving an expression for $\Mc_L$, we are then able to obtain a quantization condition for the finite-volume spectrum. This will also help us derive expressions satisfied by the physical infinite-volume amplitude, which we present in section~\ref{sec:scattering_amps}. As in ref.~\cite{Raposo:2023oru}, we restrict our attention to the kinematic region bounded from below by the nearest left-hand cut associated with two or more particle exchanges (across all channels considered) and bounded from above by the threshold for production of $n$-particle states, where $n\geq 3$.

Power-law finite-volume corrections to observables result from singularities in the physical region. In a finite volume, these singularities can only be poles arising from intermediate states going on shell. Other finite-volume corrections are exponentially suppressed as $\mathcal O\left( e^{-\mu L}\right)$, where $\mu$ is typically the mass of the lightest particle that can be exchanged, and are generally neglected. The goal throughout this derivation is to isolate all possible pole contributions and thus track the power-law volume dependence.

\begin{figure}[t]
\centering
\includegraphics[width=.9\textwidth]{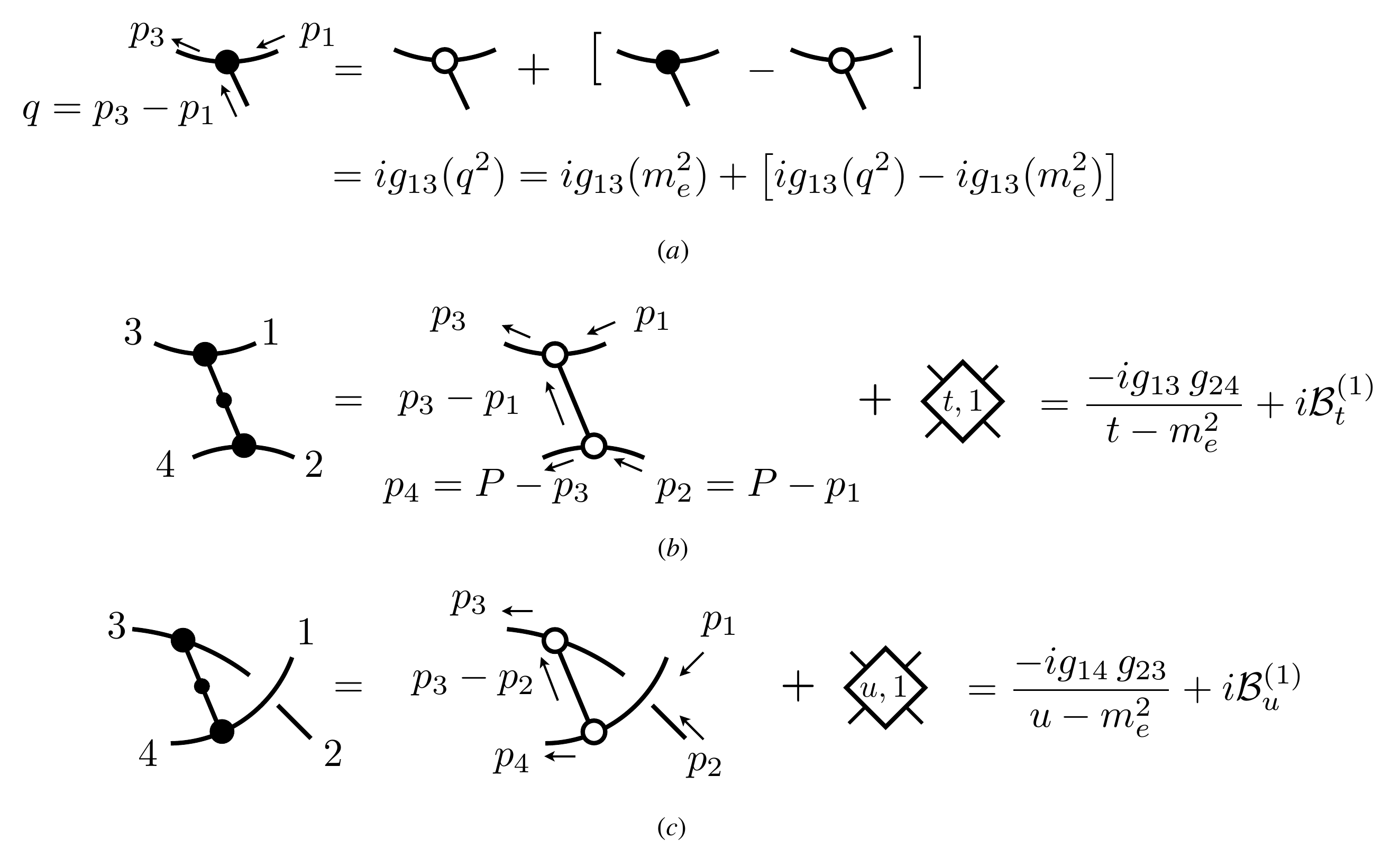}
\caption{($a$) Shown is the on-shell projection of the vertex function for scalar particles, as discussed in the main text. ($b$) and ($c$) show the diagrammatic representations of the on-shell projection for the $t$ and $u$ exchanges defined in eqs.~\eqref{eq:Box_t_on} and \eqref{eq:Box_u_on}, respectively. For all diagrams, the dark circles denote fully dressed vertices and propagators. The open circle vertices denote their on-shell limits. }
\label{fig:t-cut_OPE}
\end{figure}

For simplicity, we first assume all particles involved are spinless. This assumption is lifted in section~\ref{sec:QC_spin}. In the kinematic region considered, the only left-hand cuts of concern result from the partial-wave projection of one-particle exchanges (OPE), as depicted in figure~\ref{fig:t-cut_OPE}. In general, this will have two contributions, from $t$- and $u$-channel exchanges. These are shown in figure~\ref{fig:t-cut_OPE}($b$) and figure~\ref{fig:t-cut_OPE}($c$), respectively.

Before giving explicit expressions for these exchanges, it is crucial to unambiguously define what is meant by $t$- and $u$-channel throughout this work. For this, we ensure there is a consistent labeling scheme for all incoming and outgoing particles across all diagrams and channels considered. For a given diagram, like the exchanges shown in figure~\ref{fig:t-cut_OPE}, we choose the particles in the top incoming and outgoing legs, labeled 1 and 3 in the figure, to be the ``most like'' particles between incoming and outgoing states. This also fixes the bottom legs, labeled 2 and 4. We must then stick to this assignment for all diagrams and sub-diagrams considered.

Let us illustrate this by looking at a few examples. In a single-channel elastic process such as $DD^* \to DD^*$, the obvious choice is to use 1 and 3 to label the same species, say the heavier $D^*$, and 2 and 4 for the remaining species, in this case $D$. Note that, with this choice, we have only $u$-channel exchanges and no $t$-channel contribution. For identical particles, such as the example of $NN \to NN$ discussed in ref.~\cite{Raposo:2023oru}, all external legs correspond to the same species and the choice is trivial. In this case, we have contributions from both $t$- and $u$-channel exchanges. For coupled-channel scattering, we need to consider also cross-channel diagrams. Looking, for example, at isospin-1/2 coupled $D\pi - D\eta - D_s K$ scattering, a natural option is to pick the most similar particles across the three channels, in this case the $D$ and $D_s$ mesons, to always be on the top legs.

Specifying a labeling scheme as outlined above gives a clear definition of single $t$- and $u$-channel exchanges, which we now describe in detail. These contributions can be written in terms of two classes of functions: The first type is the fully-dressed propagator $\Delta_x (q^2)$, defined through
\begin{equation}
\Delta_x (q^2) = \frac{1}{q^2 - m_x^2 - \Pi_x(q^2) + i \epsilon} \,,
\label{eq:Deltax}
\end{equation}
where $x$ denotes the particle species, $m_x$ is its pole mass and $\Pi_x(q^2)$ its self-energy. In a finite volume, the self-energy will carry exponentially suppressed $L$ dependence. We neglect this and always use the infinite-volume object since the finite-volume corrections are exponentially suppressed. We use on-shell renormalization, demanding that $\Pi_x(q^2)$ satisfies $\Pi_x(m_x^2) = d\Pi_x/dq^2 (m_x^2) = 0$, such that $m_x$ corresponds to the physical pole mass and the single-particle pole has unit residue. The second class of function we need to consider is the dressed vertex function, represented diagrammatically in figure~\ref{fig:t-cut_OPE}($a$). Because we are considering spinless particles in this section, these vertex functions must be Lorentz scalars. We write it in full generality in terms of Lorentz invariants as $g_{xy}(p_y^2,q^2,p_x^2)$, where $q = p_y - p_x$ is the momentum transfer and $x$ and $y$ label the incoming and outgoing particle species that couple to the exchanged particle.

Let us look first at an off-shell tree-level $t$-channel contribution. With momentum assignments as shown in figure~\ref{fig:t-cut_OPE}$(b)$, this can be written as
\begin{align}
i\Mc_{t;\, \rm off,off}^{(1)} = ig_{13}(p_3^2, t,p_1^2)\,i\Delta_e(t) \,ig_{24}(p_4^2,t,p_2^2) \,,
\end{align}
where the numeric labels on the vertex functions denote the particle types on the external legs and $e$ refers to the exchanged particle. Note that momentum conservation dictates that $p_2 = P-p_1$ and $p_4=P-p_3$, where $P = (E, \mathbf P)$ is the total four-momentum and $t$ is the usual Mandelstam invariant defined through $t = (p_3 - p_1)^2 $. In the limit where the external legs are placed on-shell, the vertex functions depend solely on $t$, and we can suppress the dependence on the external masses to write
\begin{align}
i\Mc_{t;\, \rm off,off}^{(1)} = ig_{13}(t)\,i\Delta_e (t) \,ig_{24}(t).
\end{align}
This quantity is represented in figure~\ref{fig:t-cut_OPE}(b).

Next, we isolate the singular contribution from this diagram in terms of on-shell quantities, for which we use two identities. The first is represented diagrammatically in figure~\ref{fig:t-cut_OPE}(a). More explicitly, for the coupling between particles $1$ and $3$ via the exchange particle, it reads
\begin{align}
g_{13}(t) &= g_{13}(m_e^2) + \left[g_{13}(t)-g_{13}(m_e^2)\right] \,,
\end{align}
where $m_e$ is the mass of the exchanged particle. Because $m_e^2$ is a constant, we can further suppress its dependence to rewrite this identity as
\begin{align}
g_{13}(t) &= g_{13} + \left[g_{13}(t)-g_{13}\right] \,.
\end{align}
A similar relation holds of course for the coupling between particles $2$ and $4$. The second identity applies to the dressed propagator and reads
\begin{align}
i\Delta_e (t)
=\frac{i}{t-m_e^2 + i\epsilon} + iS_e(t)
\equiv iD_e(t) + iS_e(t) \,,
\end{align}
where $S_e$ is a smooth function in the kinematic region considered, and the second equality defined $D_e$ as the pole piece of the propagator.

With these two identities, we can now write the $t$-channel contribution as,
\begin{align}
i\Mc_{t;\, \rm on,on}^{(1)}
& = ig_{13}\,iD_e(t) \,ig_{24} + i\Bc_{t}^{(1)} \,, \\
& \equiv i\Tc_{\rm on,on} + i\Bc_{t}^{(1)} \,,
\label{eq:Box_t_on}
\end{align}
where $\Bc_{t}^{(1)}$ is a contribution to a two-body kernel, which is by definition non-singular in the kinematic region considered. In the second line, we have introduced a notation for the pole piece. Note that this differs from the similar notation used in ref.~\cite{Raposo:2023oru} in that the couplings resulting from the vertex functions are included within this object. Following the same arguments for the $u$-channel contribution, one similarly finds
\begin{align}
i\Mc_{u;\, \rm on,on}^{(1)}
&= ig_{14}\,iD_e(u) \,ig_{23} + i\Bc_{u}^{(1)} \,,
\\
&\equiv i \, \Uc_{\rm on,on} + i\Bc_{u}^{(1)} \,,
\label{eq:Box_u_on}
\end{align}
where $u$ is also the usual Mandelstam invariant, defined through $u \equiv (p_4 - p_1)^2 = (P - p_3 - p_1)^2 $.

As we will see in section~\ref{sec:scattering_amps}, the $t$- and $u$-channel pole terms will serve as a driving term for the key integral equation obeyed by the physical infinite-volume amplitude. Note also that, in figure~\ref{fig:t-cut_OPE}, $\Bc_{t}^{(1)}$ and $\Bc_{u}^{(1)}$ are depicted as diamonds with the labels $t,1$ and $u,1$ respectively.

It is worth noting that placing the external legs on-shell only simplified the expressions by dropping the dependence on the external variables. If we simply keep these implicit, we arrive at essentially identical expressions for $\Mc_{t;\,\rm off,off}^{(1)} $ and $\Mc_{u;\,\rm off,off}^{(1)}$:
\begin{align}
i\Mc_{t;\,\rm off,off}^{(1)}
&= i\Tc_{\rm off,off}+i\Bc_{t;\,\rm off,off}^{(1)} \,,
\label{eq:Box_t_off}
\\
i\Mc_{u;\,\rm off,off}^{(1)}
&= i \, {\Uc}_{\rm off,off}+i\Bc_{u;\,\rm off,off}^{(1)} \,.
\label{eq:Box_u_off}
\end{align}
In the derivation that follows, we will start with these objects, where all external legs are off-shell. To avoid clutter of notation, we will leave the ``off,off'' subscript implicit going forward, making it explicit when placing a particle on-shell.

To consider multiple channels, each of the blocks in eqs.~\eqref{eq:Box_t_on} to \eqref{eq:Box_u_off} is promoted to a matrix in channel space. If there is no $t$-channel contribution for a particular channel combination, say the $a \to a'$ amplitude, then the element $\Tc_{a'a}$ is taken to be zero, and similarly for $u$-channel contributions.

\begin{figure}[t]
\centering
\includegraphics[width=.9\textwidth]{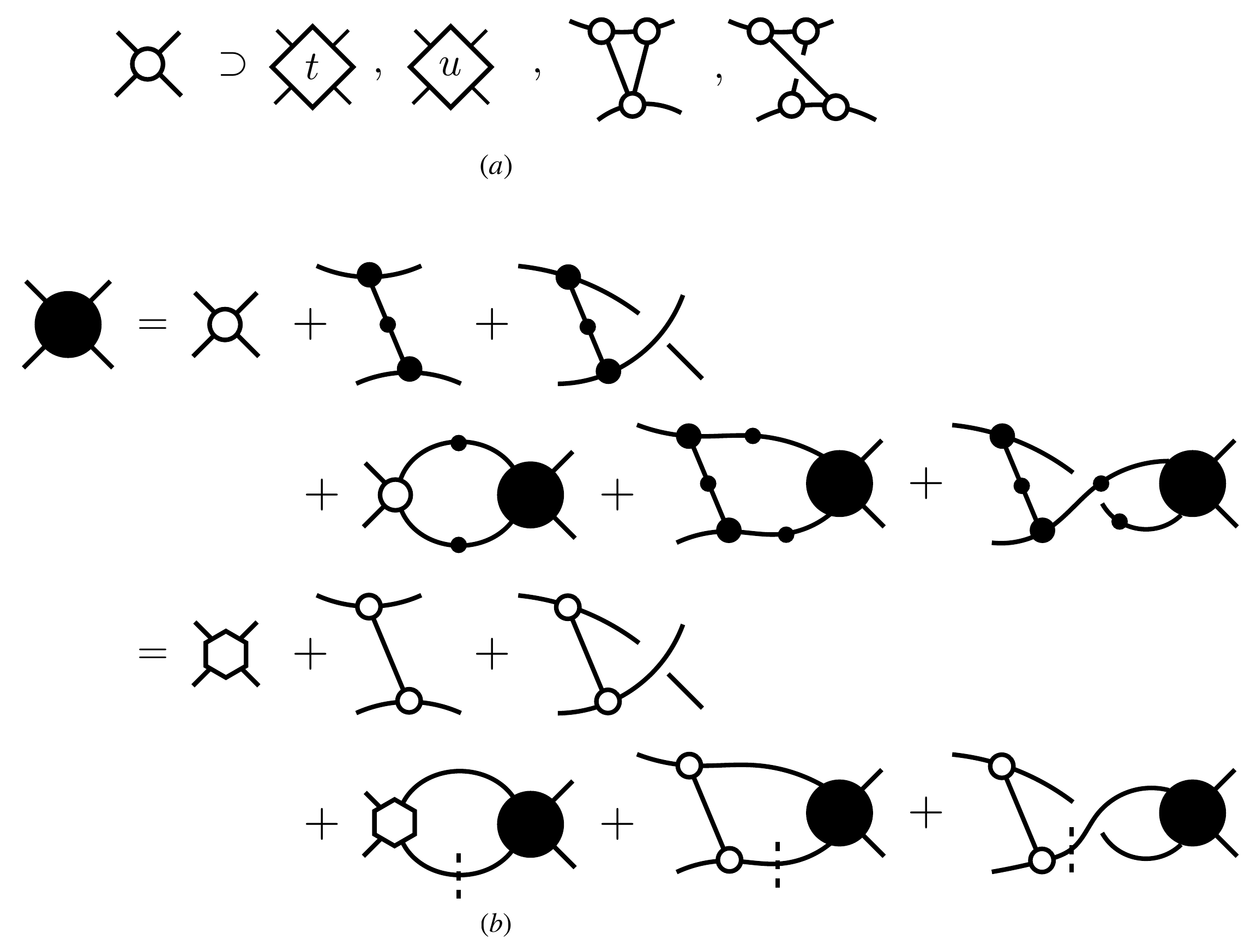}
\caption{($a$) The open circle is the Bethe-Salpeter kernel where the $t$ and $u$ exchanges have been removed. Shown are examples of contributions to this quantity.
($b$) Shown is the diagrammatic definition of the full amplitude, which includes the sum over all diagrams. The second equality refers to the partial on-shell projection written in eq.~\eqref{eq:ML_full}. The dashed line highlights the particle inside the loop that has been placed on-shell. For identical particles, the double counting in the $t$ and $u$ diagrams within loops is compensated by the symmetry factor appearing in the definition of the $\Delta_{2,L}$, in eq.~\eqref{eq:Delta2L_def}.}
\label{fig:kernel_amp}
\end{figure}

As described in ref.~\cite{Kim:2005gf}, the diagrams contributing to the amplitude can be organized into a series in terms of Bethe-Salpeter kernels and pairs of dressed propagators, as shown in figure~\ref{fig:kernel_amp}($b$). The first term is composed of a single kernel and, for each successive term, we add another kernel connected by a pair of dressed propagators. This series puts into evidence where two-particle intermediate states can arise, namely the propagator pairs, exposing the possible sources of power-like volume effects.

The Bethe-Salpeter kernel used is defined as the sum of all diagrams contributing to the amplitude which are two-particle irreducible in the $s$-channel, i.e.~those which contain no intermediate two-particle states. As a result, the infinite- and finite-volume versions of this object differ only by exponentially suppressed volume corrections. We will neglect these in the following derivation, always considering the infinite-volume kernel. Note that, according to the definition above, the Bethe-Salpeter kernel includes the contributions from $t$- and $u$-channel exchanges, $i\Mc_{t}^{(1)}$ and $i\Mc_{u}^{(1)}$. Due to their role in the generation of the nearest left-hand cut, it is convenient to define a subtracted kernel, $\Bc^{(1)}$, from which we have stripped the pole pieces of the exchanges, $i\Tc$ and $i\,\Uc$. This is shown diagrammatically in figure~\ref{fig:kernel_amp}($a$). By construction, $\Bc^{(1)}$ contains only the non-singular pieces of the exchanges, $i\Bc_t^{(1)}$ and $i\Bc_u^{(1)}$. Therefore, when placed fully on shell, it is non-singular in the kinematic region considered.

Before giving an all-orders expression for the amplitude based on the series described, let us examine second-order contributions to the finite-volume amplitude, i.e.~contributions with a single two-particle loop between two kernels. Our goal is to analyze this loop in finite volume and isolate the two-particle intermediate state singularity, which corresponds to the source of power-law volume dependence.

\begin{figure}[t]
\centering
\includegraphics[width=.9\textwidth]{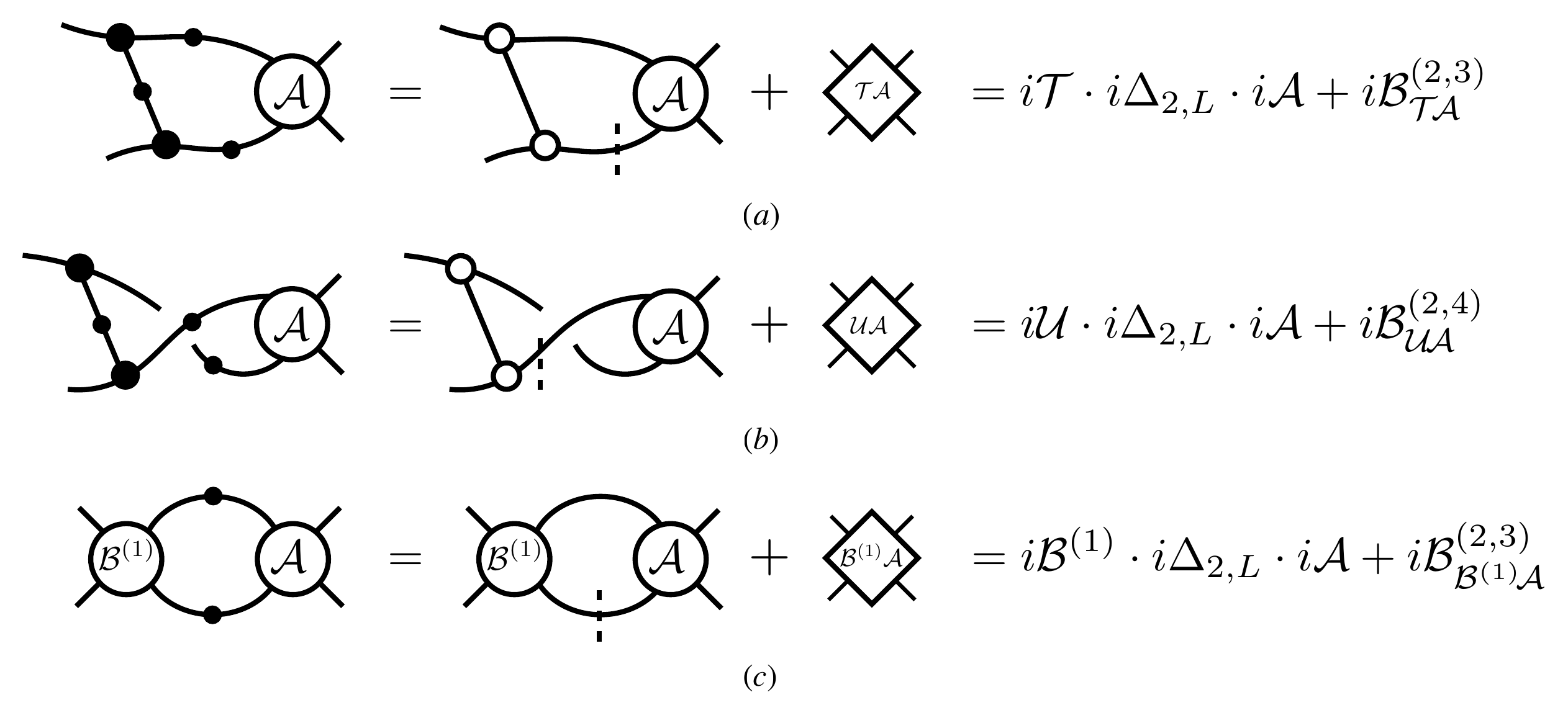}
\caption{Depicted as the partial on-shell projection appearing at the one-loop level described in the text. Here $\Ac$ is a generic vertex on the right-hand side. }
\label{fig:box_diagram}
\end{figure}

Let us consider the diagram with a $t$-channel exchange pole piece only $\Tc$ on the left-hand side and a generic kernel $\Ac$ on the right-hand side, which can stand for a $t$- or $u$-channel exchange or a non-singular kernel $\Bc^{(1)}$. This is pictured in figure~\ref{fig:box_diagram}(a). In the finite volume, we can write this contribution as
\begin{multline}
\big( i\Mc_{\Tc\!\Ac,L}^{(2)} \big)_{a'a}
= \sum_{b} \int\! \frac{{\rm d}k_0}{2\pi} \frac{1}{L^3}\sum_{\mathbf{k}}
\bigg( i\Tc_{a'b} (p_3,p_4;k,P-k) \\ \xi_b \, iD_{b1}(k^2) \, iD_{b2}((P-k)^2) \,
i\Ac_{ba} (k,P-k;p_1,p_2) \bigg)
+ \big( i\Bc_{\Tc\!\Ac}^{(2,1)} \big)_{a'a}
\,.
\label{eq:tbox_v1}
\end{multline}
In the first term, we have kept only the dressed propagator pole pieces $D$, as only these will give rise to the two-particle on-shell singularity. For clarity, we have made all kinematic variables and channel indices explicit. The index $b$ specifies the intermediate two-particle channel in the loop, with $b1$ and $b2$ identifying the particle types in each propagator and $\xi_b$ denoting the symmetry factor of the channel, set to 1/2 for identical particles and 1 otherwise. The smooth propagator pieces $S$ yield contributions that we collect into the object $\Bc_{\Tc\!\Ac}^{(2,1)}$. Strictly, this kernel and those in the following analysis carry volume dependence. We neglect this and treat them as infinite-volume objects since their finite-volume corrections are exponentially suppressed.

As discussed extensively in the literature, intermediate two-particle states that can go on shell give rise to pole singularities in the energy after integration over the zero-component of loop momentum. Performing the integral over $k^0$ in eq.~\eqref{eq:tbox_v1}, we obtain
\begin{multline}
\big( i\Mc_{\Tc\!\Ac,L}^{(2)} \big)_{a'a}
= \sum_b \frac{1}{L^3} \sum_{\mathbf{k}}
\bigg[ i\Tc_{a'b} (p_3,p_4;k,P-k) \, \xi_b \, \frac{1}{2\omega_{b1}(\mathbf{k})} \, iD_{b2} ((P-k)^2) \, \\
i\Ac_{ba} (k,P-k;p_1,p_2) \bigg]_{k^0 = \omega_{b1}({\mathbf k})}
+ \big( i\Bc_{\Tc\!\Ac}^{(2,2)} \big)_{a'a} \,,
\label{eq:tbox_v2}
\end{multline}
where we have defined $\omega_x(\mathbf{k}) \equiv \sqrt{m_x^2+\mathbf{k}^2}$. The first term corresponds to the residue of the pole at $k^0 = \omega_{b1}(\bf k)$, which we leave explicit because it contains the two-particle on-shell singularity. Note that, in this residue, the four-momentum of particle $b1$ is individually on shell, i.e.~$k^2 = m_{b1}^2$, but particle $b2$ is not itself on shell since $(P - k)^2 \neq m_{b2}^2$. Contributions from other poles in $k^0$ are absorbed into the kernel $\Bc_{\Tc\!\Ac}^{(2,1)} \to \Bc_{\Tc\!\Ac}^{(2,2)}$.

Simplifying the propagator pole piece $D_{b2} ((P-k)^2) \big\vert_{k^0 = \omega_{b1}(\bf k)}$, we find the two-particle singularity at $E = \omega_{b1}({\bf k}) + \omega_{b2}({\bf P-k})$. This condition fixes the magnitude of CM frame three-momentum to
\begin{equation}
q_b^\star
\equiv \frac{1}{2\sqrt{s}} \lambda^{1/2} (s, m_{b1}^2, m_{b2}^2)
\,,
\end{equation}
where $\lambda(x,y,z)=x^2+y^2+z^2-2(xy+xz+yz)$ is the K\"all\'en triangle formula and the subscript $b$ label emphasizes that this magnitude is channel-specific. To rewrite the pole in terms of CM frame quantities, we introduce a function that is closely related to the function denoted by $S(P,L)$ in ref.~\cite{Raposo:2023oru}:
\begin{align}
\label{eq:Delta2L_def}
(\Delta_{2,L})_{b'b} (\mathbf k', \mathbf k)
&\equiv
\frac{1}{L^3} \, \xi_b \, \frac{\omega_{b1}^\star}{\omega_{b1}(\mathbf k) }
\frac{\delta_{b'b} \, \delta_{\mathbf k' \mathbf k} \, H_b (\mathbf k^\star) }{2 E^\star \big[ (q_b^\star)^2 - (\mathbf k^\star)^2 + i\epsilon \big] } \,,
\end{align}
where we have defined $\omega_{b1}^\star \equiv \sqrt{(q_b^\star)^2 + m_{b1}^2}$ and introduced an analytic cut-off function $H_b(\mathbf k^\star)$ that equals $1$ at the pole $\vert {\mathbf k}^\star \vert = q_b^\star$ and goes to zero exponentially as $\vert \mathbf k^\star \vert \to \infty$. The dependence on $P$ is omitted. Using this definition, we can rewrite eq.~\eqref{eq:tbox_v2} as
\begin{align} \nonumber
\big( i\Mc_{\Tc\!\Ac,L}^{(2)} \big)_{a'a}
& =
\sum_{b',b} \sum_{\mathbf k',\mathbf k} \Big( \big[ i\Tc_{a'b'} (p_3, p_4; k', P-k') \big]_{k'^0= \omega_{b'1} (\mathbf k')} \, (i\Delta_{2,L})_{b'b} (\mathbf k', \mathbf k) \\
& \hspace{100pt} \big[i\Ac_{ba} (k,P-k; p_1,p_2) \big]_{k^0= \omega_{b1}(\bf k)} \Big) + \big( i\Bc_{\Tc\!\Ac}^{(2,3)} \big)_{a'a} \,,
\label{eq:tbox_v3} \\[8pt]
& \equiv \big( i\Tc \cdot i\Delta_{2,L} \cdot i\Ac \big)_{a'a} + \big( i\Bc_{\Tc\!\Ac}^{(2,3)} \big)_{a'a} \,,
\label{eq:tbox_v4}
\end{align}
where the difference is absorbed into a new kernel via $\Bc_{\Tc\!\Ac}^{(2,2)} \to \Bc_{\Tc\!\Ac}^{(2,3)}$. In the second line, we define a shorthand notation for the first term, where each object can be interpreted as a matrix in momentum and channel space, with summations over finite-volume momentum and channel index resulting from standard matrix multiplication.

An identical analysis can be carried out for contributions with a $u$-channel pole piece $\Uc$ or a kernel $\Bc^{(1)}$ on the left side of the loop, which we denote by $\Mc_{\Uc\!\Ac}^{(2)}$ and $\Mc_{\Bc^{(1)}\!\Ac}^{(2)}$, respectively. These are pictured in figure~\ref{fig:box_diagram}(b) and (c). Using the same shorthand notation, we write the latter as
\begin{align} \nonumber
i\Mc_{\Bc^{(1)}\!\Ac, L}^{(2)}
& \equiv
i\Bc^{(1)} \cdot i\Delta_{2,L} \cdot i\Ac + i\Bc_{\Bc^{(1)}\!\Ac}^{(2,3)} \,.
\end{align}
For $\Mc_{\Uc\!\Ac, L}^{(2)}$, however, we define this notation with a subtle difference. The analog of eq.~\eqref{eq:tbox_v3} in this case reads
\begin{align} \nonumber
\big( i\Mc_{\Uc\!\Ac, L}^{(2)} \big)_{a'a}
&=
\sum_{b',b} \sum_{\mathbf k',\mathbf k} \Big( \big[ i\, \Uc_{a'b'} (p_3, p_4; k', P-k') \big]_{k'^0= \omega_{b'1} (\mathbf k')} \,(i\Delta_{2,L})_{b'b} (\mathbf k', \mathbf k) \\
& \hspace{100pt} \times \big[i\Ac_{ba} (k,P-k; p_1,p_2) \big]_{k^0= \omega_{b1}(\bf k)} \Big) + \big( i\Bc_{\Uc\!\Ac}^{(2,3)} \big)_{a'a} \,.
\label{eq:ubox_v1}
\end{align}
Noting that the quantities $\omega_{b1}(\mathbf k)$ and ${ E - \omega_{b2}(\mathbf P - \mathbf k)}$ must coincide at the two-particle pole, we make the replacement
\begin{equation}
\big[ i\,\Uc (p_3,p_4;k',P-k') \big]_{k'^0= \omega_{b1}(\bf k')} \to \big[ i\,\Uc (p_3,p_4;k',P-k') \big]_{k'^0= E - \omega_{b2}(\bf P - \bf k')} \,,
\label{eq:u_replace}
\end{equation}
with the difference between the two objects being absorbed through $\Bc_{\Uc\!\Ac}^{(2,3)} \to \Bc_{\Uc\!\Ac}^{(2,4)}$. We define the shorthand notation to take this modification into account, such that
\begin{align}
i\Mc_{\Uc\!\Ac, L}^{(2)}
&\equiv i\,\Uc \cdot i\Delta_{2,L} \cdot i\Ac + i\Bc_{\Uc\!\Ac}^{(2,4)} \,.
\label{eq:ubox_v4}
\end{align}
The motivation for this replacement is ultimately to avoid the presence of a spurious singularity in the energy in the partially on-shell $u$-channel exchange. This is explained in more detail later in this section.

Putting all ingredients together, the overall second-order contribution to the finite-volume amplitude, $i\Mc_L^{(2)}$, can be written compactly as
\begin{equation}
i\mathcal M_L^{(2)}
= \big( i\Bc^{(1)} + i\Tc + i\,\Uc \big) \cdot i\Delta_{2,L} \cdot \big( i\Bc^{(1)} + i\Tc + i\,\Uc \big)
+ i\Bc^{(2)} \,,
\label{eq:ML_2}
\end{equation}
where we take into account all possibilities for the objects on the left and right sides of the loop. The kernel $\Bc^{(2)}$ collects all non-singular kernels arising from the analysis described above.

This expression can then be applied to all two-particle loops at all orders of the amplitude series. By doing so, we can obtain the following all-orders expression for the finite-volume amplitude:
\begin{align}
i\Mc_L
&= i\Bc + i\Tc + i\,\Uc
+ \sum_{n=0}^\infty \big( i\Bc + i\Tc + i\,\Uc \big) \left[ \,\cdot\, i\Delta_{2,L} \cdot \big( i\Bc + i\Tc + i\,\Uc \big) \right]^n \,.
\label{eq:ML_series}
\end{align}
The kernel $\Bc$ here has been obtained by starting from the non-singular kernel $\Bc^{(1)}$ (shown in figure~\ref{fig:kernel_amp}(a)) and subsequently absorbing all non-singular kernels arising from the two-particle loops at each order of the amplitude series, e.g.~the second-order kernel $\Bc^{(2)}$ in eq.~\eqref{eq:ML_2}. This absorption is depicted in figure~\ref{fig:kernel_amp}(c).

An important point to consider is that the ``$\,\cdot\, i\Delta_{2,L}\,\cdot\,$'' operation places the neighboring objects in specific partially off-shell kinematic configurations. More concretely, for $\Bc$, $\Tc$ and $\Uc$ appearing between these operations in the series, we will have the placements
\begin{align}
\Bc_{a'a} (k',P-k';k,P-k) &\to \left[ \Bc_{a'a} (k',P-k';k,P-k) \right]_{k'^0 = \omega_{a'1} (\mathbf k'),\, k^0 = \omega_{a1} (\mathbf k)} \,,
\label{eq:B_replacement}
\\[10pt]
%%%%%%%%%%%%%%%%%%%%%%%%%%%%%%%%%%%%%%%%%%
\Tc_{a'a} (k',P-k';k,P-k) &\to \left[ \Tc_{a'a} (k',P-k';k,P-k) \right]_{k'^0 = \omega_{a'1} (\mathbf k'),\, k^0 = \omega_{a1} (\mathbf k)} \,,
\nn\\
&\hspace{60pt} = \frac{-g_{13}g_{24}}{(\omega_{a'1} (\mathbf k')-
\omega_{a1} (\mathbf k))^2
-( \mathbf k - \mathbf k')^2
-m_e^2+i\epsilon}\,,
\label{eq:T_replacement}
\\[10pt]
%%%%%%%%%%%%%%%%%%%%%%%%%%%%%%%%%%%%%%%%%%
\Uc_{a'a} (k',P-k';k,P-k) &\to \left[ \Uc_{a'a} (k',P-k';k,P-k) \right]_{k'^0 = \omega_{a'1} (\mathbf k'),\, k^0 = E - \omega_{a2} (\mathbf P - \mathbf k)} \,
\nn\\
&\hspace{20pt}= \frac{-g_{23}g_{14}}{(\omega_{a'1} (\mathbf k')-
\omega_{a2} (\mathbf P - \mathbf k))^2
-(\mathbf P - \mathbf k - \mathbf k')^2
-m_e^2+i\epsilon}\,
.
\label{eq:U_replacement}
\end{align}
For $\Tc_{a'a}$, we have set $k = (\omega_{a1} (\mathbf k),\mathbf k)$ and $k' = (\omega_{a'1} (\mathbf k'),\mathbf k')$, and hence one can show that $t = (k' - k)^2 \leq (m_{a'1} - m_{a1})^2$ for any choice of real $\mathbf k,\mathbf k'$. With these kinematics, we cannot reach the single-exchange pole at $t = m_e^2$, in contrast to the fully on-shell $t$-channel exchange. The value of $u = (P-k'-k)^2 $, on the other hand, depends explicitly on the total energy $E$ and is not bounded. The situation is reversed for $\Uc_{a'a}$, where we have set $k = (\omega_{a1} (\mathbf k),\mathbf k)$ and $k' = (E - \omega_{a2} (\mathbf P - \mathbf k'), \mathbf k')$. We can then show that $u \leq (m_{a'1} - m_{a2})^2$ for real $\mathbf k,\mathbf k'$, while $t$ is unconstrained. This means we cannot hit the $u = m_e^2$ pole and was the motivation for our earlier kinematic replacement in eq.~\eqref{eq:u_replace}.

In previous derivations of the standard L\"uscher quantization condition, e.g.~ref.~\cite{Kim:2005gf}, the Bethe-Salpeter kernel, including the $t$- and $u$-channel exchange poles, is placed fully on shell at this stage. This is done by setting the magnitudes of the CM spatial momenta to their values at the two-particle poles in the relevant channels, e.g.~$\vert \mathbf k^\star \vert = q_a^\star$ and $\vert \mathbf k'^\star \vert = q_{a'}^\star$ above. As discussed in ref.~\cite{Raposo:2023oru}, doing this for the $\Tc$ and $\Uc$ exchanges, would introduce unphysical poles. As a result, we leave the $\Tc$ and $\Uc$ exchanges as is, and we introduce a new exchange ($\Ec$) function that includes the sum of both exchanges,
\begin{multline}
\Ec_{a'a} (\mathbf k'; \mathbf k)
\equiv
\left[ \Tc_{a'a} (k',P-k';k,P-k) \right]_{k'^0 = \omega_{a'1} (\mathbf k'),\, k^0 = \omega_{a1} (\mathbf k)}
\\
 +
\left[ \Uc_{a'a} (k',P-k';k,P-k) \right]_{k'^0 = \omega_{a'1} (\mathbf k'),\, k^0 = E - \omega_{a2} (\mathbf P - \mathbf k)}.
\label{eq:Ec_def}
\end{multline}
The kernel $\Bc$, on the other hand, does not contain these poles and can be placed fully on shell without further issues.

The contributions arising from the difference between off- and on-shell kernels generate terms that do not contain two-particle poles. These can be absorbed into $\Bc$ and the resulting object can then be set on shell. This iterative process can be applied to each order of the amplitude series of eq.~\eqref{eq:ML_series}. We are then able to reorganize the series and, hence, write down an equation that must be satisfied by the finite-volume amplitude:
\begin{align}
i\Mc_L
&= i\widetilde\Bc + i\Ec
+ \sum_{n=0}^\infty \big( i\widetilde\Bc + i\Ec \big) \left[ {}\cdot i\Delta_{2,L} \cdot \big( i\widetilde\Bc + i\Ec \big) \right]^n \,, \\
& = i\widetilde\Bc + i\Ec
+ \big( i\widetilde\Bc + i\Ec \big) \cdot i\Delta_{2,L} \cdot i\Mc_L \,.
\label{eq:ML_full}
\end{align}
Here, $\widetilde \Bc$ is the on-shell kernel obtained from $\Bc$ through the absorption of successive contributions arising from on-shell placements as described above. This object contains no singularities nor branch points in the kinematic region considered. For identical particles, the $\Ec$ function double counts the contributions from exchanges in loops. This over counting is compensated by the $\xi$ symmetry factors in the definition of $\Delta_{2,L}$ in eq.~\eqref{eq:Delta2L_def}.

\begin{figure}[t]
\centering
\includegraphics[width=.9\textwidth]{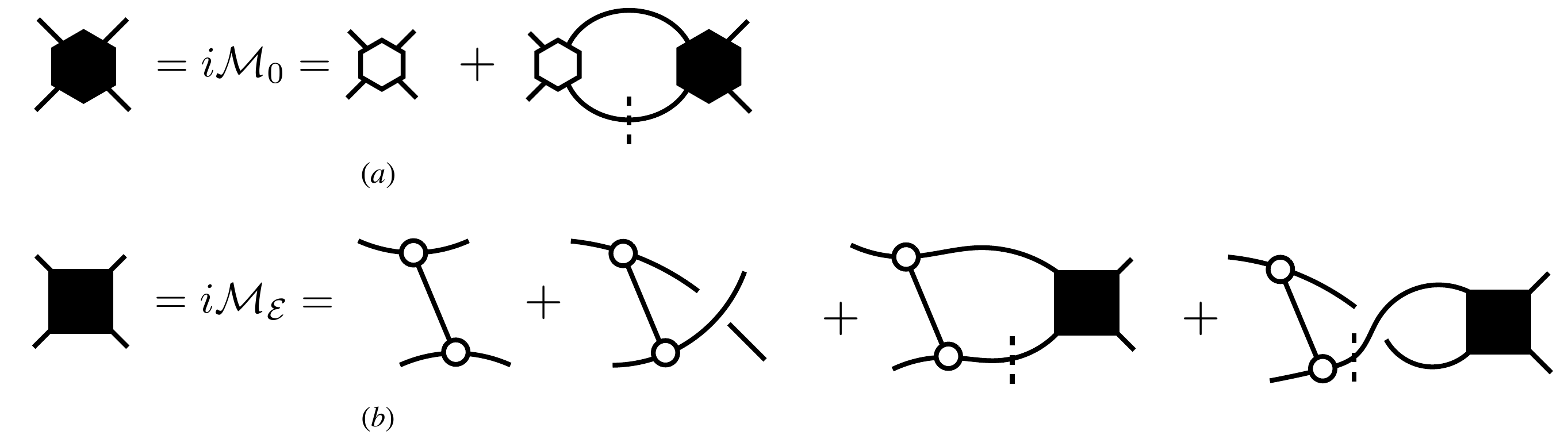}
\caption{Diagrammatic definitions for ($a$) $\Mc_{0,L}$ and ($b$) $\Mc_{\Ec,L}$ defined in eqs.~\eqref{eq:M0L} and \eqref{eq:M0tL}, respectively. }
\label{fig:M0_Mcut}
\end{figure}

To derive a closed-formed expression for $\Mc_L$, and consequently, of its poles, we propose introducing two intermediate quantities, $\Mc_{0,L}$ and $\Mc_{\Ec,L}$. These are defined so that they satisfy the equations depicted in figure~\ref{fig:M0_Mcut}, which can be written compactly as
\begin{align}
i\Mc_{0,L}
& = i\widetilde{\Bc}
+ i\widetilde{\Bc} \cdot i\Delta_{2,L} \cdot i\Mc_{0,L} \,,
\label{eq:M0L} \\
i\Mc_{\Ec,L}
& = i\Ec
+ i\Ec \cdot i\Delta_{2,L} \cdot i\Mc_{\Ec,L} \,.
\label{eq:M0tL}
\end{align}
Generally, these amplitudes have different pole structures and neither of which matches $\Mc_L$. In the limit where the OPE couplings are set to zero, $\Mc_L$ reduces to $\Mc_{0,L}$ when evaluated at on-shell external momenta and its poles are governed by the standard L\"uscher quantization condition.

Before proceeding to make use of eqs.~\eqref{eq:M0L} and \eqref{eq:M0tL}, let us review how the full amplitude, $\Mc_L$, can be recovered from these two building blocks. For this, we can rewrite the right-hand side of eq.~\eqref{eq:ML_full} exclusively in terms of $\Mc_{0,L}$ and $\Mc_{\Ec,L}$:
\begin{multline}
i\Mc_L =
i\Mc_{\Ec,L}
+ \big[ 1 + i\Mc_{\Ec,L} \cdot i\Delta_{2,L} \cdot {} \big] \, i\Mc_{0,L} \\
\times \sum_{n=0}^\infty \left({} \cdot i\Delta_{2,L} \cdot i\Mc_{\Ec,L} \cdot i\Delta_{2,L} \cdot i\Mc_{0,L}
\right)^n \big[{} \cdot i\Delta_{2,L} \cdot i\Mc_{\Ec,L} + 1 \big] \,.
\label{eq:ML_full_v2}
\end{multline}
From this form, it can be deduced that the finite-volume poles are generated from the all-orders sum shown explicitly. To make use of this, it is necessary to give more explicit definitions of both $\Mc_{0,L}$ and $\Mc_{\Ec,L}$.

Let us start by looking at $\Mc_{0,L}$. This object satisfies eq.~\eqref{eq:M0L} involving only the non-singular on-shell kernel $\widetilde{\Bc}$. The standard trick of writing the finite-volume correction of the two-particle loops in terms of on-shell quantities and a single finite-volume function $F$, that depends only on kinematics, can be applied, following the procedure given in ref.~\cite{Kim:2005gf}. This allows us to relate $\Mc_{0,L}$ to its infinite-volume counterpart, $\Mc_0$, defined to satisfy the infinite-volume version of eq.~\eqref{eq:M0L}, algebraically.

To do so, we first need to express $\Mc_{0,L}$ and $\Mc_0$ as matrices in the angular momentum basis. We decompose their angular dependence in the CM frame into spherical harmonics. For $\Mc_{0,L}$, we can write
\begin{align}
(\Mc_{0,L})_{a'a} (\mathbf{p}', \mathbf{p})
=
\sum_{\ell' m',\ell m}
\Yc_{\ell' m'} (\mathbf{p}'^\star, q_{a'}^\star) \, (\Mc_{0,L})_{a'\ell'm';a\ell m} \,
\Yc^T_{\ell m}(\mathbf{p}^\star, q_a^\star) \,,
\label{eq:M0L_PWA}
\end{align}
where $\mathbf{p}', \mathbf{p}$ are one of the outgoing and incoming spatial momenta, respectively, and we suppress the dependence on the total energy $E$ and momentum $\mathbf P$. We have also introduced
\begin{align}
\Yc_{\ell m}(\mathbf{p}^\star,q_a^\star)
&\equiv \sqrt{4\pi} \left(\frac{\vert \mathbf{p}^\star \vert}{q_a^\star}\right)^{\ell} Y_{\ell m}(\hat{\mathbf{p}}^\star) \,,
\label{eq:Yharm_def}
\\
\Yc_{\ell m}^T(\mathbf{p}^\star,q_a^\star)
&\equiv \sqrt{4\pi} \left(\frac{\vert \mathbf{p}^\star \vert}{q_a^\star}\right)^{\ell} Y^*_{\ell m}(\hat{\mathbf{p}}^\star) \,,
\label{eq:YharmT_def}
\end{align}
with $\hat{\mathbf{p}}^\star \equiv \mathbf{p}^\star/\vert \mathbf{p}^\star \vert$. Identical decompositions apply also to the infinite-volume object $\Mc_0$ or the kernel $\widetilde{\Bc}$.

We must address two important aspects here: Firstly, $\Mc_{0,L}$ or $\Mc_0$ have all external momentum arguments on shell, meaning that there is a redundancy in their dependence on $\mathbf{p}', \mathbf{p}$. We can see this by expressing these objects in terms of CM-frame-boosted momenta $\mathbf{p}'^\star, \mathbf{p}^\star$, whose magnitude is fixed to $q_{a'}^\star$ and $q_a^\star$ ($a',a$ specify the corresponding channels), leaving only their directions, i.e.~$\hat{\mathbf{p}}'^\star$ and $\hat{\mathbf{p}}^\star$, unspecified. As such, we can look at eq.~\eqref{eq:M0L_PWA} as providing an off-shell extension of $\Mc_{0,L}$ in momentum space, and similarly for $\Mc_0$ or $\widetilde{\Bc}$. Secondly, $\Mc_{0,L}$ is a finite-volume object and, in principle, only valued in the set of discretized finite-volume momenta, while projection to harmonics relies on dependence on continuum CM spatial momentum. This is resolved by noting that the dependence on the external momenta is carried by infinite-volume objects only, namely the kernels $\widetilde{\Bc}$, for which the extension to continuous infinite-volume spatial momentum is straightforward.

Using the spherical harmonic projections and combining the finite-volume eq.~\eqref{eq:M0L} and its infinite-volume analog, we can express $\Mc_{0,L}$ in terms of $\Mc_0$ through the following matrix equation:
\begin{align}
\Mc_{0,L}
&= \left[{\Mc}_{0}^{-1} + F\right]^{-1} \,.
\label{eq:M0L_vf}
\end{align}
The $F$ matrix, as standard in the literature, encodes the difference between the finite- ($\Delta_{2,L}$) and infinite-volume ($\Delta_{2,\infty}$) two-particle loops and has elements $F_{\ell'm';\ell m}$ given by
\begin{align}
F_{a'\ell'm'; a\ell m} (E, \mathbf{P})
& \equiv \delta_{a'a} \, \xi_a \left[ \frac{1}{L^3} \sum_{\mathbf k} - \int \! \frac{{\rm d}^3 \mathbf k}{(2\pi)^3} \right] \frac{\omega_{a1}^\star}{\omega_{a1}(\mathbf k) }
\frac{ \Yc^T_{\ell' m'} (\mathbf k^\star, q^\star) \, \Yc_{\ell m}(\mathbf k^\star, q^\star) \, H_a (\mathbf k^\star) }{2 E^\star \big[ (q_a^\star)^2 - (\mathbf k^\star)^2 + i\epsilon \big] } \,.
\label{eq:F_def}
\end{align}
Using the standard technology to evaluate the imaginary part of two-particle loops, it is also straightforward to show that ${\Mc}_{0}$ satisfies
\begin{align}
{\Mc}_{0}^{-1} = {\Kc}_{0}^{-1} - i\rho \,,
\label{eq:M0_uni}
\end{align}
where ${\Kc}_{0}$ is a real-valued non-singular function and $\rho$ is the standard relativistic two-body phase space factor, represented here as a diagonal matrix in angular momentum and channel space with elements
\begin{align}
\rho_{a'\ell'm';\,a\ell m} \equiv \delta_{a'a} \, \delta_{\ell'\ell} \, \delta_{m'm} \, \xi_a \frac{q_a^\star}{8\pi \sqrt{s}} \,.
\end{align}

We turn now to the other building block for the amplitude, $\Mc_{\Ec, L}$, which satisfies eq.~\eqref{eq:M0tL}. Unlike $\Mc_{0,L}$, this object does not have its external legs on shell. Consequently, it is most conveniently expressed as a matrix in the momentum basis, as opposed to the angular momentum basis. For a given cutoff function $H$, appearing in the definition of $i\Delta_{2,L}$ in eq.~\eqref{eq:Delta2L_def}, the number momenta contributing to a sum can be truncated.~\footnote{Of course, this can introduce a systematic error, as a result, one needs to do convergence tests to ensure that the final results are not sensitive to the given truncation.} Given this, one can write eq.~\eqref{eq:M0tL}, as a matrix equation in this space:
\begin{equation}
i\Mc_{\Ec,L}
= i\Ec + i\Ec \, i\Delta_{2,L} \, i\Mc_{\Ec,L} \,,
\end{equation}
where we now interpret $\Ec$ as the matrix with elements $\Ec_{a'\mathbf k';\, a \mathbf k} \equiv \Ec_{a'a} (\mathbf k', \mathbf k)$, as defined in eq.~\eqref{eq:Ec_def}, and similarly for $\Delta_{2,L}$ and $\Mc_{\Ec,L}$. We can subsequently solve this equation for $\Mc_{\Ec,L}$:
\begin{equation}
\Mc_{\Ec,L} = \Ec \left[ 1 + \Delta_{2,L} \, \Ec
\right]^{-1} \,.
\label{eq:MctL_sol}
\end{equation}
Both $\Ec$ and $\Delta_{2,L}$ are known functions and this expression provides a closed-form solution for the finite-volume object $\Mc_{\Ec,L}$. \footnote{eq.~\eqref{eq:MctL_sol} can be written as $ \Mc_{\Ec,L} = \left[ \Ec^{-1} + \Delta_{2,L}
\right]^{-1} $ when $\Ec$ is invertible, which is not necessarily true. $\Ec$ is singular, for example, when there is no OPE in at least one of the channels considered, in which case $\Ec_{aa} = 0$ for the given channels.}

Given the expressions for $\Mc_{0,L}$ and $\Mc_{\Ec,L}$ above, we can now proceed to rewrite eq.~\eqref{eq:ML_full_v2} and find the poles of $\Mc_L$. To do so, we will first need to evaluate three types of sums of $\Mc_{\Ec,L}$. For this, we also re-interpret the harmonics defined in eq.~\eqref{eq:Yharm_def} as a change of basis matrix from the angular momentum basis to the momentum basis, with elements $\Yc_{a'\mathbf k'; \, a\ell m} \equiv \delta_{a'a} \, \Yc_{\ell m}(\mathbf{k}'^\star, q_{a'}^\star)$ and $\Yc^T_{a'\ell' m'; \, a \mathbf k} \equiv \delta_{a'a} \, \Yc^T_{\ell' m'}(\mathbf{k}^\star, q_a^\star)$. The three types of sum read
\begin{align}
(i\Rc_L)_{a'\ell' m';\,a\mathbf k}
&\equiv
\left( \Yc^T \, i\Delta_{2,L} \, i\Mc_{\Ec,L} \right)_{a'\ell' m';\,a\mathbf k} \,,
\label{eq:RL}
\\
%%%%%%%%%%%%%%%%%%%%%%%%%%%%%%%%%%%%%%%%%%%%%%%%
(i\Lc_L)_{a'\mathbf k';\,a\ell m}
& \equiv
\left( i\Mc_{\Ec,L} \, i\Delta_{2,L} \, \Yc \right)_{a'\mathbf k';\,a\ell m} \,,
\label{eq:LL}
\\
%%%%%%%%%%%%%%%%%%%%%%%%%%%%%%%%%%%%%%%%%%%%%%%%
(i\Cc_L)_{a'\ell' m';\,a\ell m}
&\equiv
\left( \Yc^T \, i\Delta_{2,L} \, i\Mc_{\Ec,L} \, i\Delta_{2,L} \, \Yc \right)_{a'\ell' m';\, a\ell m} \,.
\label{eq:CL}
\end{align}
With these in mind, we find that eq.~\eqref{eq:ML_all_vf} can be rewritten as
\begin{align}
i\Mc_L &= i\Mc_{\Ec,L}
+ \left(\Yc + i\Lc_L \right) \, i\big[ \Mc_{0,L}^{-1} + \Cc_{L} \big]^{-1} \left(\Yc^T + i\Rc_L \right) \,,
\label{eq:ML_all_vf}
\end{align}
where both sides of this equation are in momentum space overall. However, the internal block $\big[ \Mc_{0,L}^{-1} + \Cc_{L} \big]^{-1}$ is to be considered in angular momentum space. This will provide the most straightforward generalization of the standard L\"uscher quantization condition, as we now discuss.

Upon further investigation of eq.~\eqref{eq:ML_all_vf}, we can see that the poles of $\Mc_L$ can arise only if
\begin{align}
\det_{a\ell m} \big[ \Mc_{0,L}^{-1} + \Cc_L \big] = 0 \,,
\end{align}
at the given energy, where the $a\ell m$ subscript is used to emphasize that the determinant is taken over the space of channel and angular momentum indices. To reach this conclusion, we rely on the fact that poles in $\Mc_{0,L}$ and $\Mc_{\Ec,L}$ are canceled by other terms in $\Mc_L$, provided neither of these blocks is zero, which can be shown explicitly. Using eq.~\eqref{eq:M0L_vf}, we can then write the equivalent condition
\begin{align}
\det_{a\ell m} \big[\Mc_{0}^{-1} + F + \Cc_L \big] = \det_{a\ell m} \big[\Kc_{0}^{-1} - i \rho + F + \Cc_L \big] = 0
\label{eq:QC_cc} \,.
\end{align}
This is the main result of this subsection and is quite similar in structure to the standard L\"uscher condition. In the absence of single-particle exchanges, the matrix $\Cc_L$ vanishes, while the infinite-volume amplitude $\Mc$ reduces to $\Mc_0$, reproducing the standard result. We have also included the middle expression, in which the quantization condition is expressed in terms of the K-matrix analog to emphasize that this form is most useful for analyzing lattice data.

\subsection{Generalization to particles with spin}
\label{sec:QC_spin}

In deriving eq.~\eqref{eq:QC_cc}, we allowed for any number of intermediate two-particle channels to go on shell, our only assumption being that the particles considered had no intrinsic spin. This appeared in two places: the propagator structure and the analytic structure of the vertex functions. These two points are, of course, closely related.

In this section, we lift this assumption and allow for particles of arbitrary spin. Because our main interest is QCD, where the left-hand cuts are primarily generated by one-pion exchanges, we assume that the exchange particle is still spinless. This assumption only simplifies the details when partial-wave projecting the resulting one-particle exchange, as will be done in section~\ref{sec:OPE}, and has no impact on the steps discussed in this section.

We build largely on the work of refs.~\cite{Briceno:2014oea, Briceno:2015csa, Raposo:2023oru}, which have previously considered two-body systems with non-zero spin. The first step is to note that the propagator for a particle $x$ with arbitrary spin $s$ can be written as
\begin{align}
\Delta_x(q) = \sum_{m_s} \frac{\eta_{x,m_s}(q) \, \bar{\eta}_{x,m_s}(q)}{q^2-m_x^2+i\epsilon}
+ S_x(q),
\label{eq:Deltax_spin}
\end{align}
where $\eta_{x, m_s}(q)$ is the spin wave function for the particle $x$ with spin quantum numbers $(s,m_s)$, and $S_x$ is a smooth function of momentum in the kinematic region considered. Here we assume that the spin is quantized along a fixed $z$-axis, but, as discussed in section~\ref{sec:OPE}, one can equivalently do this using the helicity basis.

In general, $\eta_{x,m_s}(q)$ is a column vector in the appropriate representation space of the Lorentz group. For example, for spin-1/2 particles, $\eta_{x, m_s}(q)$ would be a spinor in Dirac space. Meanwhile, $\bar{\eta}_{x, m_s}(q)$ is a row vector in this same space. For clarity, let us explicitly write eq.~\eqref{eq:Deltax_spin} with indices $\alpha$ and $\alpha'$ in the representation space,
\begin{align}
\left[\Delta_x(q^2)\right]_{\alpha' \alpha}
&= \sum_{m_s} \frac{\left[ \eta_{x, m_s}(q) \right]_{\alpha'}
\left[ \bar{\eta}_{x, m_s}(q) \right]_\alpha }{q^2-m_x^2+i\epsilon}
+\left[S_x(q^2)\right]_{\alpha' \alpha} \,,
\nn\\
&= \sum_{m_{s'}, m_s}
\left[\eta_{x,m_{s'}}(q)\right]_{\alpha'}
\left[D_x(q^2)\right]_{m_{s'}m_s}
\left[\bar{\eta}_{x,m_s}(q)\right]_\alpha
+\left[S_x(q)\right]_{\alpha' \alpha} \,,
\label{eq:Deltax_spin_rep}
\end{align}
where we introduced
\begin{align}
\left[D_x(q^2)\right]_{m_{s'}m_s} \equiv \frac{\delta_{m_{s'}m_s}}{q^2-m_x^2+i\epsilon} \,.
\end{align}
This explains all modifications to the dressed propagators necessary for analyzing the skeleton expansion.

We turn now to the kernels appearing between loops in that expansion. As before, we isolate the OPE poles to define the objects $\Tc$ and $\Uc$. All non-singular contributions to the Bethe-Salpeter kernel are collected into the object $\Bc^{(1)}$, as described in the previous section. Unlike before, these functions are now tensors in some non-trivial representation space of the Lorentz group. In other words, a generic kernel function $\Ac$ will acquire indices associated with the appropriate representation space, $\Ac \to \Ac_{\beta_1\beta_2;\,\alpha_1\alpha_2}$. Note that, in general, the four external particles can carry different spin and, as a result, their spin wave functions can live in different representation spaces. For $\Tc$ and $\Uc$ specifically, the tensor structure will be embedded into the vertex couplings $g$. To make the distinction between the spinful and spinless cases clear, we replace the $g$'s with $\Gamma$'s to emphasize their tensor nature, which requires some care.

To understand how these two modifications change the derivation given in the previous section, let us consider a loop contribution to the finite-volume amplitude, analogous to eq.~\eqref{eq:tbox_v1}. Here we make one slight change. Instead of having $\Tc$ on the left, we consider a generic $\Ac'$ vertex. Because we are interested in a contribution to the amplitude, we need to include the spin wave functions for the external particles explicitly. Furthermore, we will make the Lorentz representation indices explicit. To slightly reduce the clutter, we define
\begin{align}
\Ac_{\beta;\,\alpha}\equiv
\Ac_{\beta_1\beta_2;\,\alpha_1\alpha_2} \,,
\end{align}
and, to accommodate any number of channels, we make channel indices $a,b$ explicit:
\begin{align}
\Ac_{\beta; \, \alpha}\to
\Ac_{b\beta; \, a\alpha} \,.
\end{align}
To further simplify the notation in what follows, we also write
\begin{align}
\left[\eta(p) \eta(k) \right]_{a m_s \alpha} \equiv
\left[\eta_{a1, m_{s_1}}(p) \right]_{\alpha_1}
\left[{\eta}_{a2, m_{s_2}}(k)
\right]_{\alpha_2}
.
\label{eq:two_eta}
\end{align}

Using this notation, we finally write a single-loop contribution to the finite-volume amplitude as
\begin{align}
\big( i\Mc_{\Ac'\!\Ac,L}^{(2)} \big)_{a'm_{s'}; \,am_s}
& = \big( i\Bc_{\Ac'\!\Ac}^{(2,1)} \big)_{a'm_{s'};\,am_s} + \sum_b \sum_{\alpha',\beta',\alpha,\beta} \sum_{m_{r'},m_r} \int\! \frac{{\rm d}k_0}{2\pi} \frac{1}{L^3}\sum_{\mathbf{k}}
\nn \\
%%%%%%%%%%%%%%%%%%%%%%%%%%%%%%%%%%%%%%%%%%%%
& \times \Big( \left[\bar{\eta}(p_3)\bar{\eta}(p_4)\right]_{a'm_{s'}\alpha'} i\Ac'_{a'\alpha';\,b\beta'} (p_3,p_4;k,P-k) \left[\eta(k) \eta(P-k)\right]_{bm_{r'}\beta'}
\nn \\[5pt]
%%%%%%%%%%%%%%%%%%%%%%%%%%%%%%%%%%%%%%%%%%%%
& \times \xi_b \left[iD_{b1}(k^2)\right]_{m_{r'_1}m_{r_1}} \left[iD_{b2}((P-k)^2)\right]_{m_{r'_2}m_{r_2}}
\nn \\
%%%%%%%%%%%%%%%%%%%%%%%%%%%%%%%%%%%%%%%%%%%%
& \times \left[\bar{\eta}(k) \bar{\eta}(P-k)\right]_{bm_r\beta} i\Ac_{b\beta ;\,a\alpha} (k,P-k;p_1,p_2)
\left[{\eta}(p_1) {\eta}(p_2)\right]_{am_s\alpha} \Big)
%%%%%%%%%%%%%%%%%%%%%%%%%%%%%%%%%%%%%%%%%%%%
\,.
\label{eq:Abox_v1}
\end{align}
Note that sums over $m_r$, $\alpha$, and so on, are simply short-hand for sums over the underlying index sets.
We further simplify this by introducing a spin-basis for the endcaps. Explicitly,
\begin{equation}
\Ac_{bm_r ;\, am_s} (k_1, k_2; p_1, p_2)
=
\sum_{\beta,\alpha}
\left[\bar{\eta}(k_1) \bar{\eta}(k_2)\right]_{bm_r\beta}
\Ac_{b\beta ;\,a\alpha} (k_1, k_2; p_1, p_2)
\left[{\eta}(p_1){\eta}(p_2)\right]_{am_s\alpha} \,,
\end{equation}
where repeated channel indices are not being summed over. With this, eq.~\eqref{eq:Abox_v1} becomes
\begin{align}
\big( i\Mc_{\Ac'\!\Ac,L}^{(2)} \big)_{a'm_{s'};\,am_s}
&=\sum_b \sum_{m_{r'},m_r}
\int\! \frac{{\rm d}k_0}{2\pi} \frac{1}{L^3}\sum_{\mathbf{k}}
\Big(
i\Ac'_{a'm_{s'};\,bm_{r'}} (p_3,p_4;k,P-k) \,
\nn\\
%%%%%%%%%%%%%%%%%%%%%%%%%%%%%%%%%%%%%%%%%%%%%
&
\hspace{50pt} \times \xi_b
\left[iD_{b1}(k^2)\right]_{m_{r'_1}m_{r_1}}
\left[iD_{b2}((P-k)^2)\right]_{m_{r'_2}m_{r_2}}
\nn \\
&
\hspace{50pt} \times i\Ac_{bm_r ;\,am_s} (k,P-k;p_1,p_2)
\Big)
+\big( i\Bc_{\Ac'\!\Ac}^{(2,1)} \big)_{a'm_{s'};\,am_s}
%%%%%%%%%%%%%%%%%%%%%%%%%%%%%%%%%%%%%%%%%%%%%
\,.
\label{eq:Abox_v2}
\end{align}
Writing this in this form, we can follow the same steps as in section~\ref{sec:QC_derivation}, to arrive at the same form to eq.~\eqref{eq:tbox_v4}, by expanding the definition $\Delta_{2,L}$ in eq.~\eqref{eq:Delta2L_def} to be a diagonal matrix in spin-space,
\begin{align}
\label{eq:Delta2L_def_ms}
(\Delta_{2,L})_{b'm_{s'};\,bm_s} (\mathbf k', \mathbf k)
&\equiv
\frac{1}{L^3} \, \xi_b \, \frac{\omega_{b1}^\star}{\omega_{b1}(\mathbf k) }
\frac{\delta_{m_{s_1}'m_{s_1}} \delta_{m_{s_2}'m_{s_2}} \delta_{b'b} \, \delta_{\mathbf k' \mathbf k} \, H_b (\mathbf k^\star) }{2 E^\star \big[ (q_b^\star)^2 - (\mathbf k^\star)^2 + i\epsilon \big] } \,.
\end{align}
Next, we can write $\Delta_{2,L}$ and the endcaps using the total spin $(S,m_S)$ of the two particles using Clebsch-Gordan coefficients, $\braket{s_1 m_{s_1},s_2 m_{s_2} | S m_S}$. Because $\Delta_{2,L}$ is diagonal in spin, we simply have
\begin{align}
\label{eq:Delta2L_def_SmS}
(\Delta_{2,L})_{b'S'm_{S'};\, bSm_S} (\mathbf k', \mathbf k)
&\equiv
\frac{1}{L^3} \, \xi_b \, \frac{\omega_{b1}^\star}{\omega_{b1}(\mathbf k) }
\frac{\delta_{S'S}\,\delta_{m_{S'} m_S}\,\delta_{b'b} \, \delta_{\mathbf k' \mathbf k} \, H_b (\mathbf k^\star) }{2 E^\star \big[ (q_b^\star)^2 - (\mathbf k^\star)^2 + i\epsilon \big] } \,.
\end{align}
For the endcaps, we get
\begin{multline}
\Ac_{bS'm_{S'};\,aS m_S} (k_1,k_2; p_1,p_2) = \\
\sum_{m_{s'},m_s}
\braket{S' m_{S'} | s_1' m_{s'_1},s_2' m_{s'_2} } \Ac_{bm_{s'};\,am_s} (k_1,k_2;p_1,p_2)
\braket{s_1 m_{s_1},s_2 m_{s_2} | S m_S} \,.
\end{multline}
Performing these basis transformations, not only for the internal states but also the external states, and isolating the pole contribution, we arrive at an expression equivalent in spirit to eq.~\eqref{eq:tbox_v4},
\begin{align}
\begin{split}
 \big( i\Mc_{\Ac'\!\Ac,L}^{(2)} \big)_{a'Sm_{S'};\, aSm_S}
 &=
 \big( i\Bc_{\Ac'\!\Ac}^{(2,2)} \big)_{a'S'm_{S'};\,aSm_S}
 \\[3pt]
 &
 \hspace{-50pt}
 +
 \sum_{b',b}\sum_{R',m_{R'},R,m_R}
 \frac{1}{L^3}\sum_{\mathbf k',\mathbf k} \Big(
 \big[i\Ac'_{a'S'm_{S'};\,b'R'm_{R'}} (k', P-k') \big]_{k'^0= \omega_{b'1} (\mathbf k')}
 \\[3pt]
 %%%%%%%%%%%%%%%%%%%%%%%%%%%%%%%%%%%%%%%%%%%%%
 & \hspace{-20pt}
 \times (i\Delta_{2,L})_{b'R'm_{R'};\,bRm_R} (\mathbf{k}',\mathbf{k}) \,
 \big[i\Ac_{bRm_R;\,aSm_S} (k,P-k) \big]_{k^0= \omega_{b1}(\bf k)} \, \Big)
 %%%%%%%%%%%%%%%%%%%%%%%%%%%%%%%%%%%%%%%%%%%%%
 \,,
 \label{eq:Abox_v3}
\end{split}
\\[10pt]
&\equiv
\left[i\Ac' \cdot i\Delta_{2,L} \cdot i\Ac \right]_{a'S'm_{S'};\,aSm_S} +
\big( i\Bc_{\Ac'\!\Ac}^{(2,2)} \big)_{a'S'm_{S'};\,aSm_S} \,.
\end{align}
This expression holds if $\Ac'$ is either the $\Tc$ or $\Bc^{(1)}$ type. If $\Ac'$ arises from a $\Uc$ exchange, one must perform the kinematic replacement introduced in eq.~\eqref{eq:u_replace}.

In summary, we get that by incorporating spin, the kernel-replacement rules in eqs.~\eqref{eq:B_replacement}, \eqref{eq:T_replacement} and \eqref{eq:U_replacement}, must be replaced with
\begin{align}
\Bc_{a'S'm_{S'};\,aSm_S} (k', P-k'; k, P-k) &\to \Big[\Bc_{a'S'm_{S'};\, aSm_S} (k', P-k'; k, P-k) \Big]_{\substack{k'^0 = \omega_{a'1} (\mathbf k') \\[1pt] \hspace{-3pt} k^0 = \omega_{a1} (\mathbf k)}} \,,
\label{eq:B_replacemnt_spin}
\\[0pt]
%%%%%%%%%%%%%%%%%%%%%%%%%%%%%%%%%%%%%%%%%%
\Tc_{a'S'm_{S'};\,aSm_S} (k', P-k'; k, P-k) &\to \Big[\Tc_{a'S'm_{S'};\, aSm_S} (k', P-k'; k, P-k) \Big]_{\substack{k'^0 = \omega_{a'1} (\mathbf k') \\[1pt] \hspace{-3pt} k^0 = \omega_{a1} (\mathbf k)}}
\nn\\
&\hspace{5pt}= \frac{-\left[\Gamma_{13}\Gamma_{24}\right]_{a'S'm_{S'};\,aSm_S}}
{\left( \omega_{a'1} (\mathbf k') - \omega_{a1} (\mathbf k) \right)^2 - (\mathbf k -\mathbf k')^2 - m_e^2 + i\epsilon} \,,
\label{eq:T_replacement_spin}
\\[12pt]
%%%%%%%%%%%%%%%%%%%%%%%%%%%%%%%%%%%%%%%%%%
\Uc_{a's'm_S';\,asm_S} (k',P-k';k,P-k) &\to \big[ \Uc_{a'S'm_{S'};\, aSm_S} (k', P-k'; k, P-k) \Big]_{\substack{\hspace{-19pt} k'^0 = \omega_{a'1} (\mathbf k') \\[1pt] \hspace{3pt} k^0 = E - \omega_{a2} (\mathbf P - \mathbf k)}}
\nn\\
&\hspace{-30pt} = \frac{-\left[\Gamma_{23}\Gamma_{14}\right]_{a's'm_{S'};\,asm_S}}{(\omega_{a'1} (\mathbf k')-
\omega_{a2} (\mathbf P - \mathbf k))^2
-(\mathbf P - \mathbf k - \mathbf k')^2
-m_e^2+i\epsilon} \,,
\label{eq:U_replacement_spin}
\end{align}
where
\begin{align}
\left[\Gamma_{13}\Gamma_{24}\right]_{a'S'm_{S'};\,aSm_S}
& \equiv \sum_{m_{s'},m_s}
\braket{S' m_{S'} | s_1' m_{s'_1}, s_2' m_{s'_2}}
\bigg[\left[\bar{\eta}_{a'1,m_{s'_1}}(k') \, \Gamma_{13} \, {\eta}_{a1, m_{s_1}}(k) \right]
\nn \\
& \hspace{-50pt} \times \left[ \bar{\eta}_{a'2,m_{s_2'}} (P-k') \,\Gamma_{24} \, {\eta}_{a2,m_{s_2}}(P-k) \right]
\bigg]_{\substack{k'^0 = \omega_{a'1} (\mathbf k') \\[1pt] \hspace{-3pt} k^0 = \omega_{a1} (\mathbf k)}}
\braket{s_1 m_{s_1},s_2 m_{s_2} | S m_S} \,, \\
%%%%%%%%%%%%%%%%%%%%%%%%%%%%%%%%%%%%%%%%%%%%%
\left[\Gamma_{23}\Gamma_{14}\right]_{a'S'm_{S'};\,aSm_S}
&\equiv
\sum_{m_{s'},m_s}
\braket{S' m_{S'} | s_1' m_{s'_1}, s_2' m_{s'_2}}
\bigg[\left[\bar{\eta}_{a'1, m_{s_1'}}(k') \, \Gamma_{23} \, {\eta}_{a2, m_{s_2}} (P-k) \right]
\nn\\
& \hspace{-50pt} \times \left[\bar{\eta}_{a'2, m_{s'_2}} (P-k') \, \Gamma_{14} \, {\eta}_{a1 , m_{s_1}} (k) \right] \bigg]_{\substack{\hspace{-19pt} k'^0 = \omega_{a'1} (\mathbf k') \\[1pt] \hspace{3pt} k^0 = E - \omega_{a2} (\mathbf P - \mathbf k)}}
\braket{s_1 m_{s_1},s_2 m_{s_2} | S m_S} \,.
\end{align}

Given these various building blocks, we are now at a stage to define how the building blocks in the finite-volume amplitude change. The exchange function, defined in eq.~\eqref{eq:Ec_def} is now a matrix in channel and spin space,
\begin{align}
\Ec_{a'a} (k'_1,k'_2;k_1,k_2)
&\to
\Ec_{a'S'm_{S'};\,aSm_S} (k'_1,k'_2;k_1,k_2),
\label{eq:Ec_def_spin}
\end{align}
as is $\Mc_{\Ec,L}$. Formally, we can still write the solutions to the matrix equation for $\Mc_{\Ec,L}$, given in eq.~\eqref{eq:MctL_sol} as a matrix in this enhanced space of momentum and spin,
\begin{equation}
\Mc_{\Ec,L} = \Ec \left[ 1 + \Delta_{2,L} \Ec\right]^{-1} \,.
\label{eq:MctL_sol_spin}
\end{equation}

Similarly, we write $\Mc_{0,L}$ as a matrix in spin space. We can still use eq.~\eqref{eq:M0L_PWA} for its matrix elements,
\begin{align}
(\Mc_{0,L})_{a'S'm_{S'};\,aSm_S} (\mathbf{p}', \mathbf{p})
= \sum_{\ell' m',\ell m}
\Yc_{\ell' m'} (\mathbf{p}'^\star, q_{a'}^\star) \, (\Mc_{0,L})_{a'\ell'm'S'm_{S'};\,a\ell mSm_S} \,
\Yc^T_{\ell m}(\mathbf{p}^\star, q_a^\star) \,.
\label{eq:M0L_PWA_spin}
\end{align}
We can also write the $F$ function in eq.~\eqref{eq:F_def} as a matrix in spin, but because of $\Delta_{2,L}$ is diagonal in $S$, see eq.~\eqref{eq:Delta2L_def_SmS}, so is $F$,~\cite{Briceno:2014oea, Briceno:2015csa}
\begin{align}
F_{a'\ell'm'S'm_{S'};\, a\ell mSm_S} (E, \mathbf{P})
\equiv \delta_{S'S}\delta_{m_{S'}m_S}
F_{a'\ell'm';\, a\ell m} (E, \mathbf{P}) .
\end{align}
Following the same reasoning, we rewrite $\Rc_L$, $\Lc_L$ and $\Cc_L$ matrices defined in eqs.~\eqref{eq:RL} to \eqref{eq:CL} to explicitly include spin indices,
\begin{align}
(i\Rc_L)_{a'\ell' m'S'm_{S'};\,a\mathbf k Sm_S}
&\equiv
\left( \Yc^T \, i\Delta_{2,L} \, i\Mc_{\Ec,L} \right)_{a'\ell' m'S'm_{S'};\, a\mathbf k Sm_S},
\label{eq:RL_spin}
\\[10pt]
%%%%%%%%%%%%%%%%%%%%%%%%%%%%%%%%%%%%%%%%%%%%%%%%
(i\Lc_L)_{a'\mathbf k'S'm_{S'};\,a\ell mSm_S}
& \equiv
\left( i\Mc_{\Ec,L} \, i\Delta_{2,L} \, \Yc \right)_{a'\mathbf k'S'm_{S'};\,a\ell mSm_S} \,,
\label{eq:LL_spin}
\\[10pt]
%%%%%%%%%%%%%%%%%%%%%%%%%%%%%%%%%%%%%%%%%%%%%%%%
(i\Cc_L)_{a'\ell' m'S'm_{S'};\,a\ell mSm_S}
&\equiv
\left( \Yc^T \, i\Delta_{2,L} \, i\Mc_{\Ec,L} \, i\Delta_{2,L} \, \Yc \right)_{a'\ell' m'S'm_{S'};\,a\ell mSm_S} \,.
\label{eq:CL_spin}
\end{align}

Having written down every building block in $\Mc_{2,L}$ in the $\ell mSm_S$ basis, we could proceed to rewrite the quantization condition in this basis. That said, it is more convenient to take one final step and project these to the $Jm_JS\ell$ basis. Because we have chosen to quantize both the spin and orbital angular momentum along the $z$-axis, this is now a simple exercise. Because the only quantities that appear in the quantization condition, eq.~\eqref{eq:QC_cc}, are $\Mc_0$, $F$ and, $\Cc_L$, we project these here,
\begin{align}
& \left(\Mc_0\right)_{a' J' m_{J'} \ell'S';\, a J m_J\ell S } \nn\\
&\hspace{1.0cm}= \sum_{m',m_{S'},m,m_S} \braket{J' m_{J'} | \ell' m', S' m_{S'}} \left( \Mc_0 \right)_{a' \ell' m' S' m_{S'};\, a\ell mSm_S}
\braket{\ell m, S m_S | J m_J} \,, \\[15pt]
%%%%%%%%%%%%%%%%%%%%%%%%%%%%%%%%%%%%%%%%%%%%%%%%%
& F_{a' J' m_{J'} \ell'S';\, a J m_J\ell S } \nn \\
& \hspace{1.0cm} = \delta_{S' S} \sum_{m',m,m_S} \braket{J' m_{J'} | \ell' m', S m_S} F_{a'\ell' m';\, a \ell m}
\braket{\ell m, S m_S | J m_J} \,, \\
%%%%%%%%%%%%%%%%%%%%%%%%%%%%%%%%%%
& (\Cc_L)_{a'J'm_{J'}\ell'S';\, a J m_J\ell S } \nn\\
&\hspace{1.0cm}= \sum_{m',m_{S'},m,m_S} \braket{J' m_{J'} | \ell' m', S' m_{S'}} (\Cc_L)_{a'\ell' m' S' m_{S'};\, a\ell mSm_S}
\braket{\ell m, S m_S | J m_J} \,.
\end{align}
Having these quantities in the desired basis, we can now rewrite the quantization condition to include an arbitrary number of channels and spin,
\begin{align}
\det_{aJ m_J \ell S} \big[\Mc_{0}^{-1} + F + \Cc_L \big] = 0
\label{eq:QC_cc_spin} \,,
\end{align}
where the determinant now runs over the larger index space. Note that, in the limit where the exchanges couplings are set to $0$, the $\Cc_L$ term vanishes and one recovers the quantization condition for two-body systems with arbitrary spin, first derived in refs.~\cite{Briceno:2014oea, Briceno:2015csa}.

\subsection{Equivalence to the quantization condition in ref.~\cite{Raposo:2023oru}}
\label{sec:equivalence}

In ref.~\cite{Raposo:2023oru}, quantization conditions are provided for single-channel systems of two identical particles of arbitrary spin, taking into account the nearest left-hand cut arising from one-particle exchanges. In this section, we show their equivalence to the results presented in this work, namely eqs.~\eqref{eq:QC_cc} (spinless particles) and \eqref{eq:QC_cc_spin} (arbitrary spin).

The condition for identical spinless particles, given in eq.~(5.3) of ref.~\cite{Raposo:2023oru}, can be stated in the form
\begin{equation}
\det_{\ell m} \Big[ \big(\overline{\mathcal K}{}^{\sf os} \big)^{-1} + F^{\Ec} \Big] = 0 \,,
\label{eq:QC_prev}
\end{equation}
where $\overline\Kc{}^{\sf os}$ is a scheme-dependent, non-singular, infinite-volume function and $F^\Ec$ is a finite-volume function, defined as
\begin{align}
F^\Ec \equiv v^T S \frac{1}{1 + \widetilde\Ec S} \, v
= v^T \sum_{n=0}^\infty S \big(-\widetilde\Ec S \big)^n v \,,
\label{eq:FE_def}
\end{align}
corresponding to eq.~(5.4) of ref.~\cite{Raposo:2023oru}. The objects $S$ and $\widetilde\Ec$ are matrices in a mixed finite-volume momentum and angular momentum basis, with elements given by
\begin{align}
S_{\mathbf k' \ell' m';\, \mathbf k \ell m} &\equiv \frac{1}{L^3} \frac12 \frac{\omega^\star}{\omega(\mathbf k)} \frac{ 4\pi \, \vert \mathbf k^\star \vert^{\ell + \ell'} \, Y^*_{\ell' m'}(\hat{\mathbf k}^\star) \, Y_{\ell m}(\hat{\mathbf k}^\star) \, \delta_{\mathbf k' \mathbf k} \, H(\mathbf k^\star)}{2E^\star \big [ (q^\star)^2 - (\mathbf k^\star)^2 + i\epsilon \big ]} \,,
\label{eq:SL_def} \\[5pt]
\widetilde\Ec_{\mathbf k' \ell' m';\, \mathbf k \ell m}
&\equiv \frac{1}{4 \pi \vert \mathbf k'^\star \vert^{\ell'} \vert \mathbf k^\star \vert^{\ell}}
\left[ \int\! {\rm d} \Omega_{\hat{\mathbf p}'^\star} \!\int\! {\rm d} \Omega_{\hat{\mathbf p}^\star} \, Y^*_{\ell' m'}(\hat{\mathbf p}'^\star) \, \Ec(\mathbf p', \mathbf p) \, Y_{\ell m} (\hat{\mathbf p}^\star) \right]_{\substack{|\mathbf p'^\star|=|\mathbf k'^\star|\\[2pt] \hspace{0pt} |\mathbf p^\star|=|\mathbf k^\star|} } \,,
\label{eq:Epw_def}
\end{align}
cf.~eqs.~(3.21) and (3.64) in ref.~\cite{Raposo:2023oru}.

Here, kinematic quantities are defined as in section~\ref{sec:QC_derivation} and we drop particle and channel labels since we are considering identical particles in a single channel. As before, $H(\mathbf k^\star)$ is a smooth regulator function, $\Ec$ is the exchange function defined in eq.~\eqref{eq:Ec_def}, and $\int {\rm d} \Omega_{\hat{\mathbf p}^\star}$ denotes integration over the angular degrees of freedom of the CM-boosted momentum $\mathbf p^\star$. Finally, $v$ and $v^T$ are trivial column and row vectors of ones, respectively, in momentum space: $v_{\mathbf k} = 1$. These contract with the momentum indices of the internal block in eq.~\eqref{eq:FE_def} to produce the matrix $F^\Ec$ which is defined in the angular momentum index space. Note that there are a few minor differences between the definitions given here versus ref.~\cite{Raposo:2023oru}, which we discuss at the end of the section. We also deviate from the original notation to align more closely with the notation and conventions used in this work.

From its definition, we see that $S$ carries the two-particle pole and is therefore closely related to the object $\Delta_{2,L}$ defined in eq.~\eqref{eq:Delta2L_def}. Similarly, $\widetilde\Ec$ is the matrix of spherical harmonic projections of the exchange function $\Ec$, evaluated at finite-volume momenta. As such, we can provide the following relations between their elements:
\begin{align}
S_{\mathbf k'\ell'm';\, \mathbf k \ell m} &= (q^\star)^{\ell + \ell'} \, \Yc^T_{\ell' m'} (\hat{\mathbf k}'^\star, q^\star) \, (\Delta_{2,L})_{\mathbf k';\,\mathbf k} \, \Yc_{\ell m} (\hat{\mathbf k}^\star, q^\star) \,,
\label{eq:Delta2_rel} \\[5pt]
\Ec_{\mathbf k';\, \mathbf k} &= \sum_{\ell' m',\, \ell m}
(q^\star)^{\ell + \ell'} \Yc_{\ell' m'} (\hat{\mathbf k}'^\star, q^\star) \, \widetilde\Ec_{\mathbf k'\ell'm';\, \mathbf k \ell m} \, \Yc^T_{\ell m} (\hat{\mathbf k}^\star, q^\star) \,,
\label{eq:Epw_rel}
\end{align}
where we omit the trivial channel indices in $\Delta_{2,L}$ and $\Ec$ and use the appropriate form of these objects for identical particles.

Looking at the series definition of $F^\Ec$ in eq.~\eqref{eq:FE_def}, we can apply these relations and move around the spherical harmonics and powers of on-shell momentum to obtain
\begin{align}
F^\Ec_{\ell'm';\, \ell m} & = (q^\star)^{\ell+\ell'} \bigg( \sum_{n=0}^\infty \Yc^T \Delta_{2,L} \left[ - \Ec \Delta_{2,L} \right]^n \Yc \bigg)_{\ell' m';\, \ell m} \,, \\
& = (q^\star)^{\ell+\ell'} \big( \Yc^T \Delta_{2,L} \, \Yc + \Yc^T \Delta_{2,L} \,\Mc_{\Ec, L} \, \Delta_{2,L} \, \Yc \big)_{\ell' m';\, \ell m} \,,
\end{align}
with $\Yc$ and $\Yc^T$ as defined in eqs.~\eqref{eq:Yharm_def} and \eqref{eq:YharmT_def}. In the second line, we use eq.~\eqref{eq:MctL_sol} to replace the infinite series with $\Mc_{\Ec,L}$. It is now straightforward to see that the second term inside the brackets corresponds to the sum $\Cc_L$ defined in eq.~\eqref{eq:CL}. The first term can be identified as the sum term of the L\"uscher finite-volume function $F$ in eq.~\eqref{eq:F_def}. Introducing the notation $I$ for the matching integral term:
\begin{equation}
I_{\ell'm';\,\ell m} \equiv \frac12 \int \! \frac{{\rm d}^3 \mathbf k}{(2\pi)^3} \frac{\omega^\star}{\omega(\mathbf k) } \frac{ \Yc^T_{\ell' m'} (\mathbf k^\star, q^\star) \, \Yc_{\ell m}(\mathbf k^\star, q^\star) \, H (\mathbf k^\star) }{2 E^\star \big[ (q^\star)^2 - (\mathbf k^\star)^2 + i\epsilon \big] } \,,
\label{eq:I_def}
\end{equation}
we are now able to rewrite the elements of $F^\Ec$ as
\begin{equation}
F^\Ec_{\ell'm';\, \ell m} = (q^\star)^{\ell+\ell'} \big( F + \Cc_L + I \big)_{\ell'm';\, \ell m} \,.
\label{eq:FE_rel}
\end{equation}
The factors of $(q^\star)^{\ell+\ell'}$ in the expressions above arise from the different normalization of the barrier factors in ref.~\cite{Raposo:2023oru}.

Let us turn now to the kernel $\overline{\mathcal K}{}^{\sf os}$. In section~4.2 of ref.~\cite{Raposo:2023oru}, it is shown that this object obeys identical relations to the finite- and infinite-volume amplitude as the kernel $\widetilde\Bc$ introduced in section~\ref{sec:QC_derivation}. More concretely, eq.~\eqref{eq:ML_full} mirrors eq.~(4.14) of ref.~\cite{Raposo:2023oru} and its infinite-volume version matches eq.~(4.18) of the earlier work. Consequently, we can identify $\overline{\mathcal K}{}^{\sf os}$ with $\widetilde\Bc$ and use this to relate it to $\Mc_0$. This allows us to write
\begin{equation}
(q^\star)^{\ell+\ell'} (\Mc_0^{-1})_{\ell'm';\,\ell m} = \big[\big(\overline{\mathcal K}{}^{\sf os} \big)^{-1} \big]_{\ell'm';\,\ell m} + (q^\star)^{\ell+\ell'} I_{\ell'm';\,\ell m} \,.
\label{eq:M0_Kos_rel}
\end{equation}

We can now substitute eqs.~\eqref{eq:FE_rel} and \eqref{eq:M0_Kos_rel} into the quantization condition eq.~\eqref{eq:QC_prev} and to remove the extra factors of on-shell momentum, yielding the expected condition eq.~\eqref{eq:QC_cc} in the case of a single scattering channel with identical spinless particles.

For the arbitrary spin case, we proceed almost identically to the spinless case. The relevant objects $\overline{\mathcal K}{}^{\sf os}$, $F^\Ec$, $\widetilde\Ec$ and $S$ now acquire spin indices, and the determinant in eq.~\eqref{eq:QC_prev} is performed over this extended index space. We may keep these in the basis of orbital momentum and individual particle spins $\ell m m_{s_1} m_{s_2}$, which arises naturally in the analysis of two-particle loops, or change to more convenient bases, such as total angular momentum $J m_J \ell S$, as described in section~\ref{sec:QC_spin}. Besides accommodating the additional spin indices, eqs.~\eqref{eq:Delta2_rel} to \eqref{eq:M0_Kos_rel} then require no further modification and we reach eq.~\eqref{eq:QC_cc_spin}.

We close this section by commenting on key differences between the expressions given in eqs.~\eqref{eq:QC_prev}--\eqref{eq:Epw_def} and the original expressions in ref.~\cite{Raposo:2023oru}: Firstly, a minor detail is that the we use the finite-volume momentum $\mathbf k$ here as an index, instead of the CM-frame-boosted $\mathbf k^\star$. This is not an issue, as the Lorentz transformations provide a one-to-one mapping between the boosted and unboosted momenta. Secondly, the finite-volume function $F^\Ec$ above is defined using the projections $\widetilde\Ec$ of the exchange function $\Ec$. This replaces $F^\Tc$ in the previous article, where only the projection of the $t$-channel contribution is used:
\begin{equation}
F^\Tc \equiv v^T S \frac{1}{1 + 2\widetilde\Tc S} v \,,
\label{eq:FT_def}
\end{equation}
with $\widetilde\Tc$ being the $t$-channel exchange, with kinematics as shown in eq.~\eqref{eq:T_replacement}, projected to spherical harmonics as done in eq.~\eqref{eq:Epw_def}. The factor of 2 in front of $\widetilde\Tc$ arises from converting $u$-channel exchanges into $t$-channel exchanges, which we are able do in the case of identical particles. The symmetry properties of $\overline{\mathcal K}{}^{\sf os}$ under particle exchanges, discussed briefly at the end of Appendix C of ref.~\cite{Raposo:2023oru}, allow us to show that
\begin{equation}
\Big[ (\overline{\mathcal K}{}^{\sf os})^{-1} + F^\Ec \Big]^{-1} = \Big[ (\overline{\mathcal K}{}^{\sf os})^{-1} + F^\Tc \Big]^{-1} \,,
\end{equation}
up to the absorption of non-singular contributions into $\overline{\mathcal K}{}^{\sf os}$. As a result, the two versions of the quantization condition must also be equivalent. Thirdly, relating to the definition of $S$. The two-particle pole in eq.~\eqref{eq:SL_def} carries a factor of $\omega^\star/(2E^\star\omega(\mathbf k))$, as opposed to $1/4\omega(\mathbf k)$ in eq.~(5.3) of ref.~\cite{Raposo:2023oru}. These factors coincide at the pole $(\mathbf k^\star)^2 = (q^\star)^2$ and thus the two definitions are equivalent up to a non-singular contribution that is absorbed into a redefinition of the kernel $\overline{\mathcal K}{}^{\sf os}$.

\section{Physical scattering amplitudes}
\label{sec:scattering_amps}

The derivation of the physical amplitude $\Mc$ is relatively straightforward now that we have defined the finite-volume amplitude in Secs.~\ref{sec:QC_derivation} and \ref{sec:QC_spin}. This can be directly obtained from $\Mc_L$ in eq.~\eqref{eq:ML_all_vf} by taking the ordered double limit
\begin{align}
\Mc = \lim_{\epsilon \to 0} \lim_{L\to \infty} \Mc_L \,.
\label{eq:double_lim}
\end{align}
In practice, this means one must replace the sums over discrete momentum with integrals over continuous momentum:
\begin{align}
\frac{1}{L^3}\sum_{\mathbf k}
~ \longrightarrow ~
\int \! \frac{{\rm d}^3\mathbf k}{(2\pi)^3} \,,
\end{align}
such that finite-volume loops become infinite-volume loops. Using this prescription, the expression relating $\Mc$ to the quantities $\Mc_0$ and $\Mc_\Ec$ in momentum space will look essentially identical to eq.~\eqref{eq:ML_all_vf} without the $L$ subscripts, where $\Mc_0$ and $\Mc_\Ec$ are defined as the ordered infinite-volume limits of $\Mc_{0,L}$ and $\Mc_{\Ec,L}$, respectively. This will then provide a direct relation between $\Mc_0$, the quantity appearing in the quantization conditions derived in section~\ref{sec:QC}, to the desired physical amplitude $\Mc$.

The following subsection deals with the partial-wave-projected infinite-volume amplitude for the case of particles with zero intrinsic spin. In section~\ref{sec:amps_spin}, we generalize this to particles with non-zero spin. We close with section~\ref{sec:OPE}, which details the general procedure for projecting OPEs to definite total angular momentum and parity ($J^P$) and provides explicit expressions for the cases of spinless-spinless and vector-spinless scattering.

Because we are interested in physical amplitudes that are Lorentz invariant in this section, we restrict our attention to the CM frame, where the total four-momentum is simply $P = (\sqrt{s},0)$. As such, we will omit the $\star$ superscripts when denoting CM kinematics, to avoid notation clutter.

\subsection{Partial-wave projected amplitudes for spinless particles}
\label{sec:amps_nospin}

The goal of this subsection is to show that, after partial-wave projecting the amplitude to a definite $J^P$, it can be written as
\begin{align}
\Mc^{J^P} &=
\Mc^{J^P}_{\Ec}
+
\left(1+i\Lc^{J^P}
\right) \, \left[\left(\Mc^{J^P}_{0}\right)^{-1}+
\Cc^{J^P} \right]^{-1}
\left(1
+
i\Rc^{J^P}\right),
\label{eq:M_all_vf}
\end{align}
where we have suppressed the dependence on the total energy of the system, as well as indices over channel space. The objects appearing in this equation, which we define in detail below, are all matrices in channel space. As the reader may note, this equation closely resembles the expression for the finite-volume amplitude in terms of the auxiliary amplitudes $\Mc_{0,L}$ and $\Mc_{\Ec,L}$ given in eq.~\eqref{eq:ML_all_vf}.

Let us now proceed to define each one of the building blocks above, starting with the partial-wave-projected auxiliary amplitudes $\Mc_{0}^{J^P}$ and $\Mc_{\Ec}^{J^P}$. The functions $\Rc^{J^P}$, $\Lc^{J^P}$, and $\Cc^{J^P}$ are projections of the infinite-volume analogs of $\Rc_L$, $\Lc_L$, and $\Cc_L$ defined in eqs.~\eqref{eq:RL}, \eqref{eq:LL}, and \eqref{eq:CL}, respectively. Since these depend on $\Mc_\Ec$, we will address them last.

In order to define $\Mc_{0}^{J^P}$ and $\Mc_{\Ec}^{J^P}$, we first write down the integral equations satisfied by $\Mc_0$ and $\Mc_\Ec$ and then partial-wave project. These equations are the infinite-volume limit, as defined in eq.~\eqref{eq:double_lim}, of those satisfied by $\Mc_{0,L}$ and $\Mc_{\Ec,L}$, presented in eqs.~\eqref{eq:M0L} and \eqref{eq:M0tL}, respectively. Explicitly,
\begin{align}
i\Mc_{0} &=
i\widetilde{\Bc}
+ i\widetilde{\Bc}\,\cdot i \Delta_{2,\infty} \cdot \, \, i\Mc_{0} \,,
\label{eq:M0}
%%%%%%%%%%%%%%%%%%%%%%%%%%%%%%%%%%%%
\\
i\Mc_{\Ec}
&
=
i\Ec
+ i\Ec\,
\cdot i \Delta_{2,\infty} \cdot \,
i\Mc_{\Ec} \,,
\label{eq:M0t}
\end{align}
where $\widetilde{\Bc}$ and $\Ec$ are the same driving terms appearing in the definition of the finite-volume amplitudes. Here, we have also introduced $\Delta_{2,\infty}$ and a dot product notation. These are jointly defined via
\begin{align}
\left[i\Ac'\,\cdot i \Delta_{2,\infty} \cdot \, i\Ac\right]_{a'a}
&\equiv
\sum_{b}
\int \! \frac{{\rm d}^3 \mathbf k}{(2\pi)^3} ~
i\Ac'_{a'b}(\mathbf k) \,
(i\Delta_{2,\infty})_b (\mathbf k)\,
i\Ac_{ba}(\mathbf k) \,, \\
%%%%%%%%%%%%%%%%%%%%%%%%%%%%%%%%%%%%%%%%%%%%%
&\equiv
\sum_{b}
\int \! \frac{{\rm d}^3 \mathbf k}{(2\pi)^3} ~
i\Ac'_{a'b}(\mathbf k) \, \xi_b \, \frac{\omega_1 (q_b \hat{\mathbf k})}{\omega_1 (\mathbf k)}
\frac{ i \, H_b (\mathbf k)}{2 E \big[ q_b^2 - \mathbf k^2 + i\epsilon \big] } \,
i\Ac_{ba}(\mathbf k) \,,
\label{eq:Delta2inf}
\end{align}
where $\Ac$ and $\Ac'$ are generic infinite-volume quantities, and we omit their dependence on external momentum argument. Note that $\Delta_{2,\infty}$ is the infinite-volume version of $\Delta_{2,L}$, defined in eq.~\eqref{eq:Delta2L_def}.

The integral equation for $\Mc_0$ can be solved algebraically, by first writing it in the partial-wave basis. By considering an arbitrary number of open channels composed of spinless distinguishable particles, the partial wave basis is obtained using eq.~\eqref{eq:M0L_PWA},
\begin{align}
\left( \Mc_0 \right)_{a'a}(\mathbf p'; \mathbf p) & =
\sum_{\ell'm', \ell m} \Yc_{\ell' m'}(\mathbf p', q_{a'}) \, \left( \Mc_0 \right)_{a'\ell' m';\, a\ell m} \, \Yc^T_{\ell m} (\mathbf{p}, q_a) \,,
\label{eq:M0_PWA}
\end{align}
where again we have suppressed the dependence on the total energy of the system for the projected amplitudes, which are fixed to be on-shell. Since $\Mc_0$ does not depend on the azimuthal quantum number $m$, and the orbital angular momentum $\ell$ is conserved in channels with two spinless bosons, we can write $\left( \Mc_0 \right)_{a'\ell' m';\, a\ell m} \equiv \delta_{\ell'\ell} \, \delta_{m'm} \left( \Mc_0 \right)_{a'\ell;\,a\ell} $, simplifying the equation above to
\begin{align}
\left( \Mc_0 \right)_{a'a} (\mathbf p', \mathbf p) = \sum_{\ell m} \Yc_{\ell m} (\mathbf p', q_{a'}) \, \left( \Mc_0 \right)_{a'\ell;\,a\ell} \, \Yc^T_{\ell m} (\mathbf{p}, q_a) \,.
\label{eq:M0_PWA}
\end{align}
The orbital angular momentum $\ell$ fixes the total angular momentum ($J$) and parity ($P$) of the system,
\begin{align}
\big( \Mc_0^{J^P} \big)_{a'a}
= \left( \Mc_0 \right)_{a'\ell;\,a\ell} \delta_{J\ell} \,,
\end{align}
and $P=(-1)^{\ell} \, \pi_{a1}\pi_{a2}$, where $\pi_{a1}$ and $\pi_{a2}$ are the intrinsic parity of the two particles in channel $a$.

There is a subtle point worth addressing about the partial-wave projection used in eq.~\eqref{eq:M0_PWA}. In particular, this partial-wave projection is slightly unconventional due to the explicit presence of the barrier factors embedded in the definition of $\Yc$ and $\Yc^T$ in eqs.~\eqref{eq:Yharm_def} and \eqref{eq:YharmT_def}, respectively. This definition is forced upon us from the partial-wave projection of $\Mc_{0,L}$ in eq.~\eqref{eq:M0L_PWA}. The need for this is to avoid unphysical singularities in finite-volume functions, such as $F$ in eq.~\eqref{eq:F_def}, at threshold. Because we have used these barrier factors in writing our quantization condition in terms of partial-wave projected quantities, this is propagated onto the definition of the partial-wave projected $\widetilde{\Bc}$ kernels. This is to say, this is yet another way in which the $\widetilde{\Bc}$ kernels are scheme-dependent. Consequently, the integral equation relating $\widetilde{\Bc}^{\,J^P}$ to $\Mc_0^{J^P}$ depends on these barrier factors, a dependence subsequently inherited by the projections $\Mc_0^{J^P}$.

In contrast to $\Mc_0$, we are not aware of an algebraic solution for the integral equation for $\Mc_\Ec$, eq.~\eqref{eq:M0t}, that gives the correct analytic structure. In this case, however, we know the driving term of this integral equation, namely the OPE function $\Ec$, explicitly. Consequently, we can attempt to solve this equation numerically. Remarkably, since it closely resembles the integral equations for three-body systems, we may adopt the techniques already in use to solve these~\cite{Jackura:2020bsk, Dawid:2023jrj,Jackura:2023qtp,Briceno:2024ehy, Dawid:2024dgy}. The first step towards this goal is to write the integral equation for the partial-wave-projected $\Mc_\Ec^{J^P}$. The only quantity that makes this step non-trivial is $\Ec$, since it carries angular dependence in the exchange propagator, as well as in the couplings for systems with non-zero spin. In section~\ref{sec:OPE}, we give a detailed description of how the OPE can be projected to definite $J^P$ in general. For spinless particles, this procedure is relatively straightforward.

As discussed above, when partial-wave projecting $\Mc_0$, we are ``forced'' to introduce the barrier factors. Since we did not project $\Ec$ in the derivation of the finite-volume formalism, we have more freedom in defining this projection now. In turn, we choose to define a more standard projection without barrier factors:
\begin{align}
\Ec_{a'a}(\mathbf p', \mathbf p)
= \sum_{\ell m} 4\pi\,Y_{\ell m}(\hat{\mathbf p}')
\, \Ec_{a'\ell;\,a\ell} (|\mathbf p'|, |\mathbf p|) \, Y^*_{\ell m}(\hat{\mathbf p}) \,,
\label{eq:Ec_PWA}
\end{align}
where we again drop azimuthal quantum number indices in the projections and use the fact that orbital angular momentum is conserved for spinless particles. Another notable distinction between this equation and eq.~\eqref{eq:M0_PWA} is the fact that the projections depend on the spatial momentum magnitudes, $|\mathbf p'|$ and $|\mathbf p|$. This is because $\Ec$ is not fully on shell (note $(P-p)^2 \neq m_2^2$ and $(P-p')^2 \neq m_4^2$), unlike $\Mc_0$ above.\footnote{It is worth commenting that because $\Ec$ is a partially off-shell quantity, we could have very well introduced barrier factors when partial-wave projecting. This difference would just be absorbed into the definition of the projections $\Ec_{a'\ell;\,a\ell}$, as done in ref.~\cite{Raposo:2023oru}.} Again, we have that
\begin{align}
\Ec_{a'a}^{J^P} = \Ec_{a'\ell;\,a\ell} \, \delta_{J\ell} \,.
\label{eq:Ec_PWA_v2}
\end{align}
Using similar arguments, we can write analogous expressions for $\Mc_\Ec$, namely
\begin{align}
\left( \Mc_{\Ec} \right)_{a'a} (\mathbf{p}', \mathbf{p})
&= \sum_{\ell m} 4\pi \, Y_{\ell m}(\hat{\mathbf p}') \, (\Mc_{\Ec})_{a'\ell;\,a\ell} (|\mathbf p'|, |\mathbf p|) \, Y^*_{\ell m} (\hat{\mathbf p}) \,,
\label{eq:Mc_t_PWA} \\
\big( \Mc_\Ec^{J^P} \big)_{a'a} &= \left( \Mc_\Ec \right)_{a'\ell;\,a\ell} \delta_{J\ell} \,.
\label{eq:Mc_t_PWA_v2}
\end{align}

Inserting eqs.~\eqref{eq:Ec_PWA} and \eqref{eq:Mc_t_PWA} into the integral equation for $\Mc_\Ec$, eq.~\eqref{eq:M0t}, and performing the integration over the angular degrees of freedom, we obtain an integral equation for $\Mc_\Ec^{J^P}$:
\begin{align}
\big( i \Mc_{\Ec}^{J^P} \big)_{a'a} (p', p)
&= i \Ec_{a'a}^{J^P} (p', p)
+ \sum_{b} \int_0^\infty \! \frac{k^2 {\rm d}k}{2\pi^2} ~ i\Ec_{a'b}^{J^P}(p', k) \, (i\Delta_{2,\infty})_b (k) \,
\big( i\Mc_\Ec^{J^P} \big)_{ba} (k, p) \,,
\label{eq:Mt_int}
\end{align}
where the integral is now one-dimensional, over the magnitude of spatial momentum $k$.\footnote{We point the reader to ref.~\cite{Briceno:2024ehy}, where a nearly identical projection of integral equations was done for three-particle systems.} Note that all arguments here are spatial momentum magnitudes, not four-momenta.

Although the integral runs to $\infty$, the cut-off function in the definition $\Delta_{2,\infty}$ allows one to truncate the integral at some value of $k$, which we denote $k_{\rm max}$. Given a value for $k_{\rm max}$, one can discretize the momentum, and write eq.~\eqref{eq:Mt_int} as a matrix equation over generally off-shell momenta. This is most easily solved by fixing $p$ and $p'$ to be in the set of discrete momenta. As discussed in great detail in ref.~\cite{Dawid:2023jrj}, for singular integrands and driving terms one may need to deform the contour of integration to avoid singularities. This is certainly the case for eq.~\eqref{eq:Mt_int}, where there are two classes of singularities, logarithmic ones present in $\Ec^{J^P}$ and the pole singularity present in $\Delta_{2,\infty}$. After solving for the amplitude along the discrete values of the momenta defined along the contour, one can use the integral equation, eq.~\eqref{eq:Mt_int}, to evaluate $\Mc^{J^P}_\Ec$ for on-shell momenta. For more details on this procedure, we point the reader to ref.~\cite{Dawid:2023jrj} and references within.

To briefly summarize, given the finite-volume spectrum, we can determine $\big( \Mc_0^{J^P} \big)_{a'a}$ from the quantization condition, eq.~\eqref{eq:QC_cc}. Given the $\Ec^{J^P}_{a'a}$, we know how to obtain $\big( \Mc_\Ec^{J^P} \big)_{a'a}$ from its integral equation, eq.~\eqref{eq:Mt_int}. These are the two key ingredients to obtain the full amplitude.

The last three quantities we need to examine before arriving at the expression for $\Mc^{J^P}$ given in eq.~\eqref{eq:M_all_vf} are $\Rc^{J^P}$, $\Lc^{J^P}$, and $\Cc^{J^P}$. As mentioned, these are projections of the infinite-volume limits of the objects $\Rc_L$, $\Lc_L$, and $\Cc_L$, respectively defined in eqs.~\eqref{eq:RL}, \eqref{eq:LL}, and \eqref{eq:CL}. We denote the infinite-volume objects by dropping the $L$ subscript and write the continuous momentum dependence as an argument instead of an index. We then decompose the remaining angular dependence in $\Rc$ and $\Lc$ using spherical harmonics:
\begin{align}
\Rc_{a'\ell' m';\, a}(\mathbf p) &
\equiv
\sum_{\ell m}
\delta_{\ell' \ell} \,
\delta_{m' m} \Rc_{a'\ell;\, a\ell}(|\mathbf p|) \, \sqrt{4\pi} \, Y^*_{\ell m}(\hat{\mathbf p})
=
\Rc_{a'\ell';\,a\ell'}(|\mathbf p|) \, \sqrt{4\pi} \, Y^*_{\ell' m'}(\hat{\mathbf p}) \,, \\
\Lc_{a';\, a\ell m}(\mathbf p') & \equiv \sum_{\ell'm'} \sqrt{4\pi} \, Y_{\ell' m'}(\hat{\mathbf p}') \, \delta_{\ell'\ell}\, \delta_{m'm} \,\Lc_{a'\ell;\,a\ell} (|\mathbf p'|) = \sqrt{4\pi} \, Y_{\ell m}(\hat{\mathbf p}') \, \Lc_{a'\ell;\,a\ell} (|\mathbf p'|) \,.
\end{align}
Again, we use the fact that these are infinite-volume objects, and thus independent of $m$ and diagonal in $\ell$ for spinless bosons, to simplify the relations above. The same is true for $\Cc$, which is written already in the angular momentum basis: $\Cc_{a'\ell m;\, a\ell m} \equiv \delta_{\ell'\ell} \, \delta_{m'm} \, \Cc_{a'\ell;\,a\ell}$. The projections to definite $J^P$ then take the usual form:
\begin{equation}
\Rc^{J^P}_{a'a} = \Rc_{a'\ell;\,a\ell} \, \delta_{J\ell} \,, \qquad
\Lc^{J^P}_{a'a} = \Lc_{a'\ell;\,a\ell} \, \delta_{J\ell} \,, \qquad \Cc^{J^P}_{a'a} = \Cc_{a'\ell;\,a\ell} \, \delta_{J\ell} \,.
\end{equation}
Using eq.~\eqref{eq:Mc_t_PWA} to project $\Mc_\Ec$ within the definitions of $\Rc$, $\Lc$ and $\Cc$, and using the orthogonality relation between spherical harmonics, we can arrive at
\begin{align}
i\Rc^{J^P}_{a'a} (p)
&= \int_{0}^\infty \! \frac{k^2 {\rm d}k}{2\pi^2} \left[\frac{k}{q_{a'}} \right]^{J} \! (i\Delta_{2,\infty} )_{a'} (k) \, \big( i\Mc_\Ec^{J^P} \big)_{a'a} (k, p) \,,
\label{eq:Rinf} \\
%%%%%%%%%%%%%%%%%%%%%%%%%%%%%%%%%%%%%%%%%%%%%%%%
i\Lc^{J^P}_{a'a} (p') &= \int_{0}^\infty \! \frac{k^2 {\rm d}k}{2\pi^2} \big( i\Mc_\Ec^{J^P} \big)_{a'a} (p',k) \, (i\Delta_{2,\infty})_{a} (k) \left[\frac{k}{q_a} \right]^{J} \,,
\label{eq:Linf} \\
%%%%%%%%%%%%%%%%%%%%%%%%%%%%%%%%%%%%%%%%%%%%%%%%
i\Cc^{J^P}_{a'a} &=
\int_{0}^\infty \! \frac{k'^2 {\rm d}k'}{2\pi^2}
\int_{0}^\infty \! \frac{k^2 {\rm d}k}{2\pi^2} \left[\frac{k'}{q_{a'}} \right]^{J} \! (i\Delta_{2,\infty})_{a'} (k')
\, \big( i\Mc_\Ec^{J^P} \big)_{a'a} (k', k) \, (i\Delta_{2,\infty} )_{a} (k) \left[\frac{k}{q_a} \right]^{J} \,.
\label{eq:Cinf}
\end{align}

As alluded to at the start of the section, the infinite-volume limit of eq.~\eqref{eq:ML_all_vf} now allows us to relate the partial-wave projections of the auxiliary amplitudes $\Mc_0$ and $\Mc_\Ec$, and those of the derived auxiliary quantities $\Rc$, $\Lc$ and $\Cc$, to those of the full amplitude $\Mc$. With the partial-wave decompositions given above for each of these objects, it is straightforward to derive eq.~\eqref{eq:M_all_vf} for on-shell momenta.

\subsection{Generalization to particles with spin}
\label{sec:amps_spin}

At this stage, including the possibility that the particles have intrinsic spin is conceptually straightforward. As discussed in section~\ref{sec:QC_spin}, by introducing spin, each of the building blocks considered becomes a matrix in spin space. Take, for example, the exchange functions, which acquire indices associated with the spin of the initial and final particles:
\begin{align}
\Ec_{a'a} (k'_1,k'_2;k_1,k_2)
& \to \Ec_{a'm_{s_1}'m_{s_2}';\,am_{s_1}m_{s_2}} (k'_1,k'_2;k_1,k_2) \,.
\label{eq:Ec_def_spin_v2}
\end{align}
Given this, one can use the steps described in section~\ref{sec:QC_spin} to change these objects to the $Jm_J\ell S$ basis and, thus, to definite $J^P$. Let us label the resulting projections by $\Ec_{a'\ell'S';\, a\ell S}^{J^P}(|\mathbf p'|,|\mathbf p|)$. Note that, since the particles now have intrinsic spin, these are no longer diagonal in orbital angular momentum.

Performing the appropriate projections and changes of basis, one can rewrite the integral equation for the projections of $\Mc_{\Ec}$, eq.~\eqref{eq:Mt_int}, in the form
\begin{multline}
\big( \Mc_\Ec^{J^P} \big)_{a'\ell'S';\,a\ell S} (p',p)
= i\Ec_{a'\ell'S';\,a\ell S}^{J^P} (p',p) \\
+ \sum_{b} \sum_{\ell''S''} \int_0^\infty \frac{k^2 {\rm d}k}{2\pi^2}\, \,\,
i\Ec_{a'\ell'S';\,b\ell'' S''}^{J^P}(p',k)
\,
(i\Delta_{2,\infty})_{b}(k)
\,
\big( i\Mc_\Ec^{J^P} \big)_{b\ell'' S'';\,a\ell S}(k,p) \,.
\label{eq:Mt_int_spin}
\end{multline}
This can be solved, as was done for three-body systems in ref.~\cite{Briceno:2024ehy}, by solving this as a matrix equation in momentum and $\ell S$ space.

The rest of the equations in the previous section can be immediately be generalized to systems with spin by promoting everything to include $\ell S$ indices,
\begin{align}
i\Rc^{J^P}_{a'\ell'S';\,a\ell S } (p)
&= \int_{0}^\infty \! \frac{k^2 {\rm d}k}{2\pi^2}
\left[\frac{k}{q_{a'}} \right]^{\ell'} \!(i\Delta_{2,\infty})_{a'}(k) \, (i{\Mc}^{J^P}_\Ec)_{a'\ell'S';\,a\ell S} (k, p) \,,
\label{eq:Rinf_spin} \\
%%%%%%%%%%%%%%%%%%%%%%%%%%%%%%%%%%%%%%%%%%%%%%%%
i\Lc^{J^P}_{a'\ell'S';\,a\ell S} (p')
&= \int_{0}^\infty \! \frac{k^2 {\rm d}k}{2\pi^2} \, (i{\Mc}^{J^P}_{\Ec})_{a'\ell'S';\,a\ell S} (p',k)
\, (i\Delta_{2,\infty})_{a} (k) \left[\frac{k}{q_a} \right]^{\ell} \,,
\label{eq:Linf_spin} \\
%%%%%%%%%%%%%%%%%%%%%%%%%%%%%%%%%%%%%%%%%%%%%%%%
i\Cc^{J^P}_{a'\ell'S';\,a\ell S}
&= \int_{0}^\infty \! \frac{k'^2 {\rm d}k'}{2\pi^2}
\int_{0}^\infty \! \frac{k^2 {\rm d}k}{2\pi^2}
\left[\frac{k'}{q_{a'}} \right]^{\ell'} \!
(i\Delta_{2,\infty})_{a'} (k')
\nn \\
&\hspace{110pt} \times (i\Mc^{J^P}_\Ec)_{a'\ell'S';\,a\ell S}(k', k) \, (i\Delta_{2,\infty})_{a}(k)
\left[\frac{k}{q_a} \right]^\ell \,.
\label{eq:Cinf_spin}
\end{align}
Given this enhanced matrix space, the main equation for $\Mc^{J^P}$, eq.~\eqref{eq:M_all_vf}, remains unchanged.

Of course, this hinges upon having an expression for $\Ec^{J^P}$. In the subsequent section we explain how this can be done. In particular, we use the helicity basis to arrive at some key examples.

\subsection{Evaluating the one-particle exchange}
\label{sec:OPE}

In this section, we explain how the OPE can be partial-wave projected onto the $\ell S$ basis. Given that in QCD the nearest left-hand cuts are due to pseudoscalar exchanges, we will assume that the exchange particle is spinless.

An example we will emphasize is $D^*D\to D^*D$ scattering in the $T_{cc}$ channel, where the nearest left-hand cut is due to $u$ channel $\pi$-exchange. Given this motivating example, let us look at a generic $u$-channel OPE, which, following section~\ref{sec:QC_spin}, can generically be written in the form
\begin{align}
\label{eq:ope_generic}
i\,\Uc = \left(\bar{\eta}_4 \, i\Gamma_{14} \, \eta_1 \right) \, \frac{i}{u - m_e^2 + i\epsilon} \, \left(\bar{\eta}_3 \, i\Gamma_{23} \, \eta_2 \right) \, ,
\end{align}
where $u = (p_1 - p_4)^2$ is the usual Mandelstam invariant, $m_e$ is the mass of the exchange particle and the $\Gamma$ tensors result from the on-shell reduction of the vertex functions. We use the shorthand notation $\eta_j \equiv \eta_{j,\lambda_j}(p_j)$ for the spin wave-functions, where $j$ labels the particle, $p_j$ is its four-momentum and $\lambda_j$ is its helicity. We omit channel indices as these will not be relevant for the following. We find that working in the helicity basis is particularly advantageous when writing explicit expressions for the vertex functions for amplitude analyses and projecting to partial waves. To express the amplitude in terms of fixed-$z$ spin projections, as is needed for the form of the QC presented in eq.~\eqref{eq:QC_cc_spin}, one must apply the following unitary transformations on the wave functions (see refs.~\cite{Jacob:1959at,Martin:1970hmp}):
\begin{align}
\eta_{m_s}(p) = \sum_{\lambda = -s}^{s} \eta_\lambda(p)\,D_{\lambda m_s}^{(s)}(\hat{\mathbf p}) \, ,
\end{align}
where $D_{\lambda m}^{(s)}$ is the Wigner $D$ rotation matrix element and $\hat{\mathbf p}$ is the orientation of the particle momentum which can be expressed in terms of the polar, $\theta$, and azimuthal, $\varphi$, angles of the particle, $\hat{\mathbf p} = (\sin\theta\cos\varphi,\sin\theta\sin\varphi,\cos\theta)$. The specific form of the wave functions and vertex factors depends on the particles involved in the reaction. Generally, for integer spin particles, the wave functions are (cf.~refs.~\cite{Auvil:1966eao, Durand:1962zza,Lorce:2009bs})
\begin{align}
\eta_\lambda(p) \Big\rvert_{s\in \mathbb{Z}^+} \longrightarrow \varepsilon^{\mu_1\cdots\mu_s}(p,\lambda) = \sum_{\lambda',\lambda''} \braket{1\lambda',(s-1) \lambda'' | s\lambda} \, \varepsilon^{\mu_1}(p,\lambda') \, \varepsilon^{\mu_2\cdots\mu_s}(p,\lambda'') \, ,
\end{align}
where $\braket{j_1m_1,j_2m_2 | jm}$ is the Clebsch-Gordan coefficient and $\varepsilon^{\mu}(p,\lambda)$ is the standard helicity basis spin-1 polarization four-vector. For half-integer spin particles, the wave functions are
\begin{align}
\eta_\lambda(p) \Big\rvert_{2s\in \mathbb{Z}^+} \longrightarrow u^{\mu_1\cdots\mu_s}(p,\lambda) = \sum_{\lambda',\lambda''} \braket{{\textstyle \frac{1}{2}} \lambda', (s - {\textstyle \frac{1}{2}}) \lambda'' | s \lambda} \, u(p,\lambda') \, \varepsilon^{\mu_1\cdots\mu_s}(p,\lambda'') \, ,
\end{align}
where $u(p,\lambda)$ is the standard helicity basis Dirac spinor.

The OPE in eq.~\eqref{eq:ope_generic} can be projected into a partial wave via the following procedure: First, we write the propagator as a function of the CM frame scattering angle $\theta$, defined via $\mathbf{p}_1 \cdot \mathbf{p}_3 = \lvert\mathbf{p}_1 \rvert\lvert \mathbf{p}_3\rvert \cos\theta$,
\begin{align}
u = (p_1 - p_4)^2 & = m_1^2 + m_4^2 - 2\omega_1\omega_4 - 2\lvert\mathbf{p}_1 \rvert\lvert \mathbf{p}_3\rvert \cos\theta \, , \nn \\[5pt]
& = u_0 - 2\lvert\mathbf{p}_1 \rvert\lvert \mathbf{p}_3\rvert \, (1 + \cos\theta) \, ,
\end{align}
where $u_0 = m_1^2 + m_4^2 - 2\omega_1\omega_4 + 2\lvert\mathbf{p}_1 \rvert\lvert \mathbf{p}_3\rvert$ is the backward limit of $u$. We can therefore write eq.~\eqref{eq:ope_generic} as
\begin{align}
\label{eq:ope_generic_v2}
i\,\Uc & = \frac{i\mathcal{H}_{\lambda_1,\lambda_2 , \lambda_3 \lambda_4}(\theta)}{2\lvert\mathbf{p}_1 \rvert\lvert \mathbf{p}_3\rvert(\zeta+i\epsilon - \cos\theta)} \, ,
\end{align}
where $\zeta$ corresponds to the value of $\cos\theta$ when the propagator goes on mass-shell, $u = m_e^2$,
\begin{align}
\zeta \equiv \cos\theta \rvert_{u = m_e^2} = -1 - \frac{m_e^2 - u_0}{2\lvert\mathbf{p}_1 \rvert\lvert \mathbf{p}_3\rvert} \, ,
\end{align}
and $\mathcal{H}_{\lambda_1,\lambda_2 , \lambda_3 \lambda_4}(\theta)$ is the product of vertex functions. In general, $\mathcal{H}$ is a complicated function of the scattering angle $\theta$. However, the dependence can be easily written for low-lying spins, see ref.~\cite{Jackura:2023qtp}, which we follow.

Following that work, here we assume that the reaction plane is oriented such that the azimuthal angle is zero, which we are free to do. Given this, the partial wave projection of eq.~\eqref{eq:ope_generic_v2} over the CM frame scattering angle is
\begin{align}
\Uc^J_{\substack{\lambda_1,\lambda_2 \\ \lambda_3, \lambda_4}} = \frac{1}{2}\int_{0}^{\pi}\!\mathrm{d}\theta \, \sin\theta \, \frac{d_{\lambda\lambda'}^{(J)}(\theta)}{2\lvert\mathbf{p}_1 \rvert\lvert \mathbf{p}_3\rvert(\zeta - \cos\theta)} \, \mathcal{H}_{\lambda_1,\lambda_2 , \lambda_3 \lambda_4}(\theta) \, ,
\end{align}
where $\lambda = \lambda_1 - \lambda_2$ and $\lambda' = \lambda_3 - \lambda_4$ and $d_{\lambda\lambda'}^{(J)}$ is the Wigner little $d$ matrix element.~\footnote{We use the usual convention for rotation matrices, $e^{-i \mathbf{J}\cdot \mathbf{\theta}}$, to define the Wigner $d$ matrix. In using our results, one should note that some software may use different conventions, for example, \textsc{Mathematica} uses a sign convention opposite of ours.} While the helicity basis is useful in constructing exchange amplitudes of definite $J$, our target amplitude should be of definite spin-parity $J^P$. Therefore, we apply a unitary transformation on the helicity partial wave amplitudes to those in the spin-orbit, or $\ell S$, basis~\cite{Jacob:1959at}
\begin{align}
\Uc_{\ell 'S';\,\ell S}^{J} = \sum_{\lambda_1,\lambda_2}\sum_{\lambda_3,\lambda_4} \mathcal{P}_{\lambda_3\lambda_4}({}^{2S'+1}\ell'_J) \, \Uc^J_{\substack{\lambda_1,\lambda_2 \\ \lambda_3, \lambda_4}} \, \mathcal{P}_{\lambda_1\lambda_2}({}^{2S+1}\ell_J) \, ,
\end{align}
where the coefficients
\begin{align}
\mathcal{P}_{\lambda_1\lambda_2}({}^{2S+1}\ell_J) = \sqrt{\frac{2\ell+1}{2J+1}} \, \braket{J\lambda | \ell 0 , S \lambda} \, \braket{S \lambda | s_1\lambda_1, s_2 \, (-\lambda_2)} \, ,
\end{align}
where the total angular momentum $J$ is restricted as $\lvert S - \ell \rvert \le J \le \ell + S$. Similar relations hold for the final state. Upon projecting to definite $J$, and recoupling to the $\ell S$ basis, the partial wave amplitudes take the form~\cite{Jackura:2023qtp}
\begin{align}
\label{eq:ope_pw_proj}
\Uc^{J^P} = \mathcal{C}_1^{J^P} + \mathcal{C}_2^{J^P}\,Q_0(\zeta + i\epsilon) \, ,
\end{align}
where $Q_0$ is the Legendre function of the second kind, the coefficients $\mathcal{C}_1$ and $\mathcal{C}_2$ which are matrices in $\ell S$ space to be tabulated for a target $J^P$. The $Q_0$ function has branch points at $\zeta = \pm 1$, which yields a branch cut, the so-called left-hand cut, in the complex $s$ plane of the scattering amplitude. The remainder $\mathcal{C}_1$ and $\mathcal{C}_2$ coefficients are assured to be non-singular functions in the kinematic region considered.

We do not attempt writing a generic amplitude and partial wave projection for any spin, but instead follow the procedure of ref.~\cite{Jackura:2023qtp} to construct partial wave OPE amplitudes and illustrate it by examining some cases of interest to the community.

\subsubsection{Spinless-Spinless scattering via Spinless Exchange}
Let us consider a generic process involving spinless particles in the external states. Since the wave functions are trivial, the OPE is
\begin{align}
i\,\Uc = ig' \, \frac{i}{u - m_e^2 + i\epsilon} \, ig \, ,
\end{align}
where $g$ and $g'$ are the couplings of particles 1 and 4 to $e$ and particles 2 and 3 to $e$, respectively. In this case, it is easy to show that the partial wave projected OPE is
\begin{align}
\Uc^{J} = \frac{g'g}{2\lvert\mathbf{p}_1 \rvert\lvert \mathbf{p}_3\rvert}\,Q_J(\zeta + i\epsilon) \, .
\end{align}
One can show that this is equivalent to eq.~\eqref{eq:ope_pw_proj} using an identity implemented in ref.~\cite{Jackura:2023qtp}. This follows from the fact that $Q_J$ can be written as an integral of the Legendre polynomial $P_J$. Then, one can use the recursion relation for the $P_J$, to find
\begin{align}
Q_J(\zeta) = P_J(\zeta) \, Q_0(\zeta) - W_{J - 1}(\zeta) \, ,
\end{align}
where $ W_{J - 1}$ is a polynomial in $\zeta$
\begin{align}
W_{\ell - 1}(\zeta) = \sum_{n=1}^{\ell} \frac{1}{n} P_{n-1}(\zeta) \, P_{\ell - n}(\zeta) \, .
\end{align}

\subsubsection{Vector-Spinless scattering via Spinless Exchange}
\label{sec:vector_scalar_ope}
Next, we consider the $u$-channel exchange in a vector-spinless reaction, e.g., $D^*D \to D^*D$ via $\pi$ exchange. In particular, we consider a generic $0^{\mp} + 1^{-} \to 1^{-} + 0^{\mp}$ process via a $0^{\mp}$ exchange. Since the vertex is proportional to a polarization vector (see ref.~\cite{Durand:1962zza}), which must satisfy the transversity condition $p_3 \cdot \varepsilon(p_3, \lambda_3) = 0$, one can write the most general OPE for such processes in the form,
\begin{align}
i\,\Uc = i [g' \, p_2 \cdot \varepsilon^*(p_3,\lambda_3) ] \, \frac{i}{u - m_\pi^2 + i\epsilon} \, i [g\,p_4 \cdot \varepsilon(p_1,\lambda_1) ] \, .
\end{align}
The polarization vector in the helicity basis (in the reaction plane) is defined as
\begin{align}
\varepsilon(p,\pm) & = \frac{1}{\sqrt{2}} (0, \mp \cos\theta, -i, \pm \sin\theta) \, , \\[5pt]
\varepsilon(p,0) & = \frac{1}{m}(\lvert\mathbf{p}\rvert, \omega\sin\theta, 0, \omega \cos\theta) \, .
\end{align}
By direct evaluation, the $\mathcal{H}$ function $\mathcal{H}_{\lambda_1\lambda_3} \equiv - g'g\,[p_2 \cdot \varepsilon^*(p_3,\lambda_3) ] [p_4 \cdot \varepsilon(p_1,\lambda_1) ]$ is
\begin{align}
\mathcal{H}_{\pm\pm} = -\mathcal{H}_{\pm\mp} & = g'g\,\frac{1}{2} \lvert\mathbf{p}_1 \rvert\lvert \mathbf{p}_3\rvert \sin^2\theta \, , \\[5pt]
\mathcal{H}_{\pm 0} & = \mp g'g \, \frac{\lvert\mathbf{p}_1\rvert }{\sqrt{2} \, m_1} \sin\theta \, \left(\omega_4 \lvert\mathbf{p}_1\rvert + \omega_1 \lvert\mathbf{p}_3\rvert \,\cos\theta \right) \, , \\[5pt]
\mathcal{H}_{0\pm} & = \pm g'g\, \frac{\lvert\mathbf{p}_3\rvert }{\sqrt{2} \, m_3 }\sin\theta \, \left(\omega_2 \lvert\mathbf{p}_3\rvert + \omega_3 \lvert\mathbf{p}_1\rvert \,\cos\theta \right) \, , \\[5pt]
\mathcal{H}_{0 0} & = -\frac{g'g}{m_1 m_3} \left(\omega_4 \lvert\mathbf{p}_1\rvert + \omega_1 \lvert\mathbf{p}_3\rvert \,\cos\theta \right) \left(\omega_2 \lvert\mathbf{p}_3\rvert + \omega_3 \lvert\mathbf{p}_1\rvert \,\cos\theta \right)\, .
\end{align}
Let $\pi$ and $\pi'$ be the product of the intrinsic parities for the initial and final state, respectively. For the case of $D^*D\to D^*D$, we have $\pi = \pi' = +$. Thus, the allowed parity of the system is $P = (-1)^\ell = (-1)^{\ell'}$. A particularly interesting case is the $J^P = 1^+$ channel, where the $T_{cc}$ has been discovered by the LHCb collaboration. Here, both $S$ and $D$ waves contribute. For example, following ref.~\cite{Jackura:2023qtp}, we find the $\Cc_1$ and $\Cc_2$ coefficients for the ${}^3S_1\to{}^3S_1$ channel are
\begin{align}
\Cc_1^{1^+}({}^3S_1\to{}^3S_1) & = -\frac{g'g}{6}\left[ \frac{2}{3}(1-\gamma_3)(1-\gamma_1) - \gamma_3\gamma_1 + \zeta f - f' \right] \, , \\[5pt]
\Cc_2^{1^+}({}^3S_1\to{}^3S_1) & = -\frac{g'g}{6}\left[ (1-\zeta^2) f + \zeta f' \right] \, ,
\end{align}
where $\gamma_j = 1 / \sqrt{1 - \beta_j^2}$ with $\beta_j = \lvert\mathbf{p}_j\rvert/\omega_j$ for $j=1,\ldots,4$, and
\begin{align}
f = \gamma_3 \left( \frac{\beta_3}{\beta_2} + \zeta \right) + \gamma_1
\left( \frac{\beta_1}{\beta_4} + \zeta \right) - \zeta \, ,
\end{align}
and
\begin{align}
f' = \gamma_3\gamma_1 \zeta \left( \frac{\beta_3}{\beta_2} + \frac{\beta_1}{\beta_4} + \zeta \right) + \gamma_3\gamma_1 \frac{\beta_3\beta_1 }{\beta_4\beta_2} \, .
\end{align}
Further cases can be evaluated similarly. Note that one can relate the results presented in ref.~\cite{Jackura:2023qtp} by identifying $\mathcal{H} = (-g'g / 3) \, q_p^\star q_k^\star \,\mathcal{H}'$, where $q_k^\star$, $q_p^\star$ and $\mathcal{H}'$ are as defined in that work (the latter without the prime).

\section{Conclusions}

We have derived a generic non-perturbative framework that relates finite-volume energy spectra to two-body scattering amplitudes, including nearby left-hand singularities arising from single-particle exchanges. The formalism allows for an arbitrary number of two-body channels in which the particles can have arbitrary spin. Applications of this framework proceed in two steps: first, the new quantization condition is used to constrain the intermediate quantity $\mathcal M_0$ that is insensitive to the left-hand cut; second, the full scattering amplitude is reconstructed from $\mathcal M_0$ via a set of integral equations that also incorporate all single-particle exchanges.

It is straightforward to generalize this work to include branch cuts due to two- or more particles using models due to these exchanges, as was done recently in ref.~\cite{Dawid:2024oey}. While higher-particle exchanges can be incorporated similarly, we expect their effects will be suppressed compared with OPE, thus we leave their investigation for future applications.

We conclude by emphasizing two key messages. First, this formalism is a conceptually simple modification of previously published two-particle formalisms that do not account for the presence of the one-particle exchange. Second, the formalism is complete and ready for immediate application.

%%%%%%%%%%%%%%%%%%%%%%%%%%%%%%%%%%%%
%	SECTION - Acknowledgements
%%%%%%%%%%%%%%%%%%%%%%%%%%%%%%%%%%%%
\section*{Acknowledgements}

ABR is supported by the European Research Council (ERC) consolidator grant StrangeScatt-101088506. RAB was supported in part by the U.S. Department of Energy, Office of Science, Office of Nuclear Physics under Awards No. DE-AC02-05CH11231. AWJ acknowledges the support of the USDOE ExoHad Topical Collaboration, contract DE-SC0023598. MTH is supported in part by UK STFC grants ST/T000600/1 and ST/X000494/1 and additionally by UKRI Future Leader Fellowship MR/T019956/1.

\bibliographystyle{jhep}
\bibliography{ref.bib}

\providecommand{\href}[2]{#2}\begingroup\raggedright\begin{thebibliography}{100}

\bibitem{Raposo:2023oru}
A.~B. Raposo and M.~T. Hansen, \emph{{Finite-volume scattering on the left-hand
  cut}}, \href{https://doi.org/10.1007/JHEP08(2024)075}{\emph{JHEP} {\bfseries
  08} (2024) 075} [\href{https://arxiv.org/abs/2311.18793}{{\ttfamily
  2311.18793}}].

\bibitem{Pelaez:2021dak}
J.~R. Pel\'aez, A.~Rodas and J.~R. de~Elvira, \emph{{Precision dispersive
  approaches versus unitarized chiral perturbation theory for the lightest
  scalar resonances $\sigma /f_0(500) $ and $\kappa /K_0^*(700) $}},
  \href{https://doi.org/10.1140/epjs/s11734-021-00142-9}{\emph{Eur. Phys. J.
  ST} {\bfseries 230} (2021) 1539}
  [\href{https://arxiv.org/abs/2101.06506}{{\ttfamily 2101.06506}}].

\bibitem{Accardi:2023chb}
A.~Accardi et~al., \emph{{Strong interaction physics at the luminosity frontier
  with 22 GeV electrons at Jefferson Lab}},
  \href{https://doi.org/10.1140/epja/s10050-024-01282-x}{\emph{Eur. Phys. J. A}
  {\bfseries 60} (2024) 173}
  [\href{https://arxiv.org/abs/2306.09360}{{\ttfamily 2306.09360}}].

\bibitem{Luscher:1986pf}
M.~Luscher, \emph{{Volume Dependence of the Energy Spectrum in Massive Quantum
  Field Theories. 2. Scattering States}},
  \href{https://doi.org/10.1007/BF01211097}{\emph{Commun. Math. Phys.}
  {\bfseries 105} (1986) 153}.

\bibitem{Huang:1957im}
K.~Huang and C.~N. Yang, \emph{{Quantum-mechanical many-body problem with
  hard-sphere interaction}},
  \href{https://doi.org/10.1103/PhysRev.105.767}{\emph{Phys. Rev.} {\bfseries
  105} (1957) 767}.

\bibitem{Rummukainen:1995vs}
K.~Rummukainen and S.~A. Gottlieb, \emph{{Resonance scattering phase shifts on
  a nonrest frame lattice}},
  \href{https://doi.org/10.1016/0550-3213(95)00313-H}{\emph{Nucl. Phys. B}
  {\bfseries 450} (1995) 397}
  [\href{https://arxiv.org/abs/hep-lat/9503028}{{\ttfamily hep-lat/9503028}}].

\bibitem{Feng:2004ua}
X.~Feng, X.~Li and C.~Liu, \emph{{Two particle states in an asymmetric box and
  the elastic scattering phases}},
  \href{https://doi.org/10.1103/PhysRevD.70.014505}{\emph{Phys. Rev. D}
  {\bfseries 70} (2004) 014505}
  [\href{https://arxiv.org/abs/hep-lat/0404001}{{\ttfamily hep-lat/0404001}}].

\bibitem{He:2005ey}
S.~He, X.~Feng and C.~Liu, \emph{{Two particle states and the S-matrix elements
  in multi-channel scattering}},
  \href{https://doi.org/10.1088/1126-6708/2005/07/011}{\emph{JHEP} {\bfseries
  07} (2005) 011} [\href{https://arxiv.org/abs/hep-lat/0504019}{{\ttfamily
  hep-lat/0504019}}].

\bibitem{Christ:2005gi}
N.~H. Christ, C.~Kim and T.~Yamazaki, \emph{{Finite volume corrections to the
  two-particle decay of states with non-zero momentum}},
  \href{https://doi.org/10.1103/PhysRevD.72.114506}{\emph{Phys. Rev. D}
  {\bfseries 72} (2005) 114506}
  [\href{https://arxiv.org/abs/hep-lat/0507009}{{\ttfamily hep-lat/0507009}}].

\bibitem{Kim:2005gf}
C.~h. Kim, C.~T. Sachrajda and S.~R. Sharpe, \emph{{Finite-volume effects for
  two-hadron states in moving frames}},
  \href{https://doi.org/10.1016/j.nuclphysb.2005.08.029}{\emph{Nucl. Phys. B}
  {\bfseries 727} (2005) 218}
  [\href{https://arxiv.org/abs/hep-lat/0507006}{{\ttfamily hep-lat/0507006}}].

\bibitem{Lage:2009zv}
M.~Lage, U.-G. Meissner and A.~Rusetsky, \emph{{A Method to measure the
  antikaon-nucleon scattering length in lattice QCD}},
  \href{https://doi.org/10.1016/j.physletb.2009.10.055}{\emph{Phys. Lett. B}
  {\bfseries 681} (2009) 439}
  [\href{https://arxiv.org/abs/0905.0069}{{\ttfamily 0905.0069}}].

\bibitem{Bernard:2010fp}
V.~Bernard, M.~Lage, U.~G. Meissner and A.~Rusetsky, \emph{{Scalar mesons in a
  finite volume}}, \href{https://doi.org/10.1007/JHEP01(2011)019}{\emph{JHEP}
  {\bfseries 01} (2011) 019} [\href{https://arxiv.org/abs/1010.6018}{{\ttfamily
  1010.6018}}].

\bibitem{Fu:2011xz}
Z.~Fu, \emph{{Rummukainen-Gottlieb's formula on two-particle system with
  different mass}},
  \href{https://doi.org/10.1103/PhysRevD.85.014506}{\emph{Phys. Rev. D}
  {\bfseries 85} (2012) 014506}
  [\href{https://arxiv.org/abs/1110.0319}{{\ttfamily 1110.0319}}].

\bibitem{Leskovec:2012gb}
L.~Leskovec and S.~Prelovsek, \emph{{Scattering phase shifts for two particles
  of different mass and non-zero total momentum in lattice QCD}},
  \href{https://doi.org/10.1103/PhysRevD.85.114507}{\emph{Phys. Rev. D}
  {\bfseries 85} (2012) 114507}
  [\href{https://arxiv.org/abs/1202.2145}{{\ttfamily 1202.2145}}].

\bibitem{Briceno:2012yi}
R.~A. Briceno and Z.~Davoudi, \emph{{Moving multichannel systems in a finite
  volume with application to proton-proton fusion}},
  \href{https://doi.org/10.1103/PhysRevD.88.094507}{\emph{Phys. Rev. D}
  {\bfseries 88} (2013) 094507}
  [\href{https://arxiv.org/abs/1204.1110}{{\ttfamily 1204.1110}}].

\bibitem{Hansen:2012tf}
M.~T. Hansen and S.~R. Sharpe, \emph{{Multiple-channel generalization of
  Lellouch-Luscher formula}},
  \href{https://doi.org/10.1103/PhysRevD.86.016007}{\emph{Phys. Rev. D}
  {\bfseries 86} (2012) 016007}
  [\href{https://arxiv.org/abs/1204.0826}{{\ttfamily 1204.0826}}].

\bibitem{Guo:2012hv}
P.~Guo, J.~Dudek, R.~Edwards and A.~P. Szczepaniak, \emph{{Coupled-channel
  scattering on a torus}},
  \href{https://doi.org/10.1103/PhysRevD.88.014501}{\emph{Phys. Rev. D}
  {\bfseries 88} (2013) 014501}
  [\href{https://arxiv.org/abs/1211.0929}{{\ttfamily 1211.0929}}].

\bibitem{Briceno:2013lba}
R.~A. Briceno, Z.~Davoudi and T.~C. Luu, \emph{{Two-Nucleon Systems in a Finite
  Volume: (I) Quantization Conditions}},
  \href{https://doi.org/10.1103/PhysRevD.88.034502}{\emph{Phys. Rev. D}
  {\bfseries 88} (2013) 034502}
  [\href{https://arxiv.org/abs/1305.4903}{{\ttfamily 1305.4903}}].

\bibitem{Briceno:2014oea}
R.~A. Briceno, \emph{{Two-particle multichannel systems in a finite volume with
  arbitrary spin}},
  \href{https://doi.org/10.1103/PhysRevD.89.074507}{\emph{Phys. Rev. D}
  {\bfseries 89} (2014) 074507}
  [\href{https://arxiv.org/abs/1401.3312}{{\ttfamily 1401.3312}}].

\bibitem{Polejaeva:2012ut}
K.~Polejaeva and A.~Rusetsky, \emph{{Three particles in a finite volume}},
  \href{https://doi.org/10.1140/epja/i2012-12067-8}{\emph{Eur. Phys. J. A}
  {\bfseries 48} (2012) 67} [\href{https://arxiv.org/abs/1203.1241}{{\ttfamily
  1203.1241}}].

\bibitem{Hansen:2014eka}
M.~T. Hansen and S.~R. Sharpe, \emph{{Relativistic, model-independent,
  three-particle quantization condition}},
  \href{https://doi.org/10.1103/PhysRevD.90.116003}{\emph{Phys. Rev. D}
  {\bfseries 90} (2014) 116003}
  [\href{https://arxiv.org/abs/1408.5933}{{\ttfamily 1408.5933}}].

\bibitem{Hansen:2015zga}
M.~T. Hansen and S.~R. Sharpe, \emph{{Expressing the three-particle
  finite-volume spectrum in terms of the three-to-three scattering amplitude}},
  \href{https://doi.org/10.1103/PhysRevD.92.114509}{\emph{Phys. Rev. D}
  {\bfseries 92} (2015) 114509}
  [\href{https://arxiv.org/abs/1504.04248}{{\ttfamily 1504.04248}}].

\bibitem{Briceno:2017tce}
R.~A. Brice\~no, M.~T. Hansen and S.~R. Sharpe, \emph{{Relating the
  finite-volume spectrum and the two-and-three-particle $S$ matrix for
  relativistic systems of identical scalar particles}},
  \href{https://doi.org/10.1103/PhysRevD.95.074510}{\emph{Phys. Rev. D}
  {\bfseries 95} (2017) 074510}
  [\href{https://arxiv.org/abs/1701.07465}{{\ttfamily 1701.07465}}].

\bibitem{Guo:2017ism}
P.~Guo and V.~Gasparian, \emph{{A solvable three-body model in finite volume}},
  \href{https://doi.org/10.1016/j.physletb.2017.10.009}{\emph{Phys. Lett. B}
  {\bfseries 774} (2017) 441}
  [\href{https://arxiv.org/abs/1701.00438}{{\ttfamily 1701.00438}}].

\bibitem{Hammer:2017uqm}
H.-W. Hammer, J.-Y. Pang and A.~Rusetsky, \emph{{Three-particle quantization
  condition in a finite volume: 1. The role of the three-particle force}},
  \href{https://doi.org/10.1007/JHEP09(2017)109}{\emph{JHEP} {\bfseries 09}
  (2017) 109} [\href{https://arxiv.org/abs/1706.07700}{{\ttfamily
  1706.07700}}].

\bibitem{Hammer:2017kms}
H.~W. Hammer, J.~Y. Pang and A.~Rusetsky, \emph{{Three particle quantization
  condition in a finite volume: 2. general formalism and the analysis of
  data}}, \href{https://doi.org/10.1007/JHEP10(2017)115}{\emph{JHEP} {\bfseries
  10} (2017) 115} [\href{https://arxiv.org/abs/1707.02176}{{\ttfamily
  1707.02176}}].

\bibitem{Mai:2017bge}
M.~Mai and M.~D\"oring, \emph{{Three-body Unitarity in the Finite Volume}},
  \href{https://doi.org/10.1140/epja/i2017-12440-1}{\emph{Eur. Phys. J. A}
  {\bfseries 53} (2017) 240}
  [\href{https://arxiv.org/abs/1709.08222}{{\ttfamily 1709.08222}}].

\bibitem{Doring:2018xxx}
M.~D\"oring, H.~W. Hammer, M.~Mai, J.~Y. Pang, t.~A. Rusetsky and J.~Wu,
  \emph{{Three-body spectrum in a finite volume: the role of cubic symmetry}},
  \href{https://doi.org/10.1103/PhysRevD.97.114508}{\emph{Phys. Rev. D}
  {\bfseries 97} (2018) 114508}
  [\href{https://arxiv.org/abs/1802.03362}{{\ttfamily 1802.03362}}].

\bibitem{Briceno:2018mlh}
R.~A. Brice\~no, M.~T. Hansen and S.~R. Sharpe, \emph{{Numerical study of the
  relativistic three-body quantization condition in the isotropic
  approximation}},
  \href{https://doi.org/10.1103/PhysRevD.98.014506}{\emph{Phys. Rev. D}
  {\bfseries 98} (2018) 014506}
  [\href{https://arxiv.org/abs/1803.04169}{{\ttfamily 1803.04169}}].

\bibitem{Klos:2018sen}
P.~Klos, S.~K\"onig, H.~W. Hammer, J.~E. Lynn and A.~Schwenk, \emph{{Signatures
  of few-body resonances in finite volume}},
  \href{https://doi.org/10.1103/PhysRevC.98.034004}{\emph{Phys. Rev. C}
  {\bfseries 98} (2018) 034004}
  [\href{https://arxiv.org/abs/1805.02029}{{\ttfamily 1805.02029}}].

\bibitem{Guo:2018ibd}
P.~Guo, M.~D\"oring and A.~P. Szczepaniak, \emph{{Variational approach to
  $N$-body interactions in finite volume}},
  \href{https://doi.org/10.1103/PhysRevD.98.094502}{\emph{Phys. Rev. D}
  {\bfseries 98} (2018) 094502}
  [\href{https://arxiv.org/abs/1810.01261}{{\ttfamily 1810.01261}}].

\bibitem{Briceno:2018aml}
R.~A. Brice\~no, M.~T. Hansen and S.~R. Sharpe, \emph{{Three-particle systems
  with resonant subprocesses in a finite volume}},
  \href{https://doi.org/10.1103/PhysRevD.99.014516}{\emph{Phys. Rev. D}
  {\bfseries 99} (2019) 014516}
  [\href{https://arxiv.org/abs/1810.01429}{{\ttfamily 1810.01429}}].

\bibitem{Blanton:2019igq}
T.~D. Blanton, F.~Romero-L\'opez and S.~R. Sharpe, \emph{{Implementing the
  three-particle quantization condition including higher partial waves}},
  \href{https://doi.org/10.1007/JHEP03(2019)106}{\emph{JHEP} {\bfseries 03}
  (2019) 106} [\href{https://arxiv.org/abs/1901.07095}{{\ttfamily
  1901.07095}}].

\bibitem{Pang:2019dfe}
J.-Y. Pang, J.-J. Wu, H.~W. Hammer, U.-G. Mei\ss{}ner and A.~Rusetsky,
  \emph{{Energy shift of the three-particle system in a finite volume}},
  \href{https://doi.org/10.1103/PhysRevD.99.074513}{\emph{Phys. Rev. D}
  {\bfseries 99} (2019) 074513}
  [\href{https://arxiv.org/abs/1902.01111}{{\ttfamily 1902.01111}}].

\bibitem{Romero-Lopez:2019qrt}
F.~Romero-L\'opez, S.~R. Sharpe, T.~D. Blanton, R.~A. Brice\~no and M.~T.
  Hansen, \emph{{Numerical exploration of three relativistic particles in a
  finite volume including two-particle resonances and bound states}},
  \href{https://doi.org/10.1007/JHEP10(2019)007}{\emph{JHEP} {\bfseries 10}
  (2019) 007} [\href{https://arxiv.org/abs/1908.02411}{{\ttfamily
  1908.02411}}].

\bibitem{Hansen:2020zhy}
M.~T. Hansen, F.~Romero-L\'opez and S.~R. Sharpe, \emph{{Generalizing the
  relativistic quantization condition to include all three-pion isospin
  channels}}, \href{https://doi.org/10.1007/JHEP07(2020)047}{\emph{JHEP}
  {\bfseries 07} (2020) 047}
  [\href{https://arxiv.org/abs/2003.10974}{{\ttfamily 2003.10974}}].

\bibitem{Blanton:2020gha}
T.~D. Blanton and S.~R. Sharpe, \emph{{Alternative derivation of the
  relativistic three-particle quantization condition}},
  \href{https://doi.org/10.1103/PhysRevD.102.054520}{\emph{Phys. Rev. D}
  {\bfseries 102} (2020) 054520}
  [\href{https://arxiv.org/abs/2007.16188}{{\ttfamily 2007.16188}}].

\bibitem{Blanton:2020jnm}
T.~D. Blanton and S.~R. Sharpe, \emph{{Equivalence of relativistic
  three-particle quantization conditions}},
  \href{https://doi.org/10.1103/PhysRevD.102.054515}{\emph{Phys. Rev. D}
  {\bfseries 102} (2020) 054515}
  [\href{https://arxiv.org/abs/2007.16190}{{\ttfamily 2007.16190}}].

\bibitem{Guo:2020spn}
P.~Guo, \emph{{Modeling few-body resonances in finite volume}},
  \href{https://doi.org/10.1103/PhysRevD.102.054514}{\emph{Phys. Rev. D}
  {\bfseries 102} (2020) 054514}
  [\href{https://arxiv.org/abs/2007.12790}{{\ttfamily 2007.12790}}].

\bibitem{Romero-Lopez:2020rdq}
F.~Romero-L\'opez, A.~Rusetsky, N.~Schlage and C.~Urbach, \emph{{Relativistic
  $N$-particle energy shift in finite volume}},
  \href{https://doi.org/10.1007/JHEP02(2021)060}{\emph{JHEP} {\bfseries 02}
  (2021) 060} [\href{https://arxiv.org/abs/2010.11715}{{\ttfamily
  2010.11715}}].

\bibitem{Pang:2020pkl}
J.-Y. Pang, J.-J. Wu and L.-S. Geng, \emph{{$DDK$ system in finite volume}},
  \href{https://doi.org/10.1103/PhysRevD.102.114515}{\emph{Phys. Rev. D}
  {\bfseries 102} (2020) 114515}
  [\href{https://arxiv.org/abs/2008.13014}{{\ttfamily 2008.13014}}].

\bibitem{Blanton:2020gmf}
T.~D. Blanton and S.~R. Sharpe, \emph{{Relativistic three-particle quantization
  condition for nondegenerate scalars}},
  \href{https://doi.org/10.1103/PhysRevD.103.054503}{\emph{Phys. Rev. D}
  {\bfseries 103} (2021) 054503}
  [\href{https://arxiv.org/abs/2011.05520}{{\ttfamily 2011.05520}}].

\bibitem{Muller:2020vtt}
F.~M\"uller, T.~Yu and A.~Rusetsky, \emph{{Finite-volume energy shift of the
  three-pion ground state}},
  \href{https://doi.org/10.1103/PhysRevD.103.054506}{\emph{Phys. Rev. D}
  {\bfseries 103} (2021) 054506}
  [\href{https://arxiv.org/abs/2011.14178}{{\ttfamily 2011.14178}}].

\bibitem{Muller:2020wjo}
F.~M\"uller and A.~Rusetsky, \emph{{On the three-particle analog of the
  Lellouch-L\"uscher formula}},
  \href{https://doi.org/10.1007/JHEP03(2021)152}{\emph{JHEP} {\bfseries 03}
  (2021) 152} [\href{https://arxiv.org/abs/2012.13957}{{\ttfamily
  2012.13957}}].

\bibitem{Hansen:2021ofl}
M.~T. Hansen, F.~Romero-L\'opez and S.~R. Sharpe, \emph{{Decay amplitudes to
  three hadrons from finite-volume matrix elements}},
  \href{https://doi.org/10.1007/JHEP04(2021)113}{\emph{JHEP} {\bfseries 04}
  (2021) 113} [\href{https://arxiv.org/abs/2101.10246}{{\ttfamily
  2101.10246}}].

\bibitem{Blanton:2021mih}
T.~D. Blanton and S.~R. Sharpe, \emph{{Three-particle finite-volume formalism
  for \ensuremath{\pi}+\ensuremath{\pi}+K+ and related systems}},
  \href{https://doi.org/10.1103/PhysRevD.104.034509}{\emph{Phys. Rev. D}
  {\bfseries 104} (2021) 034509}
  [\href{https://arxiv.org/abs/2105.12094}{{\ttfamily 2105.12094}}].

\bibitem{Muller:2021uur}
F.~M\"uller, J.-Y. Pang, A.~Rusetsky and J.-J. Wu,
  \emph{{Relativistic-invariant formulation of the NREFT three-particle
  quantization condition}},
  \href{https://doi.org/10.1007/JHEP02(2022)158}{\emph{JHEP} {\bfseries 02}
  (2022) 158} [\href{https://arxiv.org/abs/2110.09351}{{\ttfamily
  2110.09351}}].

\bibitem{Blanton:2021eyf}
T.~D. Blanton, F.~Romero-L\'opez and S.~R. Sharpe, \emph{{Implementing the
  three-particle quantization condition for
  \ensuremath{\pi}$^{+}$\ensuremath{\pi}$^{+}$K$^{+}$ and related systems}},
  \href{https://doi.org/10.1007/JHEP02(2022)098}{\emph{JHEP} {\bfseries 02}
  (2022) 098} [\href{https://arxiv.org/abs/2111.12734}{{\ttfamily
  2111.12734}}].

\bibitem{Muller:2022oyw}
F.~M\"uller, J.-Y. Pang, A.~Rusetsky and J.-J. Wu, \emph{{Three-particle
  Lellouch-L\"uscher formalism in moving frames}},
  \href{https://doi.org/10.1007/JHEP02(2023)214}{\emph{JHEP} {\bfseries 02}
  (2023) 214} [\href{https://arxiv.org/abs/2211.10126}{{\ttfamily
  2211.10126}}].

\bibitem{Severt:2022jtg}
D.~Severt, M.~Mai and U.-G. Mei\ss{}ner, \emph{{Particle-dimer approach for the
  Roper resonance in a finite volume}},
  \href{https://doi.org/10.1007/JHEP04(2023)100}{\emph{JHEP} {\bfseries 04}
  (2023) 100} [\href{https://arxiv.org/abs/2212.02171}{{\ttfamily
  2212.02171}}].

\bibitem{Jackura:2022xml}
A.~W. Jackura, R.~A. Brice\'no and M.~T. Hansen, \emph{{Three-pion effects in
  $K^0-\bar{K}^0$ mixing}},
  \href{https://doi.org/10.22323/1.430.0062}{\emph{PoS} {\bfseries LATTICE2022}
  (2023) 062} [\href{https://arxiv.org/abs/2212.09951}{{\ttfamily
  2212.09951}}].

\bibitem{Baeza-Ballesteros:2023ljl}
J.~Baeza-Ballesteros, J.~Bijnens, T.~Husek, F.~Romero-L\'opez, S.~R. Sharpe and
  M.~Sj\"o, \emph{{The isospin-3 three-particle K-matrix at NLO in ChPT}},
  \href{https://doi.org/10.1007/JHEP05(2023)187}{\emph{JHEP} {\bfseries 05}
  (2023) 187} [\href{https://arxiv.org/abs/2303.13206}{{\ttfamily
  2303.13206}}].

\bibitem{Draper:2023xvu}
Z.~T. Draper, M.~T. Hansen, F.~Romero-L\'opez and S.~R. Sharpe, \emph{{Three
  relativistic neutrons in a finite volume}},
  \href{https://doi.org/10.1007/JHEP07(2023)226}{\emph{JHEP} {\bfseries 07}
  (2023) 226} [\href{https://arxiv.org/abs/2303.10219}{{\ttfamily
  2303.10219}}].

\bibitem{Bubna:2023oxo}
R.~Bubna, F.~M\"uller and A.~Rusetsky, \emph{{Finite-volume energy shift of the
  three-nucleon ground state}},
  \href{https://doi.org/10.1103/PhysRevD.108.014518}{\emph{Phys. Rev. D}
  {\bfseries 108} (2023) 014518}
  [\href{https://arxiv.org/abs/2304.13635}{{\ttfamily 2304.13635}}].

\bibitem{Dudek:2012xn}
{\scshape Hadron Spectrum} collaboration, J.~J. Dudek, R.~G. Edwards and C.~E.
  Thomas, \emph{{Energy dependence of the $\rho$ resonance in $\pi\pi$ elastic
  scattering from lattice QCD}},
  \href{https://doi.org/10.1103/PhysRevD.87.034505}{\emph{Phys. Rev. D}
  {\bfseries 87} (2013) 034505}
  [\href{https://arxiv.org/abs/1212.0830}{{\ttfamily 1212.0830}}].

\bibitem{Dudek:2014qha}
{\scshape Hadron Spectrum} collaboration, J.~J. Dudek, R.~G. Edwards, C.~E.
  Thomas and D.~J. Wilson, \emph{{Resonances in coupled $\pi K -\eta K$
  scattering from quantum chromodynamics}},
  \href{https://doi.org/10.1103/PhysRevLett.113.182001}{\emph{Phys. Rev. Lett.}
  {\bfseries 113} (2014) 182001}
  [\href{https://arxiv.org/abs/1406.4158}{{\ttfamily 1406.4158}}].

\bibitem{Fahy:2014jxa}
B.~Fahy, J.~Bulava, B.~H\"orz, K.~J. Juge, C.~Morningstar and C.~H. Wong,
  \emph{{Pion-pion scattering phase shifts with the stochastic LapH method}},
  \href{https://doi.org/10.22323/1.214.0077}{\emph{PoS} {\bfseries LATTICE2014}
  (2015) 077} [\href{https://arxiv.org/abs/1410.8843}{{\ttfamily 1410.8843}}].

\bibitem{Wilson:2014cna}
D.~J. Wilson, J.~J. Dudek, R.~G. Edwards and C.~E. Thomas, \emph{{Resonances in
  coupled $\pi K, \eta K$ scattering from lattice QCD}},
  \href{https://doi.org/10.1103/PhysRevD.91.054008}{\emph{Phys. Rev. D}
  {\bfseries 91} (2015) 054008}
  [\href{https://arxiv.org/abs/1411.2004}{{\ttfamily 1411.2004}}].

\bibitem{Green:2014dea}
J.~Green, A.~Francis, P.~Junnarkar, C.~Miao, T.~Rae and H.~Wittig,
  \emph{{Search for a bound H-dibaryon using local six-quark interpolating
  operators}}, \href{https://doi.org/10.22323/1.214.0107}{\emph{PoS} {\bfseries
  LATTICE2014} (2014) 107} [\href{https://arxiv.org/abs/1411.1643}{{\ttfamily
  1411.1643}}].

\bibitem{Wilson:2015dqa}
D.~J. Wilson, R.~A. Briceno, J.~J. Dudek, R.~G. Edwards and C.~E. Thomas,
  \emph{{Coupled $\pi\pi, K\bar{K}$ scattering in $P$-wave and the $\rho$
  resonance from lattice QCD}},
  \href{https://doi.org/10.1103/PhysRevD.92.094502}{\emph{Phys. Rev. D}
  {\bfseries 92} (2015) 094502}
  [\href{https://arxiv.org/abs/1507.02599}{{\ttfamily 1507.02599}}].

\bibitem{Briceno:2015dca}
R.~A. Briceno, J.~J. Dudek, R.~G. Edwards, C.~J. Shultz, C.~E. Thomas and D.~J.
  Wilson, \emph{{The resonant $\pi^+\gamma\to\pi^+\pi^0$ amplitude from Quantum
  Chromodynamics}},
  \href{https://doi.org/10.1103/PhysRevLett.115.242001}{\emph{Phys. Rev. Lett.}
  {\bfseries 115} (2015) 242001}
  [\href{https://arxiv.org/abs/1507.06622}{{\ttfamily 1507.06622}}].

\bibitem{Bolton:2015psa}
D.~R. Bolton, R.~A. Briceno and D.~J. Wilson, \emph{{Connecting physical
  resonant amplitudes and lattice QCD}},
  \href{https://doi.org/10.1016/j.physletb.2016.03.043}{\emph{Phys. Lett. B}
  {\bfseries 757} (2016) 50}
  [\href{https://arxiv.org/abs/1507.07928}{{\ttfamily 1507.07928}}].

\bibitem{Bulava:2015qjz}
J.~Bulava, B.~H\"orz, B.~Fahy, K.~J. Juge, C.~Morningstar and C.~H. Wong,
  \emph{{Pion-pion scattering and the timelike pion form factor from
  $N_{\mathrm{f}} = 2+1$ lattice QCD simulations using the stochastic LapH
  method}}, \href{https://doi.org/10.22323/1.251.0069}{\emph{PoS} {\bfseries
  LATTICE2015} (2016) 069} [\href{https://arxiv.org/abs/1511.02351}{{\ttfamily
  1511.02351}}].

\bibitem{Dudek:2016bxq}
J.~J. Dudek, \emph{{Hadron scattering and resonances in QCD}},
  \href{https://doi.org/10.1063/1.4949382}{\emph{AIP Conf. Proc.} {\bfseries
  1735} (2016) 020014}.

\bibitem{Dudek:2016wcf}
{\scshape Hadron Spectrum} collaboration, J.~J. Dudek, \emph{{Hadron Resonances
  from QCD}}, \href{https://doi.org/10.1051/epjconf/201611301001}{\emph{EPJ Web
  Conf.} {\bfseries 113} (2016) 01001}.

\bibitem{Junnarkar:2015jyf}
P.~Junnarkar, A.~Francis, J.~Green, C.~Miao, T.~Rae and H.~Wittig,
  \emph{{Search for the H-Dibaryon in two flavor Lattice QCD}},
  \href{https://doi.org/10.22323/1.253.0079}{\emph{PoS} {\bfseries CD15} (2015)
  079} [\href{https://arxiv.org/abs/1511.01849}{{\ttfamily 1511.01849}}].

\bibitem{Dudek:2016esq}
J.~Dudek, \emph{{An $a_0$ resonance in strongly coupled $\pi \eta$, $K
  \overline{K}$ scattering from lattice QCD}},
  \href{https://doi.org/10.22323/1.256.0095}{\emph{PoS} {\bfseries LATTICE2016}
  (2016) 095}.

\bibitem{Dudek:2016cru}
{\scshape Hadron Spectrum} collaboration, J.~J. Dudek, R.~G. Edwards and D.~J.
  Wilson, \emph{{An $a_0$ resonance in strongly coupled $\pi \eta$,
  $K\overline{K}$ scattering from lattice QCD}},
  \href{https://doi.org/10.1103/PhysRevD.93.094506}{\emph{Phys. Rev. D}
  {\bfseries 93} (2016) 094506}
  [\href{https://arxiv.org/abs/1602.05122}{{\ttfamily 1602.05122}}].

\bibitem{Morningstar:2016arm}
C.~Morningstar, J.~Bulava, B.~Fahy, J.~Fallica, A.~Hanlon, B.~H\"orz et~al.,
  \emph{{Lattice QCD Study of Excited Hadron Resonances}},
  \href{https://doi.org/10.5506/APhysPolBSupp.9.421}{\emph{Acta Phys. Polon.
  Supp.} {\bfseries 9} (2016) 421}.

\bibitem{Briceno:2016kkp}
R.~A. Brice\~no, J.~J. Dudek, R.~G. Edwards, C.~J. Shultz, C.~E. Thomas and
  D.~J. Wilson, \emph{{The $\pi\pi\to\pi\gamma^\star$ amplitude and the
  resonant $\rho\to\pi\gamma^\star$ transition from lattice QCD}},
  \href{https://doi.org/10.1103/PhysRevD.93.114508}{\emph{Phys. Rev. D}
  {\bfseries 93} (2016) 114508}
  [\href{https://arxiv.org/abs/1604.03530}{{\ttfamily 1604.03530}}].

\bibitem{Bolton:2016ptw}
D.~R. Bolton, R.~A. Brice\~no and D.~J. Wilson, \emph{{From QCD to Physical
  Resonances}}, \href{https://doi.org/10.1063/1.4949394}{\emph{AIP Conf. Proc.}
  {\bfseries 1735} (2016) 030011}
  [\href{https://arxiv.org/abs/1602.03800}{{\ttfamily 1602.03800}}].

\bibitem{Bulava:2016mks}
J.~Bulava, B.~Fahy, B.~H\"orz, K.~J. Juge, C.~Morningstar and C.~H. Wong,
  \emph{{$I=1$ and $I=2$ $\pi-\pi$ scattering phase shifts from $N_{\mathrm{f}}
  = 2+1$ lattice QCD}},
  \href{https://doi.org/10.1016/j.nuclphysb.2016.07.024}{\emph{Nucl. Phys. B}
  {\bfseries 910} (2016) 842}
  [\href{https://arxiv.org/abs/1604.05593}{{\ttfamily 1604.05593}}].

\bibitem{Briceno:2016mjc}
R.~A. Briceno, J.~J. Dudek, R.~G. Edwards and D.~J. Wilson, \emph{{Isoscalar
  $\pi\pi$ scattering and the $\sigma$ meson resonance from QCD}},
  \href{https://doi.org/10.1103/PhysRevLett.118.022002}{\emph{Phys. Rev. Lett.}
  {\bfseries 118} (2017) 022002}
  [\href{https://arxiv.org/abs/1607.05900}{{\ttfamily 1607.05900}}].

\bibitem{Moir:2016srx}
G.~Moir, M.~Peardon, S.~M. Ryan, C.~E. Thomas and D.~J. Wilson,
  \emph{{Coupled-Channel $D\pi$, $D\eta$ and $D_{s}\bar{K}$ Scattering from
  Lattice QCD}}, \href{https://doi.org/10.1007/JHEP10(2016)011}{\emph{JHEP}
  {\bfseries 10} (2016) 011}
  [\href{https://arxiv.org/abs/1607.07093}{{\ttfamily 1607.07093}}].

\bibitem{Doring:2016bdr}
M.~D\"oring, B.~Hu and M.~Mai, \emph{{Chiral Extrapolation of the Sigma
  Resonance}},
  \href{https://doi.org/10.1016/j.physletb.2018.05.042}{\emph{Phys. Lett. B}
  {\bfseries 782} (2018) 785}
  [\href{https://arxiv.org/abs/1610.10070}{{\ttfamily 1610.10070}}].

\bibitem{Wilson:2016rid}
D.~J. Wilson, \emph{{Resonances in Coupled-Channel Scattering}},
  \href{https://doi.org/10.22323/1.256.0016}{\emph{PoS} {\bfseries LATTICE2016}
  (2016) 016} [\href{https://arxiv.org/abs/1611.07281}{{\ttfamily
  1611.07281}}].

\bibitem{Briceno:2017max}
R.~A. Briceno, J.~J. Dudek and R.~D. Young, \emph{{Scattering processes and
  resonances from lattice QCD}},
  \href{https://doi.org/10.1103/RevModPhys.90.025001}{\emph{Rev. Mod. Phys.}
  {\bfseries 90} (2018) 025001}
  [\href{https://arxiv.org/abs/1706.06223}{{\ttfamily 1706.06223}}].

\bibitem{Briceno:2017qmb}
R.~A. Briceno, J.~J. Dudek, R.~G. Edwards and D.~J. Wilson, \emph{{Isoscalar
  $\pi\pi, K\overline{K}, \eta\eta$ scattering and the $\sigma, f_0, f_2$
  mesons from QCD}},
  \href{https://doi.org/10.1103/PhysRevD.97.054513}{\emph{Phys. Rev. D}
  {\bfseries 97} (2018) 054513}
  [\href{https://arxiv.org/abs/1708.06667}{{\ttfamily 1708.06667}}].

\bibitem{Brett:2017yhm}
R.~Brett, J.~Bulava, J.~Fallica, A.~Hanlon, B.~H\"orz, C.~Morningstar et~al.,
  \emph{{Scattering from finite-volume energies including higher partial waves
  and multiple decay channels}},
  \href{https://doi.org/10.1051/epjconf/201817505005}{\emph{EPJ Web Conf.}
  {\bfseries 175} (2018) 05005}
  [\href{https://arxiv.org/abs/1710.04169}{{\ttfamily 1710.04169}}].

\bibitem{Andersen:2017una}
C.~W. Andersen, J.~Bulava, B.~H\"orz and C.~Morningstar, \emph{{Elastic $I=3/2
  p$-wave nucleon-pion scattering amplitude and the $\Delta$(1232) resonance
  from N$_f$=2+1 lattice QCD}},
  \href{https://doi.org/10.1103/PhysRevD.97.014506}{\emph{Phys. Rev. D}
  {\bfseries 97} (2018) 014506}
  [\href{https://arxiv.org/abs/1710.01557}{{\ttfamily 1710.01557}}].

\bibitem{Woss:2018irj}
A.~Woss, C.~E. Thomas, J.~J. Dudek, R.~G. Edwards and D.~J. Wilson,
  \emph{{Dynamically-coupled partial-waves in $\rho\pi$ isospin-2 scattering
  from lattice QCD}},
  \href{https://doi.org/10.1007/JHEP07(2018)043}{\emph{JHEP} {\bfseries 07}
  (2018) 043} [\href{https://arxiv.org/abs/1802.05580}{{\ttfamily
  1802.05580}}].

\bibitem{Brett:2018jqw}
R.~Brett, J.~Bulava, J.~Fallica, A.~Hanlon, B.~H\"orz and C.~Morningstar,
  \emph{{Determination of $s$- and $p$-wave $I=1/2$ $K\pi$ scattering
  amplitudes in $N_{\mathrm{f}}=2+1$ lattice QCD}},
  \href{https://doi.org/10.1016/j.nuclphysb.2018.05.008}{\emph{Nucl. Phys. B}
  {\bfseries 932} (2018) 29}
  [\href{https://arxiv.org/abs/1802.03100}{{\ttfamily 1802.03100}}].

\bibitem{Molina:2018otc}
R.~Molina, D.~Guo, t.~A. Alexandru, t.~M. Mai and M.~D\"oring, \emph{{Sigma
  resonance parameters from a $N_f=2$ lattice QCD simulation}},  in \emph{{14th
  International Workshop on Hadron Physics}}, 4, 2018,
  \href{https://arxiv.org/abs/1804.10225}{{\ttfamily 1804.10225}}.

\bibitem{Francis:2018qch}
A.~Francis, J.~R. Green, P.~M. Junnarkar, C.~Miao, T.~D. Rae and H.~Wittig,
  \emph{{Lattice QCD study of the $H$ dibaryon using hexaquark and two-baryon
  interpolators}},
  \href{https://doi.org/10.1103/PhysRevD.99.074505}{\emph{Phys. Rev. D}
  {\bfseries 99} (2019) 074505}
  [\href{https://arxiv.org/abs/1805.03966}{{\ttfamily 1805.03966}}].

\bibitem{Brett:2018fdb}
R.~Brett, J.~Bulava, J.~Fallica, A.~Hanlon, B.~H\"orz and C.~Morningstar,
  \emph{{$K\pi$ scattering and excited meson spectroscopy using the Stocastic
  LapH method}}, \href{https://doi.org/10.22323/1.334.0071}{\emph{PoS}
  {\bfseries LATTICE2018} (2019) 071}
  [\href{https://arxiv.org/abs/1810.11311}{{\ttfamily 1810.11311}}].

\bibitem{Andersen:2018mau}
C.~Andersen, J.~Bulava, B.~H\"orz and C.~Morningstar, \emph{{The $I=1$
  pion-pion scattering amplitude and timelike pion form factor from $N_{\rm f}
  = 2+1$ lattice QCD}},
  \href{https://doi.org/10.1016/j.nuclphysb.2018.12.018}{\emph{Nucl. Phys. B}
  {\bfseries 939} (2019) 145}
  [\href{https://arxiv.org/abs/1808.05007}{{\ttfamily 1808.05007}}].

\bibitem{Hanlon:2018yfv}
A.~Hanlon, A.~Francis, J.~Green, P.~Junnarkar and H.~Wittig, \emph{{The $H$
  dibaryon from lattice QCD with SU(3) flavor symmetry}},
  \href{https://doi.org/10.22323/1.334.0081}{\emph{PoS} {\bfseries LATTICE2018}
  (2018) 081} [\href{https://arxiv.org/abs/1810.13282}{{\ttfamily
  1810.13282}}].

\bibitem{Woss:2019hse}
A.~J. Woss, C.~E. Thomas, J.~J. Dudek, R.~G. Edwards and D.~J. Wilson,
  \emph{{$b_1$ resonance in coupled $\pi\omega$, $\pi\phi$ scattering from
  lattice QCD}}, \href{https://doi.org/10.1103/PhysRevD.100.054506}{\emph{Phys.
  Rev. D} {\bfseries 100} (2019) 054506}
  [\href{https://arxiv.org/abs/1904.04136}{{\ttfamily 1904.04136}}].

\bibitem{Wilson:2019wfr}
D.~J. Wilson, R.~A. Briceno, J.~J. Dudek, R.~G. Edwards and C.~E. Thomas,
  \emph{{The quark-mass dependence of elastic $\pi K$ scattering from QCD}},
  \href{https://doi.org/10.1103/PhysRevLett.123.042002}{\emph{Phys. Rev. Lett.}
  {\bfseries 123} (2019) 042002}
  [\href{https://arxiv.org/abs/1904.03188}{{\ttfamily 1904.03188}}].

\bibitem{Bulava:2019hpz}
J.~Bulava, \emph{{Meson-Nucleon Scattering Amplitudes from Lattice QCD}},
  \href{https://doi.org/10.1063/5.0008643}{\emph{AIP Conf. Proc.} {\bfseries
  2249} (2020) 020006} [\href{https://arxiv.org/abs/1909.13097}{{\ttfamily
  1909.13097}}].

\bibitem{Blanton:2019vdk}
T.~D. Blanton, F.~Romero-L\'opez and S.~R. Sharpe, \emph{{$I=3$ Three-Pion
  Scattering Amplitude from Lattice QCD}},
  \href{https://doi.org/10.1103/PhysRevLett.124.032001}{\emph{Phys. Rev. Lett.}
  {\bfseries 124} (2020) 032001}
  [\href{https://arxiv.org/abs/1909.02973}{{\ttfamily 1909.02973}}].

\bibitem{Erben:2019nmx}
F.~Erben, J.~R. Green, D.~Mohler and H.~Wittig, \emph{{Rho resonance, timelike
  pion form factor, and implications for lattice studies of the hadronic vacuum
  polarization}},
  \href{https://doi.org/10.1103/PhysRevD.101.054504}{\emph{Phys. Rev. D}
  {\bfseries 101} (2020) 054504}
  [\href{https://arxiv.org/abs/1910.01083}{{\ttfamily 1910.01083}}].

\bibitem{Andersen:2019ktw}
C.~W. Andersen, J.~Bulava, B.~H\"orz and C.~Morningstar, \emph{{$I=3/2N \pi$
  scattering and the $\Delta$(1232) resonance on $N_f=2+1$ CLS ensembles using
  the stochastic LapH method}},
  \href{https://doi.org/10.22323/1.363.0039}{\emph{PoS} {\bfseries LATTICE2019}
  (2019) 039} [\href{https://arxiv.org/abs/1911.10021}{{\ttfamily
  1911.10021}}].

\bibitem{Fischer:2020yvw}
{\scshape Extended Twisted Mass, ETM} collaboration, M.~Fischer, B.~Kostrzewa,
  M.~Mai, M.~Petschlies, F.~Pittler, M.~Ueding et~al., \emph{{The
  \ensuremath{\rho}-resonance from Nf=2 lattice QCD including the physical pion
  mass}}, \href{https://doi.org/10.1016/j.physletb.2021.136449}{\emph{Phys.
  Lett. B} {\bfseries 819} (2021) 136449}
  [\href{https://arxiv.org/abs/2006.13805}{{\ttfamily 2006.13805}}].

\bibitem{Cheung:2020mql}
{\scshape Hadron Spectrum} collaboration, G.~K.~C. Cheung, C.~E. Thomas, D.~J.
  Wilson, G.~Moir, M.~Peardon and S.~M. Ryan, \emph{{DK I = 0,$ D\overline{K}
  $I = 0, 1 scattering and the $ {D}_{s0}^{\ast } $(2317) from lattice QCD}},
  \href{https://doi.org/10.1007/JHEP02(2021)100}{\emph{JHEP} {\bfseries 02}
  (2021) 100} [\href{https://arxiv.org/abs/2008.06432}{{\ttfamily
  2008.06432}}].

\bibitem{Woss:2020ayi}
{\scshape Hadron Spectrum} collaboration, A.~J. Woss, J.~J. Dudek, R.~G.
  Edwards, C.~E. Thomas and D.~J. Wilson, \emph{{Decays of an exotic $1{-+}$
  hybrid meson resonance in QCD}},
  \href{https://doi.org/10.1103/PhysRevD.103.054502}{\emph{Phys. Rev. D}
  {\bfseries 103} (2021) 054502}
  [\href{https://arxiv.org/abs/2009.10034}{{\ttfamily 2009.10034}}].

\bibitem{Hansen:2020otl}
{\scshape Hadron Spectrum} collaboration, M.~T. Hansen, R.~A. Brice\~no, R.~G.
  Edwards, C.~E. Thomas and D.~J. Wilson, \emph{{Energy-Dependent $\pi^+ \pi^+
  \pi^+$ Scattering Amplitude from QCD}},
  \href{https://doi.org/10.1103/PhysRevLett.126.012001}{\emph{Phys. Rev. Lett.}
  {\bfseries 126} (2021) 012001}
  [\href{https://arxiv.org/abs/2009.04931}{{\ttfamily 2009.04931}}].

\bibitem{Johnson:2020ilc}
{\scshape Hadron Spectrum} collaboration, C.~T. Johnson and J.~J. Dudek,
  \emph{{Excited $J^{--}$ meson resonances at the SU(3) flavor point from
  lattice QCD}}, \href{https://doi.org/10.1103/PhysRevD.103.074502}{\emph{Phys.
  Rev. D} {\bfseries 103} (2021) 074502}
  [\href{https://arxiv.org/abs/2012.00518}{{\ttfamily 2012.00518}}].

\bibitem{Green:2021qol}
J.~R. Green, A.~D. Hanlon, P.~M. Junnarkar and H.~Wittig, \emph{{Weakly bound
  $H$ dibaryon from SU(3)-flavor-symmetric QCD}},
  \href{https://doi.org/10.1103/PhysRevLett.127.242003}{\emph{Phys. Rev. Lett.}
  {\bfseries 127} (2021) 242003}
  [\href{https://arxiv.org/abs/2103.01054}{{\ttfamily 2103.01054}}].

\bibitem{Gayer:2021xzv}
{\scshape Hadron Spectrum} collaboration, L.~Gayer, N.~Lang, S.~M. Ryan,
  D.~Tims, C.~E. Thomas and D.~J. Wilson, \emph{{Isospin-1/2 D\ensuremath{\pi}
  scattering and the lightest $ {D}_0^{\ast } $ resonance from lattice QCD}},
  \href{https://doi.org/10.1007/JHEP07(2021)123}{\emph{JHEP} {\bfseries 07}
  (2021) 123} [\href{https://arxiv.org/abs/2102.04973}{{\ttfamily
  2102.04973}}].

\bibitem{Blanton:2021llb}
T.~D. Blanton, A.~D. Hanlon, B.~H\"orz, C.~Morningstar, F.~Romero-L\'opez and
  S.~R. Sharpe, \emph{{Interactions of two and three mesons including higher
  partial waves from lattice QCD}},
  \href{https://doi.org/10.1007/JHEP10(2021)023}{\emph{JHEP} {\bfseries 10}
  (2021) 023} [\href{https://arxiv.org/abs/2106.05590}{{\ttfamily
  2106.05590}}].

\bibitem{Mai:2021nul}
{\scshape GWQCD} collaboration, M.~Mai, A.~Alexandru, R.~Brett, C.~Culver,
  M.~D\"oring, F.~X. Lee et~al., \emph{{Three-Body Dynamics of the a1(1260)
  Resonance from Lattice QCD}},
  \href{https://doi.org/10.1103/PhysRevLett.127.222001}{\emph{Phys. Rev. Lett.}
  {\bfseries 127} (2021) 222001}
  [\href{https://arxiv.org/abs/2107.03973}{{\ttfamily 2107.03973}}].

\bibitem{Morningstar:2021ewk}
C.~Morningstar, J.~Bulava, A.~D. Hanlon, B.~H\"orz, D.~Mohler, A.~Nicholson
  et~al., \emph{{Progress on Meson-Baryon Scattering}},
  \href{https://doi.org/10.22323/1.396.0170}{\emph{PoS} {\bfseries LATTICE2021}
  (2022) 170} [\href{https://arxiv.org/abs/2111.07755}{{\ttfamily
  2111.07755}}].

\bibitem{PadmanathMadanagopalan:2021exb}
P.~Madanagopalan, J.~Bulava, J.~R. Green, A.~D. Hanlon, B.~H\"orz, P.~Junnarkar
  et~al., \emph{{$H$ dibaryon away from the $SU(3)_f$ symmetric point}},
  \href{https://doi.org/10.22323/1.396.0459}{\emph{PoS} {\bfseries LATTICE2021}
  (2022) 459} [\href{https://arxiv.org/abs/2111.11541}{{\ttfamily
  2111.11541}}].

\bibitem{Green:2021sxb}
J.~R. Green, A.~D. Hanlon, P.~M. Junnarkar and H.~Wittig, \emph{{Continuum
  limit of baryon-baryon scattering with SU(3) flavor symmetry}},
  \href{https://doi.org/10.22323/1.396.0294}{\emph{PoS} {\bfseries LATTICE2021}
  (2022) 294} [\href{https://arxiv.org/abs/2111.09675}{{\ttfamily
  2111.09675}}].

\bibitem{Padmanath:2022cvl}
M.~Padmanath and S.~Prelovsek, \emph{{Signature of a Doubly Charm Tetraquark
  Pole in DD* Scattering on the Lattice}},
  \href{https://doi.org/10.1103/PhysRevLett.129.032002}{\emph{Phys. Rev. Lett.}
  {\bfseries 129} (2022) 032002}
  [\href{https://arxiv.org/abs/2202.10110}{{\ttfamily 2202.10110}}].

\bibitem{Radhakrishnan:2022ubg}
{\scshape Hadron Spectrum} collaboration, A.~Radhakrishnan, J.~J. Dudek and
  R.~G. Edwards, \emph{{Radiative decay of the resonant K* and the
  \ensuremath{\gamma}K\textrightarrow{}K\ensuremath{\pi} amplitude from lattice
  QCD}}, \href{https://doi.org/10.1103/PhysRevD.106.114513}{\emph{Phys. Rev. D}
  {\bfseries 106} (2022) 114513}
  [\href{https://arxiv.org/abs/2208.13755}{{\ttfamily 2208.13755}}].

\bibitem{Lang:2022elg}
{\scshape Hadron Spectrum} collaboration, N.~Lang and D.~J. Wilson,
  \emph{{Axial-Vector D1 Hadrons in D*\ensuremath{\pi} Scattering from QCD}},
  \href{https://doi.org/10.1103/PhysRevLett.129.252001}{\emph{Phys. Rev. Lett.}
  {\bfseries 129} (2022) 252001}
  [\href{https://arxiv.org/abs/2205.05026}{{\ttfamily 2205.05026}}].

\bibitem{Bulava:2022vpq}
J.~Bulava, A.~D. Hanlon, B.~H\"orz, C.~Morningstar, A.~Nicholson,
  F.~Romero-L\'opez et~al., \emph{{Elastic nucleon-pion scattering at
  m\ensuremath{\pi} = 200 MeV from lattice QCD}},
  \href{https://doi.org/10.1016/j.nuclphysb.2023.116105}{\emph{Nucl. Phys. B}
  {\bfseries 987} (2023) 116105}
  [\href{https://arxiv.org/abs/2208.03867}{{\ttfamily 2208.03867}}].

\bibitem{Green:2022rjj}
{\scshape Baryon Scattering (BaSc)} collaboration, J.~R. Green, A.~D. Hanlon,
  P.~M. Junnarkar and H.~Wittig, \emph{{Nucleon-nucleon scattering from
  distillation}}, \href{https://doi.org/10.22323/1.430.0200}{\emph{PoS}
  {\bfseries LATTICE2022} (2023) 200}
  [\href{https://arxiv.org/abs/2212.09587}{{\ttfamily 2212.09587}}].

\bibitem{Bulava:2023wrz}
J.~Bulava, B.~H\"orz, Renwick, J.~Hudspdith, C.~Johnson, D.~Mohler et~al.,
  \emph{{D meson - pion scattering on CLS 2+1 flavor ensembles}},
  \href{https://doi.org/10.22323/1.430.0068}{\emph{PoS} {\bfseries LATTICE2022}
  (2023) 068}.

\bibitem{Lyu:2023xro}
Y.~Lyu, S.~Aoki, T.~Doi, T.~Hatsuda, Y.~Ikeda and J.~Meng, \emph{{Doubly
  Charmed Tetraquark Tcc+ from Lattice QCD near Physical Point}},
  \href{https://doi.org/10.1103/PhysRevLett.131.161901}{\emph{Phys. Rev. Lett.}
  {\bfseries 131} (2023) 161901}
  [\href{https://arxiv.org/abs/2302.04505}{{\ttfamily 2302.04505}}].

\bibitem{Draper:2023boj}
Z.~T. Draper, A.~D. Hanlon, B.~H\"orz, C.~Morningstar, F.~Romero-L\'opez and
  S.~R. Sharpe, \emph{{Interactions of \ensuremath{\pi}K,
  \ensuremath{\pi}\ensuremath{\pi}K and KK\ensuremath{\pi} systems at maximal
  isospin from lattice QCD}},
  \href{https://doi.org/10.1007/JHEP05(2023)137}{\emph{JHEP} {\bfseries 05}
  (2023) 137} [\href{https://arxiv.org/abs/2302.13587}{{\ttfamily
  2302.13587}}].

\bibitem{Rodas:2023gma}
{\scshape Hadron Spectrum} collaboration, A.~Rodas, J.~J. Dudek and R.~G.
  Edwards, \emph{{Quark mass dependence of \ensuremath{\pi}\ensuremath{\pi}
  scattering in isospin 0, 1, and 2 from lattice QCD}},
  \href{https://doi.org/10.1103/PhysRevD.108.034513}{\emph{Phys. Rev. D}
  {\bfseries 108} (2023) 034513}
  [\href{https://arxiv.org/abs/2303.10701}{{\ttfamily 2303.10701}}].

\bibitem{Rodas:2023nec}
{\scshape Hadron Spectrum} collaboration, A.~Rodas, J.~J. Dudek and R.~G.
  Edwards, \emph{{Determination of crossing-symmetric
  \ensuremath{\pi}\ensuremath{\pi} scattering amplitudes and the quark mass
  evolution of the \ensuremath{\sigma} constrained by lattice QCD}},
  \href{https://doi.org/10.1103/PhysRevD.109.034513}{\emph{Phys. Rev. D}
  {\bfseries 109} (2024) 034513}
  [\href{https://arxiv.org/abs/2304.03762}{{\ttfamily 2304.03762}}].

\bibitem{BaryonScatteringBaSc:2023ori}
{\scshape Baryon Scattering (BaSc)} collaboration, J.~Bulava et~al.,
  \emph{{Lattice QCD study of
  \ensuremath{\pi}\ensuremath{\Sigma}-K\textasciimacron{}N scattering and the
  \ensuremath{\Lambda}(1405) resonance}},
  \href{https://doi.org/10.1103/PhysRevD.109.014511}{\emph{Phys. Rev. D}
  {\bfseries 109} (2024) 014511}
  [\href{https://arxiv.org/abs/2307.13471}{{\ttfamily 2307.13471}}].

\bibitem{BaryonScatteringBaSc:2023zvt}
{\scshape Baryon Scattering (BaSc)} collaboration, J.~Bulava et~al.,
  \emph{{Two-Pole Nature of the \ensuremath{\Lambda}(1405) resonance from
  Lattice QCD}},
  \href{https://doi.org/10.1103/PhysRevLett.132.051901}{\emph{Phys. Rev. Lett.}
  {\bfseries 132} (2024) 051901}
  [\href{https://arxiv.org/abs/2307.10413}{{\ttfamily 2307.10413}}].

\bibitem{Wilson:2023anv}
{\scshape Hadron Spectrum} collaboration, D.~J. Wilson, C.~E. Thomas, J.~J.
  Dudek and R.~G. Edwards, \emph{{Charmonium \ensuremath{\chi}c0 and
  \ensuremath{\chi}c2 resonances in coupled-channel scattering from lattice
  QCD}}, \href{https://doi.org/10.1103/PhysRevD.109.114503}{\emph{Phys. Rev. D}
  {\bfseries 109} (2024) 114503}
  [\href{https://arxiv.org/abs/2309.14071}{{\ttfamily 2309.14071}}].

\bibitem{Wilson:2023hzu}
{\scshape Hadron Spectrum} collaboration, D.~J. Wilson, C.~E. Thomas, J.~J.
  Dudek and R.~G. Edwards, \emph{{Scalar and Tensor Charmonium Resonances in
  Coupled-Channel Scattering from Lattice QCD}},
  \href{https://doi.org/10.1103/PhysRevLett.132.241901}{\emph{Phys. Rev. Lett.}
  {\bfseries 132} (2024) 241901}
  [\href{https://arxiv.org/abs/2309.14070}{{\ttfamily 2309.14070}}].

\bibitem{Bulava:2023uma}
J.~Bulava et~al., \emph{{Low-lying baryon resonances from lattice QCD}},
  \href{https://doi.org/10.1393/ncc/i2024-24165-1}{\emph{Nuovo Cim. C}
  {\bfseries 47} (2024) 165}
  [\href{https://arxiv.org/abs/2310.08375}{{\ttfamily 2310.08375}}].

\bibitem{Skinner:2023wwb}
S.~Skinner, J.~Bulava, D.~Darvish, A.~D. Hanlon, B.~Hoerz, C.~Morningstar
  et~al., \emph{{Lattice QCD studies of the $\Delta$ baryon resonance and the
  $K_0^\ast(700)$ and $a_0(980)$ meson resonances: the role of exotic operators
  in determining the finite-volume spectrum}},
  \href{https://doi.org/10.22323/1.453.0074}{\emph{PoS} {\bfseries LATTICE2023}
  (2024) 074} [\href{https://arxiv.org/abs/2312.10184}{{\ttfamily
  2312.10184}}].

\bibitem{Bulava:2024bsi}
J.~Bulava et~al., \emph{{The $\Lambda$(1405) pole structure from Lattice QCD: A
  coupled-channel $\pi \Sigma$ \ensuremath{-} KN study}},
  \href{https://doi.org/10.1051/epjconf/202430301004}{\emph{EPJ Web Conf.}
  {\bfseries 303} (2024) 01004}.

\bibitem{Collins:2024sfi}
S.~Collins, A.~Nefediev, M.~Padmanath and S.~Prelovsek, \emph{{Toward the quark
  mass dependence of Tcc+ from lattice QCD}},
  \href{https://doi.org/10.1103/PhysRevD.109.094509}{\emph{Phys. Rev. D}
  {\bfseries 109} (2024) 094509}
  [\href{https://arxiv.org/abs/2402.14715}{{\ttfamily 2402.14715}}].

\bibitem{Yeo:2024chk}
{\scshape Hadron Spectrum} collaboration, J.~D.~E. Yeo, C.~E. Thomas and D.~J.
  Wilson, \emph{{DK/D\ensuremath{\pi} scattering and an exotic virtual bound
  state at the SU(3) flavour symmetric point from lattice QCD}},
  \href{https://doi.org/10.1007/JHEP07(2024)012}{\emph{JHEP} {\bfseries 07}
  (2024) 012} [\href{https://arxiv.org/abs/2403.10498}{{\ttfamily
  2403.10498}}].

\bibitem{Boyle:2024hvv}
P.~Boyle, F.~Erben, V.~G\"ulpers, M.~T. Hansen, F.~Joswig, N.~P. Lachini
  et~al., \emph{{Light and strange vector resonances from lattice QCD at
  physical quark masses}},  \href{https://arxiv.org/abs/2406.19194}{{\ttfamily
  2406.19194}}.

\bibitem{Whyte:2024ihh}
{\scshape Hadron Spectrum} collaboration, T.~Whyte, D.~J. Wilson and C.~E.
  Thomas, \emph{{Near-threshold states in coupled DD*-D*D* scattering from
  lattice QCD}}, \href{https://doi.org/10.1103/PhysRevD.111.034511}{\emph{Phys.
  Rev. D} {\bfseries 111} (2025) 034511}
  [\href{https://arxiv.org/abs/2405.15741}{{\ttfamily 2405.15741}}].

\bibitem{Dudek:2024roh}
{\scshape Hadron Spectrum} collaboration, J.~J. Dudek and C.~T. Johnson,
  \emph{{Coupled-channel J-- meson resonances from lattice QCD}},
  \href{https://doi.org/10.1103/PhysRevD.110.034512}{\emph{Phys. Rev. D}
  {\bfseries 110} (2024) 034512}
  [\href{https://arxiv.org/abs/2406.07261}{{\ttfamily 2406.07261}}].

\bibitem{Boyle:2024grr}
P.~Boyle, F.~Erben, V.~G\"ulpers, M.~T. Hansen, F.~Joswig, N.~P. Lachini
  et~al., \emph{{Physical-mass calculation of $\rho(770)$ and $K^*(892)$
  resonance parameters via $\pi \pi$ and $K \pi$ scattering amplitudes from
  lattice QCD}},  \href{https://arxiv.org/abs/2406.19193}{{\ttfamily
  2406.19193}}.

\bibitem{Yan:2024gwp}
H.~Yan, M.~Mai, M.~Garofalo, U.-G. Mei\ss{}ner, C.~Liu, L.~Liu et~al.,
  \emph{{\ensuremath{\omega} Meson from Lattice QCD}},
  \href{https://doi.org/10.1103/PhysRevLett.133.211906}{\emph{Phys. Rev. Lett.}
  {\bfseries 133} (2024) 211906}
  [\href{https://arxiv.org/abs/2407.16659}{{\ttfamily 2407.16659}}].

\bibitem{Dawid:2024dgy}
S.~M. Dawid, F.~Romero-L\'opez and S.~R. Sharpe, \emph{{Finite- and
  infinite-volume study of DD\ensuremath{\pi} scattering}},
  \href{https://doi.org/10.1007/JHEP01(2025)060}{\emph{JHEP} {\bfseries 01}
  (2025) 060} [\href{https://arxiv.org/abs/2409.17059}{{\ttfamily
  2409.17059}}].

\bibitem{Vujmilovic:2024snz}
I.~Vujmilovic, S.~Collins, L.~Leskovec, E.~Ortiz-Pacheco, P.~Madanagopalan and
  S.~Prelovsek, \emph{{$T^+_{cc}$ via the plane wave approach and including
  diquark-antidiquark operators}},
  \href{https://doi.org/10.22323/1.466.0112}{\emph{PoS} {\bfseries LATTICE2024}
  (2025) 112} [\href{https://arxiv.org/abs/2411.08646}{{\ttfamily
  2411.08646}}].

\bibitem{Francis:2024fwf}
A.~Francis, \emph{{Lattice perspectives on doubly heavy tetraquarks}},
  \href{https://doi.org/10.1016/j.ppnp.2024.104143}{\emph{Prog. Part. Nucl.
  Phys.} {\bfseries 140} (2025) 104143}.

\bibitem{Erben:2025zph}
F.~Erben, \emph{{From scattering towards multi-hadron weak decays}},  in
  \emph{{41st International Symposium on Lattice Field Theory}}, 1, 2025,
  \href{https://arxiv.org/abs/2501.19302}{{\ttfamily 2501.19302}}.

\bibitem{Dawid:2025doq}
S.~M. Dawid, Z.~T. Draper, A.~D. Hanlon, B.~H\"orz, C.~Morningstar,
  F.~Romero-L\'opez et~al., \emph{{Two- and three-meson scattering amplitudes
  with physical quark masses from lattice QCD}},
  \href{https://arxiv.org/abs/2502.17976}{{\ttfamily 2502.17976}}.

\bibitem{Lang:2025pjq}
N.~Lang and D.~J. Wilson, \emph{{$D_1$ and $D_2$ resonances in coupled-channel
  scattering amplitudes from lattice QCD}},
  \href{https://arxiv.org/abs/2502.04232}{{\ttfamily 2502.04232}}.

\bibitem{Dawid:2025zxc}
S.~M. Dawid, Z.~T. Draper, A.~D. Hanlon, B.~H\"orz, C.~Morningstar,
  F.~Romero-L\'opez et~al., \emph{{QCD predictions for physical multimeson
  scattering amplitudes}},  \href{https://arxiv.org/abs/2502.14348}{{\ttfamily
  2502.14348}}.

\bibitem{Jaffe:1976yi}
R.~L. Jaffe, \emph{{Perhaps a Stable Dihyperon}},
  \href{https://doi.org/10.1103/PhysRevLett.38.195}{\emph{Phys. Rev. Lett.}
  {\bfseries 38} (1977) 195}.

\bibitem{LHCb:2021vvq}
{\scshape LHCb} collaboration, R.~Aaij et~al., \emph{{Observation of an exotic
  narrow doubly charmed tetraquark}},
  \href{https://doi.org/10.1038/s41567-022-01614-y}{\emph{Nature Phys.}
  {\bfseries 18} (2022) 751}
  [\href{https://arxiv.org/abs/2109.01038}{{\ttfamily 2109.01038}}].

\bibitem{Du:2023hlu}
M.-L. Du, A.~Filin, V.~Baru, X.-K. Dong, E.~Epelbaum, F.-K. Guo et~al.,
  \emph{{Role of Left-Hand Cut Contributions on Pole Extractions from Lattice
  Data: Case Study for Tcc(3875)+}},
  \href{https://doi.org/10.1103/PhysRevLett.131.131903}{\emph{Phys. Rev. Lett.}
  {\bfseries 131} (2023) 131903}
  [\href{https://arxiv.org/abs/2303.09441}{{\ttfamily 2303.09441}}].

\bibitem{Sato:2007ms}
I.~Sato and P.~F. Bedaque, \emph{{Fitting two nucleons inside a box:
  Exponentially suppressed corrections to the Luscher's formula}},
  \href{https://doi.org/10.1103/PhysRevD.76.034502}{\emph{Phys. Rev. D}
  {\bfseries 76} (2007) 034502}
  [\href{https://arxiv.org/abs/hep-lat/0702021}{{\ttfamily hep-lat/0702021}}].

\bibitem{Meng:2021uhz}
L.~Meng and E.~Epelbaum, \emph{{Two-particle scattering from finite-volume
  quantization conditions using the plane wave basis}},
  \href{https://doi.org/10.1007/JHEP10(2021)051}{\emph{JHEP} {\bfseries 10}
  (2021) 051} [\href{https://arxiv.org/abs/2108.02709}{{\ttfamily
  2108.02709}}].

\bibitem{Meng:2023bmz}
L.~Meng, V.~Baru, E.~Epelbaum, A.~A. Filin and A.~M. Gasparyan, \emph{{Solving
  the left-hand cut problem in lattice QCD: Tcc(3875)+ from finite volume
  energy levels}},
  \href{https://doi.org/10.1103/PhysRevD.109.L071506}{\emph{Phys. Rev. D}
  {\bfseries 109} (2024) L071506}
  [\href{https://arxiv.org/abs/2312.01930}{{\ttfamily 2312.01930}}].

\bibitem{Bubna:2024izx}
R.~Bubna, H.-W. Hammer, F.~M\"uller, J.-Y. Pang, A.~Rusetsky and J.-J. Wu,
  \emph{{L\"uscher equation with long-range forces}},
  \href{https://doi.org/10.1007/JHEP05(2024)168}{\emph{JHEP} {\bfseries 05}
  (2024) 168} [\href{https://arxiv.org/abs/2402.12985}{{\ttfamily
  2402.12985}}].

\bibitem{Yu:2025gzg}
K.~Yu, G.-J. Wang, J.-J. Wu and Z.~Yang, \emph{{Hamiltonian Effective Field
  Theory for two-particle system containing long-range potential on the
  lattice}},  \href{https://arxiv.org/abs/2502.05789}{{\ttfamily 2502.05789}}.

\bibitem{Jackura:2023qtp}
A.~W. Jackura and R.~A. Brice\~no, \emph{{Partial-wave projection of the
  one-particle exchange in three-body scattering amplitudes}},
  \href{https://doi.org/10.1103/PhysRevD.109.096030}{\emph{Phys. Rev. D}
  {\bfseries 109} (2024) 096030}
  [\href{https://arxiv.org/abs/2312.00625}{{\ttfamily 2312.00625}}].

\bibitem{Briceno:2024ehy}
R.~A. Brice\~no, C.~S.~R. Costa and A.~W. Jackura, \emph{{Partial-wave
  projection of relativistic three-body amplitudes}},
  \href{https://arxiv.org/abs/2409.15577}{{\ttfamily 2409.15577}}.

\bibitem{Briceno:2024txg}
R.~A. Brice\~no, A.~W. Jackura, D.~A. Pefkou and F.~Romero-L\'opez,
  \emph{{Electroweak three-body decays in the presence of two- and three-body
  bound states}}, \href{https://doi.org/10.1007/JHEP05(2024)279}{\emph{JHEP}
  {\bfseries 05} (2024) 279}
  [\href{https://arxiv.org/abs/2402.12167}{{\ttfamily 2402.12167}}].

\bibitem{Hansen:2024ffk}
M.~T. Hansen, F.~Romero-L\'opez and S.~R. Sharpe, \emph{{Incorporating
  DD\ensuremath{\pi} effects and left-hand cuts in lattice QCD studies of the
  T$_{cc}$(3875)$^{+}$}},
  \href{https://doi.org/10.1007/JHEP06(2024)051}{\emph{JHEP} {\bfseries 06}
  (2024) 051} [\href{https://arxiv.org/abs/2401.06609}{{\ttfamily
  2401.06609}}].

\bibitem{Briceno:2015csa}
R.~A. Brice\~no and M.~T. Hansen, \emph{{Multichannel 0 $\to$ 2 and 1 $\to$ 2
  transition amplitudes for arbitrary spin particles in a finite volume}},
  \href{https://doi.org/10.1103/PhysRevD.92.074509}{\emph{Phys. Rev. D}
  {\bfseries 92} (2015) 074509}
  [\href{https://arxiv.org/abs/1502.04314}{{\ttfamily 1502.04314}}].

\bibitem{Jackura:2020bsk}
A.~W. Jackura, R.~A. Brice\~no, S.~M. Dawid, M.~H.~E. Islam and C.~McCarty,
  \emph{{Solving relativistic three-body integral equations in the presence of
  bound states}},
  \href{https://doi.org/10.1103/PhysRevD.104.014507}{\emph{Phys. Rev. D}
  {\bfseries 104} (2021) 014507}
  [\href{https://arxiv.org/abs/2010.09820}{{\ttfamily 2010.09820}}].

\bibitem{Dawid:2023jrj}
S.~M. Dawid, M.~H.~E. Islam and R.~A. Brice\~no, \emph{{Analytic continuation
  of the relativistic three-particle scattering amplitudes}},
  \href{https://doi.org/10.1103/PhysRevD.108.034016}{\emph{Phys. Rev. D}
  {\bfseries 108} (2023) 034016}
  [\href{https://arxiv.org/abs/2303.04394}{{\ttfamily 2303.04394}}].

\bibitem{Jacob:1959at}
M.~Jacob and G.~C. Wick, \emph{{On the General Theory of Collisions for
  Particles with Spin}},
  \href{https://doi.org/10.1006/aphy.2000.6022}{\emph{Annals Phys.} {\bfseries
  7} (1959) 404}.

\bibitem{Martin:1970hmp}
A.~D. Martin and T.~D. Spearman, \emph{{Elementary Particle Theory}}.
  North-Holland Publishing Co., Amsterdam, 1970.

\bibitem{Auvil:1966eao}
P.~R. Auvil and J.~J. Brehm, \emph{{Wave Functions for Particles of Higher
  Spin}}, \href{https://doi.org/10.1103/PhysRev.145.1152}{\emph{Phys. Rev.}
  {\bfseries 145} (1966) 1152}.

\bibitem{Durand:1962zza}
L.~Durand, P.~C. DeCelles and R.~B. Marr, \emph{{Lorentz Invariance and the
  Kinematic Structure of Vertex Functions}},
  \href{https://doi.org/10.1103/PhysRev.126.1882}{\emph{Phys. Rev.} {\bfseries
  126} (1962) 1882}.

\bibitem{Lorce:2009bs}
C.~Lorce, \emph{{Electromagnetic properties for arbitrary spin particles:
  Natural electromagnetic moments from light-cone arguments}},
  \href{https://doi.org/10.1103/PhysRevD.79.113011}{\emph{Phys. Rev. D}
  {\bfseries 79} (2009) 113011}
  [\href{https://arxiv.org/abs/0901.4200}{{\ttfamily 0901.4200}}].

\bibitem{Dawid:2024oey}
S.~M. Dawid, A.~W. Jackura and A.~P. Szczepaniak, \emph{{Finite-volume
  quantization condition from the $N/D$ representation}},
  \href{https://arxiv.org/abs/2411.15730}{{\ttfamily 2411.15730}}.

\end{thebibliography}\endgroup

\end{document}